\documentclass[10pt, final, twocolumn]{IEEEtran}
\usepackage{caption}
\usepackage{subcaption}
\usepackage{enumerate}
\usepackage{graphicx}
\usepackage{enumitem}
\usepackage{psfrag}
\usepackage{acronym}
\usepackage{tikz}
\usepackage{pgfplots}
\usepackage{xparse}
\usepackage{cite}
\usepackage{amsmath,amssymb,amsfonts}
\usepackage{graphicx}
\usepackage{xcolor}
\usepackage{colortbl} 
\usepackage{hhline} 
\usepackage{soul}
\usepackage{array}
\usepackage{verbatim}
\usepackage{bm}
\usepackage{multirow}
\usepackage{bbm}
\usepackage{tabularx}
\usepackage{amsthm} 
\usepackage{comment}
\usepackage{algorithmic}
\usepackage{algorithm}

\begin{document}

\title{Holographic Imaging with XL-MIMO and RIS: Illumination and Reflection Design}

\author{\IEEEauthorblockN{G. Torcolacci,~\IEEEmembership{Graduate Student Member,~IEEE}, A. Guerra,~\IEEEmembership{Member,~IEEE},  H. Zhang,~\IEEEmembership{Member,~IEEE}, \\ F. Guidi,~\IEEEmembership{Member,~IEEE}, Q. Yang,~\IEEEmembership{Graduate Student Member,~IEEE},  Y. C. Eldar,~\IEEEmembership{Fellow,~IEEE}, and \\ D. Dardari,~\IEEEmembership{Senior Member,~IEEE}} }

\markboth{Holographic Imaging with XL-MIMO and RIS: Illumination and Reflection Design}
{XX \MakeLowercase{\textit{et al.}}}

\maketitle

\begin{abstract}
This paper addresses a near-field imaging problem utilizing \ac{XL-MIMO} antennas and \acp{RIS} already in place for wireless communications. To this end, we consider a system with a fixed transmitting antenna array illuminating a \ac{ROI} and a fixed receiving antenna array inferring the \ac{ROI}'s scattering coefficients. Leveraging \ac{XL-MIMO} and high frequencies, the \ac{ROI} is situated in the radiative near-field region of both antenna arrays, thus enhancing the \ac{DoF} (i.e., the channel matrix rank) of the illuminating and sensing channels available for imaging, here referred to as \textit{holographic imaging}. To further boost the imaging performance, we optimize the illuminating waveform by solving a min-max optimization problem having the upper bound of the \ac{MSE} of the image estimate as the objective function. Additionally, we address the challenge of \ac{NLOS} scenarios by considering the presence of a \ac{RIS} and deriving its optimal reflection coefficients.  Numerical results investigate the interplay between illumination optimization, geometric configuration (monostatic and bistatic), the \ac{DoF} of the illuminating and sensing channels, image estimation accuracy, and image complexity.  
\end{abstract}

%
%



\acrodef{$P_{EM}$}{probability of emulation, or false alarm}
\acrodef{$P_{FA}$}{probability of false alarm}
\acrodef{$P_{MD}$}{probability of missed detection}
\acrodef{$P_{D}$}{probability of detection}

\acrodef{3D}{three-dimesional}
\acrodef{5G}{5th generation}
\acrodef{6G}{6th Generation}
\acrodef{ACF}{autocorrelation function}
\acrodef{ACG}{automatic	gain control}
\acrodef{ACI}{adjacent channel interference}
\acrodef{ACK}{acknowledge}
\acrodef{AcR}{autocorrelation receiver}
\acrodef{ADC}{analog-to-digital converter}
\acrodef{AF}{amplify \& forward}
\acrodef{AFL}{anchor-free localization}
\acrodef{AGNSS}{assisted-GNSS}
\acrodef{AGPS}{assisted GPS}
\acrodef{AI}{artificial intelligence}
\acrodef{AIC}{Akaike information criterion}
\acrodef{AO}{alternating optimization}
\acrodef{AOA}{angle-of-arrival}
\acrodef{AOD}{angle-of-departure}
\acrodef{AOT}{approximate optimum threshold}
\acrodef{AP}{access point}
\acrodef{API}{application programming interface}
\acrodef{ASK}{amplitude shift keying}
\acrodef{ASNR}{accumulated signal-to-noise ratio}
\acrodef{AUB}{asymptotic union bound}
\acrodef{AWGN}{additive white Gaussian noise}
\acrodef{BAN}{body area network}
\acrodef{BAV}{balanced antipodal Vivaldi}
\acrodef{BCH}{Bose Chaudhuri Hocquenghem}
\acrodef{BEP}{bit error probability}
\acrodef{BER}{bit error rate}
\acrodef{BF}{brute force}
\acrodef{BFC}{block fading channel}
\acrodef{BIC}{Bayesian information criterion}
\acrodef{BLUE}{best linear unbiased estimator}
\acrodef{BPAM}{binary pulse amplitude modulation}
\acrodef{BPF}{bandpass filter}
\acrodef{BPPM}{binary pulse position modulation}
\acrodef{bps}{bits per second}
\acrodef{BPSK}{binary phase shift keying}
\acrodef{BPZF}{band-pass zonal filter}
\acrodef{BS}{base station}
\acrodef{BSC}{binary symmetric channel}
\acrodef{BTB}{Bellini-Tartara bound}
\acrodef{c.c.d.f.}{complementary cumulative distribution function}
\acrodef{c.d.f.}{cumulative distribution function}
\acrodef{CAD}{computer-aided design}
\acrodef{CAIC}{consistent Akaike information criterion}
\acrodef{CAP}{continuous aperture phased}
\acrodef{CCF}{cross correlation function}
\acrodef{CCI}{co-channel interference}
\acrodef{CD}{cooperative diversity}
\acrodef{CDMA}{code division multiple access}
\acrodef{CEOT}{channel ensemble optimum threshold}
\acrodef{CEP}{codeword error probability}
\acrodef{CFAR}{constant	 false alarm rate}
\acrodef{ch.f.}{characteristic function}
\acrodef{CH}{cluster head}
\acrodef{CIR}{channel impulse response}
\acrodef{CL}{centroid localization}
\acrodef{CM}{channel model}
\acrodef{CNR}{clutter-to-noise ratio}
\acrodef{CP}{ciclic prefix}
\acrodef{CPR}{channel pulse response}
\acrodef{CR}{channel response}
\acrodef{CRB}{Cram\'{e}r-Rao bound}
\acrodef{CRC}{cyclic redundancy check}
\acrodef{CRLB}{Cram\'{e}r-Rao lower bound}
\acrodef{CS}{clock skew}
\acrodef{CSCG}{circularly symmetric complex Gaussian}
\acrodef{CSI}{channel state information}
\acrodef{CSMA}{carrier sense multiple access}
\acrodef{CSS}{chirp spread spectrum}
\acrodef{CTS}{clear-to-send}
\acrodef{CW}{continuous wave}
\acrodef{DAA}{detect and avoid}
\acrodef{DAB}{digital audio broadcasting}
\acrodef{DBB}{digital base band}
\acrodef{DBPSK}{differential binary phase shift keying}
\acrodef{DCM}{dual-carrier modulation}
\acrodef{DDP}{detected direct path}
\acrodef{DF}{detect \& forward}
\acrodef{DFMS}{monopole dual feed stripline antenna}
\acrodef{DGPS}{differential GPS}
\acrodef{DLL}{delay-locked loop}
\acrodef{DNN}{deep neural network}
\acrodef{DoD}{Department of Defense}
\acrodef{DoF}{degrees of freedom}
\acrodef{DP}{direct path}
\acrodef{DR}{detection rate}
\acrodef{DRT}{distance ratio test}
\acrodef{DS-SS}{direct-sequence spread-spectrum}
\acrodef{DS}{delay spread}
\acrodef{DTR}{differential transmitted-reference}
\acrodef{DTT}{Diffraction Tomography Theory}
\acrodef{DVB-H}{digital video broadcasting\,--\,handheld}
\acrodef{DVB-T}{digital video broadcasting\,--\,terrestrial}
\acrodef{e.m.}{electromagnetic}
\acrodef{ECC}{European Community Commission}
\acrodef{ED}{energy detector}
\acrodef{EDR}{energy detector receiver}
\acrodef{EFIM}{equivalent Fisher information matrix}
\acrodef{EIRP}{effective radiated isotropic power}
\acrodef{EKF}{extended Kalman filter}
\acrodef{KKT}{Karush–Kuhn–Tucker}
\acrodef{ELP}{equivalent low-pass}
\acrodef{EM}{electromagnetic}
\acrodef{EMCB}{extended Miller Chang bound}
\acrodef{EME}{minimum eigenvalue ratio detector}
\acrodef{EMI}{electromagnetic interference}
\acrodef{ENP}{estimated noise power}
\acrodef{ESA}{European Space Agency}
\acrodef{EU}{European Union}
\acrodef{EVD}{eigenvalue decomposition}
\acrodef{FAR}{false alarm rate}
\acrodef{FCC}{Federal Communications Commission}
\acrodef{FDMA}{frequency division multiple access}
\acrodef{FDMA}{frequency division multiple access}
\acrodef{FEC}{forward error correction}
\acrodef{FEC}{forward error correction}
\acrodef{FFD}{full function device}
\acrodef{FFR}{full function reader}
\acrodef{FF}{far-field}
\acrodef{FFT}{fast Fourier transform}
\acrodef{FG}{factor graph}
\acrodef{FH-SS}{frequency-hopping spread-spectrum}
\acrodef{FH}{frequency-hopping}
\acrodef{FIM}{Fisher information matrix}
\acrodef{FLL}{Frequency-locked loop}
\acrodef{FS}{frame synchronization}
\acrodef{FT}{Fourier Transform}
\acrodef{GA}{Gaussian approximation}
\acrodef{GD}{gradient descent}
\acrodef{GDOP}{geometric dilution of precision}
\acrodef{GLR}{generalized likelihood ratio}
\acrodef{GLRT}{generalized likelihood ratio test}
\acrodef{GML}{generalized maximum likelihood}
\acrodef{GPRS}{general packet radio service}
\acrodef{GPS}{global positioning system}
\acrodef{HAP}{high altitude platform}
\acrodef{HCRB}{hybrid Cram\'{e}r-Rao bound}
\acrodef{HDSA}{high-definition situation-aware}
\acrodef{Hi-RADIAL}{High-accuracy RAdio Detection, Identification, And Localization}
\acrodef{HMM}{hidden Markov model}
\acrodef{HPA}{high-power amplifier}
\acrodef{HPBW}{half power beam width}
\acrodef{HW}{hardware}
\acrodef{i.i.d.}{independent, identically distributed}
\acrodef{ICT}{information and communication technologies}
\acrodef{IE}{informative element}
\acrodef{IEEE}{Institute of Electrical and Electronics Engineers}
\acrodef{IF}{intermediate frequency}
\acrodef{IFFT}{inverse fast Fourier transform}
\acrodef{IMF}{ideal matched filter}
\acrodef{IMU}{inertial measurement unit}
\acrodef{INR}{interference-to-noise ratio}
\acrodef{INS}{inertial navigation system}
\acrodef{IoT}{Internet of things}
\acrodef{IIoT}{industrial Internet of things}
\acrodef{INS}{inertial navigation system}
\acrodef{IR-UWB}{impulse radio UWB}
\acrodef{IR}{impulse radio}
\acrodef{IRI}{inter-reader interference}
\acrodef{IRS}{intelligent reflecting surface} 
\acrodef{ISAC}{integrated sensing and communications}
\acrodef{ISI}{inter-symbol interference} 
\acrodef{isi}{intra-symbol interference} 
\acrodef{ISM}{industrial, scientific and medical}
\acrodef{ISNR}{interference-plus-signal-to-noise-ratio}
\acrodef{ISP}{inverse scattering problem}
\acrodef{IT}{interference temperature}
\acrodef{ITC}{information theoretic criteria}
\acrodef{JBSF}{jump back and search forward}
\acrodef{JF}{just forward}
\acrodef{KF}{Kalman filter}
\acrodef{KKT}{Karush–Kuhn–Tucker}
\acrodef{LDC}{low duty cycle}
\acrodef{LDPC}{low density parity check}
\acrodef{LEO}{localization error outage}
\acrodef{LG}{Laguerre-Gaussian}
\acrodef{LIS}{large intelligent surface}
\acrodef{LLR}{log-likelihood ratio}
\acrodef{LLRT}{log-likelihood ratio test}
\acrodef{LRT}{likelihood ratio test}
\acrodef{LNA}{low-noise amplifier}
\acrodef{LOS}{line-of-sight}
\acrodef{LRT}{likelihood ratio test}
\acrodef{LS}{least square}
\acrodef{LS}{least squares}
\acrodef{M-PSK}{$M$-ary phase shift keying}
\acrodef{M-QAM}{$M$-ary quadrature amplitude modulation}
\acrodef{m.g.f.}{moment generating function}
\acrodef{MAC}{medium access control}
\acrodef{MAE}{mean absolute error}
\acrodef{MAI}{multiple access interference}
\acrodef{MAN}{metropolitan area network}
\acrodef{MAP}{maximum a posteriori}
\acrodef{MB-OFDM}{multi-band OFDM}
\acrodef{MB-UWB}{multi-band UWB}
\acrodef{MB}{multi-band}
\acrodef{MC}{multi-carrier}
\acrodef{MCB}{Miller Chang bound}
\acrodef{MCRB}{modified Cram\'{e}r-Rao bound}
\acrodef{MDD}{minimum distance distribution}
\acrodef{MDL}{minimum description length}
\acrodef{MF}{matched filter}
\acrodef{MGF}{moment generating function}
\acrodef{MI}{mutual information}
\acrodef{MIMO}{multiple-input multiple-output}
\acrodef{MISO}{multiple-input single-output}
\acrodef{ML}{maximum likelihood}
\acrodef{MM}{min-max}
\acrodef{MME}{maximum-minimum eigenvalue ratio detector}
\acrodef{mm-waves}{Millimeter Waves}
\acrodef{MMSE}{minimum mean-square error}
\acrodef{MPC}{multipath component}
\acrodef{MRC}{maximal ratio combiner}
\acrodef{MS}{mobile station}
\acrodef{MSB}{most significant bit}
\acrodef{MSE}{mean squared error}
\acrodef{NMSE}{normalized mean squared error}
\acrodef{MSK}{minimum shift keying}
\acrodef{MUI}{multi-user interference}
\acrodef{MUR}{multistatic radar}
\acrodef{MVU}{minimum variance unbiased}
\acrodef{MZZB}{modified Ziv-Zakai bound}
\acrodef{NB}{narrowband}
\acrodef{NBI}{narrowband interference}
\acrodef{NEO}{navigation error outage}
\acrodef{NFER}{near-Þeld electromagnetic ranging}
\acrodef{NF}{near-field}
\acrodef{NFF}{near-field focused}
\acrodef{NL}{nonlinear}
\acrodef{NLOS}{non-line-of-sight}
\acrodef{NP}{Neyman-Pearson}
\acrodef{NTIA}{National Telecommunications and Information Administration}
\acrodef{NTP}{network time protocol}
\acrodef{OAM}{orbital angular momentum} 
\acrodef{OC}{optimum combining}
\acrodef{OFDM}{orthogonal frequency division multiplexing}
\acrodef{OOK}{on-off keying}
\acrodef{OP}{outage probability}
\acrodef{OT}{optimum threshold}
\acrodef{P-Max}{$P$-Max}  
\acrodef{p.d.f.}{probability density function}
\acrodef{p.m.f.}{probability mass function}
\acrodef{PA}{power amplifier}
\acrodef{PAM}{pulse amplitude modulation}
\acrodef{PAN}{personal area network}
\acrodef{PAR}{peak-to-average ratio}
\acrodef{P-CRLB}{Posterior Cramer-Rao Lower Bound}
\acrodef{PCA}{principal component analysis}
\acrodef{PD}{probability of detection}
\acrodef{PDP}{power delay profile}
\acrodef{PE}{probability of emulation}
\acrodef{PEB}{position error bound}
\acrodef{PEC}{perfect electric conductor}
\acrodef{PEP}{packet error probability}
\acrodef{PF}{particle filter}
\acrodef{PFA}{probability of false alarm}
\acrodef{PHY}{physical layer}
\acrodef{PL}{path-loss}
\acrodef{PLL}{phase-locked loop}
\acrodef{PMD}{probability of missed detection}
\acrodef{PN}{pseudo-noise}
\acrodef{PSF}{point spread function}
\acrodef{ppm}{part-per-million}
\acrodef{PPM}{pulse position modulation}
\acrodef{PR}{pseudo-random}
\acrodef{PRake}{partial rake}
\acrodef{PRF}{pulse repetition frequency}
\acrodef{PRP}{pulse repetition period}
\acrodef{PSD}{power spectral density}
\acrodef{PSEP}{pairwise synchronization error probability}
\acrodef{PSNR}{peak signal to noise ratio}
\acrodef{PSK}{phase shift keying}
\acrodef{PSVD}{product singular value decomposition}
\acrodef{PSWF}{prolate spheroidal wave function}
\acrodef{PU}{primary user}
\acrodef{QAM}{quadrature amplitude modulation}
\acrodef{QoS}{quality of service}
\acrodef{QPSK}{quadrature phase shift keying}
\acrodef{R.V.}{random variable}
\acrodef{RADAR}{radar}
\acrodef{RCS}{radar cross section}
\acrodef{RDL}{"random data limit"}
\acrodef{REM}{radio environment map}
\acrodef{REO}{ranging error outage}
\acrodef{RF}{radio-frequency}
\acrodef{RFID}{radio-frequency identification}
\acrodef{RFR}{reduced function reader}
\acrodef{RFT}{reduced function tag}
\acrodef{RII}{ranging information intensity}
\acrodef{RIS}{reconfigurable intelligent surface}
\acrodef{rms}{root mean square}
\acrodef{RMSE}{root-mean-square error}
\acrodef{ROC}{receiver operating characteristic}
\acrodef{ROI}{region of interest}
\acrodef{RRC}{root raised cosine}
\acrodef{RSN}{radar sensor network}
\acrodef{RSS}{received signal strength}
\acrodef{RSSI}{received signal strength indicator}
\acrodef{RTLS}{real time locating systems}
\acrodef{RTT}{round-trip time}
\acrodef{S-V}{Saleh-Valenzuela}
\acrodef{SA}{simulated annealing}
\acrodef{SaG}{stop-and-go}
\acrodef{SAR}{synthetic aperture radar}
\acrodef{SBS}{serial backward search}
\acrodef{SBSMC}{serial backward search for multiple clusters}
\acrodef{SCM}{supply chain management}
\acrodef{SCR}{signal-to-clutter ratio}
\acrodef{SEP}{symbol error probability}
\acrodef{SIS}{small intelligent surface}
\acrodef{SFD}{start frame delimiter}
\acrodef{SIMO}{single-input multiple-output}
\acrodef{SINR}{signal-to-interference plus noise ratio}
\acrodef{SIR}{signal-to-interference ratio}
\acrodef{SISO}{single-input single-output}
\acrodef{SNR}{signal-to-noise ratio}
\acrodef{SoC}{system on chip}
\acrodef{SoO}{signal of opportunity}
\acrodef{SoP}{system on package}
\acrodef{SOT}{sub-optimum threshold}
\acrodef{SPAWN}{sum-product algorithm over a wireless network}
\acrodef{SPEB}{squared position error bound}
\acrodef{SPMF}{single-path matched filter}
\acrodef{SQNR}{signal-to-quantization-noise ratio}
\acrodef{SRE}{smart radio environment}
\acrodef{SS}{spread spectrum}
\acrodef{ST}{simple thresholding}
\acrodef{SU}{secondary user}
\acrodef{SVD}{singular value decomposition}
\acrodef{SW}{software}
\acrodef{SW}{sync word}
\acrodef{TDE}{time delay estimation}
\acrodef{TDL}{tapped delay line}
\acrodef{TDMA}{time division multiple access}
\acrodef{TDOA}{time difference-of-arrival}
\acrodef{TH}{time-hopping}
\acrodef{THz}{TeraHertz}
\acrodef{TNR}{threshold-to-noise ratio}
\acrodef{TOA}{Time-of-arrival}
\acrodef{TOF}{time-of-flight}
\acrodef{TPC}{transmit power control}
\acrodef{TR}{transmitted-reference}
\acrodef{TS}{tabu search}
\acrodef{TSVD}{truncated singular value decomposition}
\acrodef{TV}{total variation denoising}
\acrodef{UAV}{unmanned aerial vehicle}
\acrodef{UB}{union bound}
\acrodef{UCA}{uniform circular array}
\acrodef{UDP}{undetected direct path}
\acrodef{UE}{User Equipment}
\acrodef{UHF}{ultra-high frequency}
\acrodef{ULP}{user location protocol}
\acrodef{UMP}{uniformly most powerful}
\acrodef{UMPI}{uniformly most powerful invariant}
\acrodef{URA}{uniform rectangular array}
\acrodef{UT}{user terminal}
\acrodef{UTC}{coordinated universal time}
\acrodef{UTM}{universal transverse Mercator}
\acrodef{UTRA}{UMTS terrestrial radio access}
\acrodef{UAV}{unmanned aerial vehicle}
\acrodef{UUV}{unmanned underwater vehicle}
\acrodef{UWB}{ultrawide-band}
\acrodef{UWBcap}[UWB]{Ultrawide band}
\acrodef{VFIL}{virtual force iterative localization}
\acrodef{VGA}{variable-gain amplifier}
\acrodef{VNA}{vector network analyzer}
\acrodef{WAF}{wall attenuation factor}
\acrodef{WB}{wideband}
\acrodef{WBI}{wideband interference}
\acrodef{WCL}{weighted centroid localization}
\acrodef{WED}{wall extra delay}
\acrodef{WiMAX} {worldwide interoperability for microwave access}
\acrodef{WLAN}{wireless local area network}
\acrodef{WLS}{weighted least squares}
\acrodef{WMAN}{wireless metropolitan area network}
\acrodef{WPAN}{wireless personal area networks}
\acrodef{WRAPI}{wireless research application programming interface}
\acrodef{WSN}{wireless sensor network}
\acrodef{WSR}{wireless sensor radar}
\acrodef{WSS}{wide-sense stationary}
\acrodef{WWB}{Weiss-Weinstein bound}
\acrodef{WWLB}{Weiss-Weinstein lower bound}
\acrodef{ZZB}{Ziv-Zakai bound}
\acrodef{ZZLB}{Ziv-Zakai lower bound}
\acrodef{XL-MIMO}{extremely large-scale multiple-input multiple-output}


%

\newcommand{\rect}[1] {\text{rect} \left ({#1} \right )}
\newcommand{\sinc}[1] {\text{sinc} \left ({#1} \right )}
\newcommand{\argmax}[1]{\underset{{#1}}{\operatorname{argmax}}}
\newcommand{\argmin}[1]{\underset{{#1}}{\operatorname{argmin}}}
\newcommand{\E}[1] {\mathbb{E}\left\{#1\right\}}
\newcommand{\Real}[1]{\Re^{#1}}
\newcommand{\floor}[1] {f \left ({#1} \right )}
\def\erfc{{\text{erfc}}}
\def\erf{{\text{erf}}}
\def\inverfc{{\text{inverfc}}}
\newcommand{\rank}{{\rm rank}}
\newcommand{\diag}{{\rm diag}}
\newcommand{\degree}{\ensuremath{^\circ}}
\newcommand{\ra}{\rightarrow}
\newcommand{\rf}{\leftarrow}
\newcommand{\cn}{{\mathcal{CN}}} 
\newcommand{\tr}{\operatorname{tr}}
\newcommand{\minimize}[1]{\underset{{#1}}{\operatorname{minimize}}}  
\newcommand{\maximize}[1]{\underset{{#1}}{\operatorname{maximize}}}  

\newcommand{\SNR}{\text{SNR}}
\newcommand{\TNR}{\mathsf{TNR}}
\newcommand{\sigmaN} {\sigma_{\text{N}}}
\newcommand{\MSE} {\text{MSE}}

\newcommand{\bolda}{{\bf a}}
\newcommand{\boldb}{{\bf b}}
\newcommand{\boldbeta}{{\boldsymbol{\beta}}}
\newcommand{\boldc}{{\bf c}}
\newcommand{\boldd}{{\bf d}}
\newcommand{\bolde}{{\bf e}}
\newcommand{\boldf}{{\bf f}}
\newcommand{\boldg}{{\bf g}}
\newcommand{\boldh}{{\bf h}}
\newcommand{\boldi}{{\bf i}}
\newcommand{\boldj}{{\bf j}}
\newcommand{\boldk}{{\bf k}}
\newcommand{\boldl}{{\bf l}}
\newcommand{\boldm}{{\bf m}}
\newcommand{\boldn}{{\bf n}}
\newcommand{\boldo}{{\bf o}}
\newcommand{\boldp}{{\bf p}}
\newcommand{\boldq}{{\bf q}}
\newcommand{\boldr}{{\bf r}}
\newcommand{\bolds}{{\bf s}}
\newcommand{\boldsp} {{\bf s}^{\prime}}
\newcommand{\boldt}{{\bf t}}
\newcommand{\boldu}{{\bf u}}
\newcommand{\boldv}{{\bf v}}
\newcommand{\boldw}{{\bf w}}
\newcommand{\boldx}{{\bf x}}
\newcommand{\boldy}{{\bf y}}
\newcommand{\boldz}{{\bf z}}

\newcommand{\boldA}{{\bf A}}
\newcommand{\boldB}{{\bf B}}
\newcommand{\boldC}{{\bf C}}
\newcommand{\boldD}{{\bf D}}
\newcommand{\boldE}{{\bf E}}
\newcommand{\boldF}{{\bf F}}
\newcommand{\boldG}{{\bf G}}
\newcommand{\boldH}{{\bf H}}
\newcommand{\boldI}{{\bf I}}
\newcommand{\boldJ}{{\bf J}}
\newcommand{\boldK}{{\bf K}}
\newcommand{\boldL}{{\bf L}}
\newcommand{\boldM}{{\bf M}}
\newcommand{\boldN}{{\bf N}}
\newcommand{\boldO}{{\bf O}}
\newcommand{\boldP}{{\bf P}}
\newcommand{\boldQ}{{\bf Q}}
\newcommand{\boldR}{{\bf R}}
\newcommand{\boldS}{{\bf S}}
\newcommand{\boldT}{{\bf T}}
\newcommand{\boldU}{{\bf U}}
\newcommand{\boldV}{{\bf V}}
\newcommand{\boldW}{{\bf W}}
\newcommand{\boldX}{{\bf X}}
\newcommand{\boldY}{{\bf Y}}
\newcommand{\boldZ}{{\bf Z}}

\newcommand{\stx}{S_{\text{T}}}
\newcommand{\srx}{S_{\text{R}}}
\newcommand{\lt}{L_{\text{T}}}
\newcommand{\lr}{L_{\text{R}}}
\newcommand{\rhot}{\rho_{\text{T}}}
\newcommand{\phit}{\varphi_{\text{T}}}
\newcommand{\rhor}{\rho_{\text{R}}}
\newcommand{\phir}{\varphi_{\text{R}}}
\newcommand{\Pt}{P_{\text{T}}}
\renewcommand{\Pr}{P_{\text{R}}}
\newcommand{\Nt}{N_{\text{T}}}
\newcommand{\Nr}{N_{\text{R}}}
\newcommand{\Gt}{{\bf G_{\text{T}}}}
\newcommand{\Gr}{{\bf G_{\text{R}}}}

\newcommand{\lambdaA}{\lambda_{\boldA}}
\newcommand{\vA}{\mathbf{v}_{\mathbf{A}}}
\newcommand{\gammastar}{{\boldsymbol{\gamma}}^\star}

\newcommand{\Ndof}{N_{\text{DOF}}}

\newcommand{\bbeta}{{\bm {\beta}}}
\newcommand{\bchi}{{\bm{\chi}}}
\newcommand{\bsigma}{{\bm{\Sigma}}}
\newcommand{\blambda}{{\bm{\Lambda}}}
\newcommand{\invsigma}{{\bm{\Sigma}}^{-1}}
\newcommand{\pinvsigma}{{\bm{\Sigma}}^{\dagger}}

\newcommand{\tbw}{\tilde{\boldw}}

\newcommand{\tbsigma}{\tilde{\bm{\Sigma}}}
\newcommand{\hbbeta}{{\hat{\boldsymbol{\beta}}}}
\newcommand{\betan}{\beta_n}
\newcommand{\xin}{\xi_n}
\newcommand{\sumn}{\sum_{n=1}^\infty}
\newcommand{\tildexn}{\tilde{x}_n}
\newcommand{\xn}{x_n}
\newcommand{\xns}{x_n^\star}
\newcommand{\lambdas}{\lambda^\star}
\newcommand{\fz}{f_0}
\newcommand{\fo}{f_1}

\newcommand{\gtni}{g_{\text{T}, n, i}}
\newcommand{\grrn}{g_{\text{R}, r, n}}
\newcommand{\Thetain}{\boldsymbol{\Theta}_{i,n}}
\newcommand{\Thetanr}{\boldsymbol{\Theta}_{n, r}}

\newcommand{\stni}{s_{\text{T}, n, i}}
\newcommand{\srrn}{s_{\text{R}, r, n}}
\acresetall

\bstctlcite{IEEEexample:BSTcontrol}

\section{Introduction}\label{sec: Introduction}

The rapid progress of wireless communication systems has laid the foundation for the forthcoming generation of networks, referred to as \ac{6G} systems, which will integrate and synergize localization, sensing, and communications. This convergence is commonly referred to as \ac{ISAC} \cite{SarEtAl:J20, liu2023seventy}. 
These capabilities are enabled by high-frequency bands and electrically large antenna arrays, e.g., based on metasurfaces and \ac{XL-MIMO} antennas \cite{BoyEtAL:J21, FaeEtAl:J19,wang2023extremely,lu2023tutorial}. As a result, \ac{6G} systems are expected to mainly operate in the radiative near-field region, enabling unprecedented levels of communication and sensing performance, flexibility, and resolution~\cite{guidi2021radio,ZhangetAl:J22,BjoEtAl:J19}. 
In this context, significant research contributions have shed light on the potential of holographic communications \cite{Dar:J20,HaiyetAl:J22,  AnEtAl1:J23,DarDec:J20}, localization \cite{palmucci2023two,ElzEtAl:J21,GueEtAl:J21, HeEtAl:J23}, sensing \cite{liu2022integrated}, and imaging \cite{GuiGueDar:J15}.  

While the benefits of operating in the near-field propagation regime have been extensively demonstrated for localization, communication, and sensing, the potential advantages of performing holographic imaging within wireless communication networks have been largely unexplored to date. This work aims to investigate the capabilities of near-field imaging, shedding light on its potential in \ac{6G} scenarios where \acp{RIS} are also employed to cope with \ac{NLOS} channel conditions \cite{huang2019reconfigurable, renzo2019smart}. 

With \textit{holographic imaging}, we hereby refer to the possibility of estimating the reflective properties of a \ac{ROI} using \ac{XL-MIMO} systems operating in the near-field. The procedure involves initially illuminating the \ac{ROI}, represented as a pixel-based image, and then capturing its backscattered \ac{EM} field through the receiving array. Thanks to the near-field propagation regime, which allows for the extraction of both depth and angular information compared to the far-field and hence increasing the number of exploitable \ac{DoF} \cite{gong2023holographic}, more informative measurements can be collected from the \ac{ROI}, thus leading to improved imaging capabilities.

Conventional radio imaging techniques, typically employed for medicine, biology, geology, and engineering, use \ac{EM} fields to create 3D images of physical entities using dedicated infrastructures. The common approach is to address an \ac{ISP}, where the primary objective is to extract features related to potential scatterers in a given \ac{ROI} by analyzing the scattered \ac{EM} field \cite{Chen:B18, BerBocDe:B21}.
The primary challenge in such problems lies in their ill-posed nature, necessitating the incorporation of regularization techniques for their effective resolution \cite{CroEtAl:J12}.
Various technologies and transmission techniques have been systematically explored to minimize reconstruction errors and enhance target images' resolution. 
Examples include \ac{SAR}
\cite{LopFor:J00, CafetAl:J91, DodTri:C18, BroetAl:J98}, classic holography \cite{SunetAl:J14, ZhoAlCh:J18}, {MIMO} antennas \cite{GaoetAl:J18, ZhuYar:J12, MaetAL:J11}, computational imaging \cite{MaitEtAl:J18, KhareetAl:J15, CosetAl:J12}, and others \cite{MastetAl:J98, CuiTri:22, Wal:J80}. Notably, many specialized algorithms have been devised for image retrieval, each exhibiting advantages and disadvantages. For instance, a range of methodologies exist in the domain of \ac{SAR}-based methods, including back-propagation \cite{yegulalp:C99, NaetAl:C06}, range migration \cite{LopFor:J00, PratietAl:C91}, range-Doppler \cite{CafetAl:J91}, chirp-scaling \cite{RanetAl:94}, as well as inverse/stripmap/tomographic/spotlight \ac{SAR} algorithms \cite{FortunetAL:B94, WangetAl:J18, DueretAl:J15}. Furthermore, a multitude of widely adopted strategies and approximations, such as Stolt interpolation \cite{Stolt:J78}, stationary phase method \cite{AhmedetAl:B14}, and many others, assume pivotal roles in enhancing imaging precision and dependability across diverse applications.
Deep learning techniques have also recently been applied to near-field imaging in \cite{MinOrOk:23, CheIhaLiuHao:20, Wan:22}. 

Regarding the operational frequencies and devices, imaging is frequently performed using microwaves or visible light technologies, like lidars \cite{KirEtAl:C09, gong2016three}.
In this regard, only a few works have explored the use of systems primarily designed for wireless communications to perform imaging of unknown objects \cite{BroetAl:J98, MamArbMad:C19, PieSol:J98, BucCrocIse:J99, CuiTri:22, luo2024integrated, he2022high}.
Indeed, most imaging approaches proposed in the literature require illuminating and sensing the backscattered \ac{EM} field from a large set of angles through the deployment of a dedicated imaging infrastructure typically working in the far-field region. Hence, their applicability in next-generation wireless systems endowed with \ac{ISAC} capabilities is limited.

When addressing \ac{NLOS} imaging, methods have been devised to effectively illuminate stationary objects, such as walls, to enable target reconstruction \cite{OtoLinWet:J18}. In light of the foreseen \ac{6G} scenarios, we explore the possibility of using \acp{RIS} for holographic imaging in \ac{LOS} conditions by optimizing their reflective characteristics to control the propagation channel \cite{YanZha:C20} and backscatter the illumination towards the \ac{ROI}. 
In \cite{YanZha:C20}, the authors present an approach for computational imaging using a \ac{RIS}, based on distributed antenna systems and stochastic modulation of detecting signals. Their work predominantly operates within the far-field regime, thus neglecting the large number \ac{DoF} available in the near-field region facilitated by the spherical \ac{EM} wavefront propagation. 
In \cite{JiangEtAl:J23}, the authors present a method for near-field computational imaging that integrates a \ac{RIS} with holographic aperture technology, where the \ac{RIS} is used to generate multiple virtual \ac{EM} masks on the target \ac{ROI}. This \ac{RIS} operates as an active reflector, amplifying and reflecting the impinging signals, thus requiring a higher complexity and elevated hardware costs.

To the authors' best knowledge, investigation of illumination and \ac{RIS} optimization strategies when operating in the near-field region has not yet been tackled in the literature. 
To fill this gap, in our paper, we consider near-field imaging (\textit{holographic imaging}) leveraging \ac{XL-MIMO}, \ac{RIS}, and high-frequency bands within the context of next-generation wireless systems.
In particular, we propose an analytical framework that captures the distinctive features of the available \ac{DoF} of the near-field channel and the presence of a \ac{RIS} to enhance imaging performance in \ac{NLOS} scenarios. 
Our main contributions are as follows.

\begin{itemize}
\item \textit{Illumination Waveform Design}: 
We enhance holographic imaging by optimizing the transmitting waveform for improved \ac{ROI} illumination, particularly with signals tailored for wireless communications and near-field propagation. We propose an optimization approach that searches for the illumination signal minimizing the \ac{MSE} of the image estimate. To this end, we first derive a closed-form expression for the \ac{MSE} in image estimation. Then, we perform a min-max operation to minimize the \ac{MSE} over the illumination signal by considering the maximum value for the scattering coefficients characterizing the \ac{ROI}, hence determining the most effective illumination signal to employ at the transmitter.

\item \textit{RIS Configuration Design}: We extend our model to incorporate the presence of a passive \ac{RIS} enabling monostatic imaging in \ac{NLOS} situations between the transmitter/receiver and the \ac{ROI} (“\textit{see around the corner}"). 
We investigate the optimal \ac{RIS} configuration maximizing the \ac{DoF} of the cascade channel between the transmitting/receiving antenna, \ac{RIS}, and the \ac{ROI}, thereby maximizing the total cascade channel's gain and reducing the \ac{MSE} of the image estimate. Moreover, we derive a closed-form expression for the optimal \ac{RIS} phase profile configuration for performing imaging in the absence of visibility between the transmitter and the \ac{ROI}.

\item \textit{Numerical Analysis}: We corroborate the theoretical findings through numerical simulations investigating the interplay between the \ac{DoF} of the near-field illuminating and sensing channels, the geometry of the system (monostatic \textit{vs}. bistatic configuration), the illumination optimization strategy, the \ac{RIS} configuration, the image estimation accuracy, and image complexity.
We show how the performance can be significantly improved when the number of \ac{DoF} of the channel is larger than the dimensionality of the image and a suitable illuminating waveform is employed. Moreover, results indicate that optimization of the illuminating signal is effective only when the transmitting antenna is in strong near-field conditions with respect to the \ac{ROI}. Finally, we demonstrate that imaging is possible in \ac{NLOS} only when the \ac{RIS} is optimally configured.
\end{itemize}

The remainder of the paper is organized as follows. Sec.~\ref{sec: SystemModel} introduces the system model for holographic imaging. In Sec.~\ref{sec: LOSImaging}, we discuss the \ac{ISP} analytical formulation and feasible relaxation techniques, while Sec.~\ref{sec:OptimizProblem} deals with the optimization of the transmitted illuminating signal in \ac{LOS} configurations. Sec.~\ref{sec: RISaidedImag} extends the presented analysis to the case of a \ac{RIS}-aided system working in \ac{NLOS}, and Sec.~\ref{sec: NumericalResults} illustrates the obtained numerical results. Finally, Sec.~\ref{sec: Conclusion} concludes the paper.

\paragraph*{Notation}
Throughout the paper, we use the following notation. Lowercase bold variables, e.g., $\mathbf{x}$, denote vectors in the \ac{3D} space. Boldface capital letters denote matrices, e.g., $\mathbf{X}$. The identity and zero matrices with size $N \times M$ are written as $\mathbf{I}_{N \times M}$ and $\mathbf{0}_{N \times M}$. The transpose operator is indicated by $(\cdot)^{T}$, the Hermitian operator is $(\cdot)^{H}$, and the Moore-Penrose pseudoinverse operator is represented by $(\cdot)^{\dagger}$. The $\mathcal{L}_2$-norm of a vector $\mathbf{r}$ is $\|\mathbf{r}\|$, the Frobenius norm of a matrix $\mathbf{X}$ is $\left\| \mathbf{X} \right\|_{\mathrm{F}}$ , and $j$ is the imaginary unit. Calligraphic fonts are used to denote sets, i.e., $\mathcal{X}$, while $ \mathbf{x} \sim \cn (\boldsymbol{\mu}, \boldsymbol{\Sigma})$ is a complex random vector distributed according to a complex normal distribution with mean vector $\boldsymbol{\mu}$ and covariance matrix $\boldsymbol{\Sigma}$. The notation $\operatorname{diag}\left( \mathbf{x}\right)$ denotes an operator that generates a diagonal matrix whose main diagonal is given by the vector $\mathbf{x}$.
Finally, $\sigma_1(\mathbf{A})\ge \sigma_2(\mathbf{A}) \ge \ldots \ge \sigma_K(\mathbf{A})$ denote the singular values of the matrix $\mathbf{A}\in \mathbb{C}^{N \times M}$, where  $K=\min(N,M)$.

\section{System Model}
\label{sec: SystemModel}

\begin{figure*}[t!] 
  \centering
  \begin{subfigure}{0.5\textwidth} 
   \includegraphics[width=\textwidth, keepaspectratio, trim=0mm 0mm 0mm 0mm,  clip]{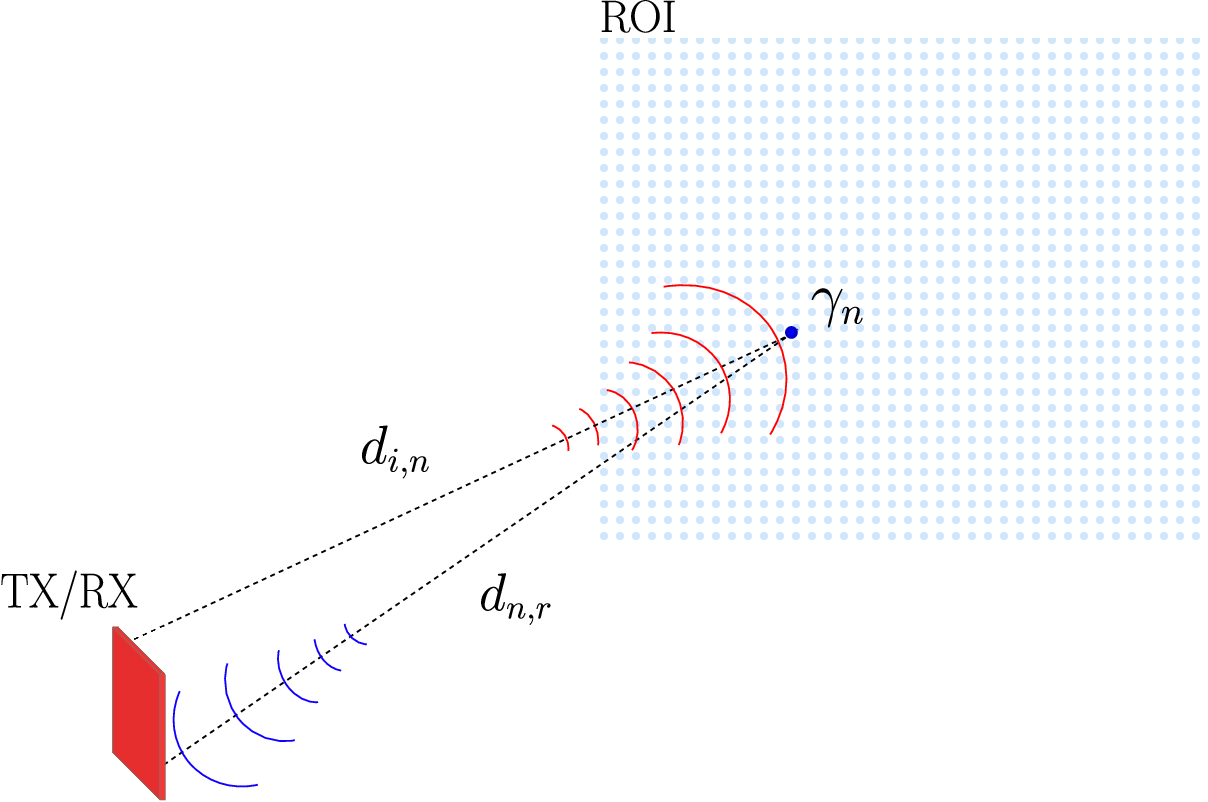} 
    \caption{}\label{fig:monostatic_NORIS}
  \end{subfigure}%
 \begin{subfigure}{0.5\textwidth} 
\includegraphics[width=\textwidth, keepaspectratio, trim=0mm 0mm 0mm 0mm,  clip]{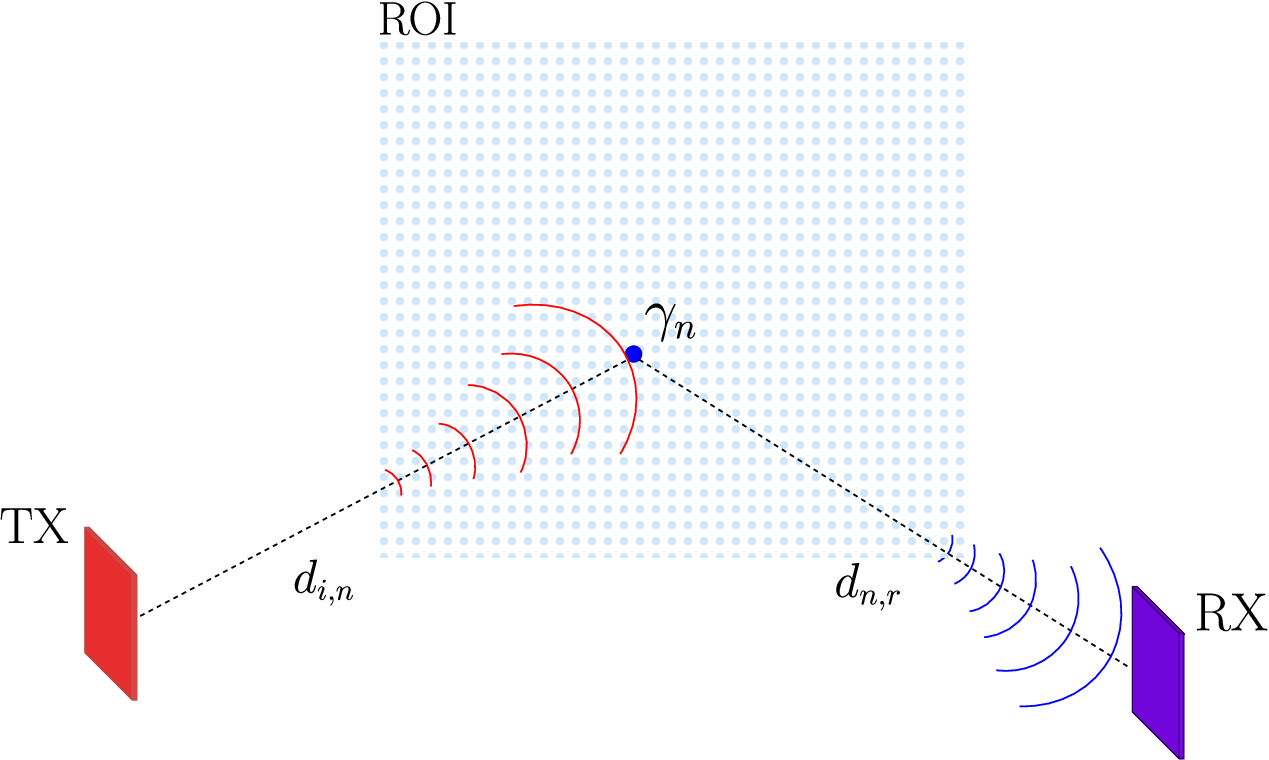}  
    \caption{}\label{fig:bistatic_LOS}
  \end{subfigure}%
  \caption{(a) Monostatic configuration with a single XL-MIMO transceiver (TX/RX) performing \ac{LOS} imaging of a \ac{ROI}. The generic target scattering point is indicated with $\gamma_n$. The distances between the XL-MIMO transceiver elements and the scattering points are indicated with $d_{i,n}$ and $d_{n,r}$. (b) Bistatic setup, where the transmitting antenna (TX) illuminates the selected \ac{ROI} and a distinct receiving antenna (RX) collects the scattered \ac{EM} signal. Both the TX and RX are modeled as XL-MIMO arrays.}\label{fig:SystemScenario}
\end{figure*}

\begin{figure}[t!]
    \centering
    \includegraphics[width=\columnwidth, keepaspectratio, trim=0mm 0mm 0mm 0mm, clip]{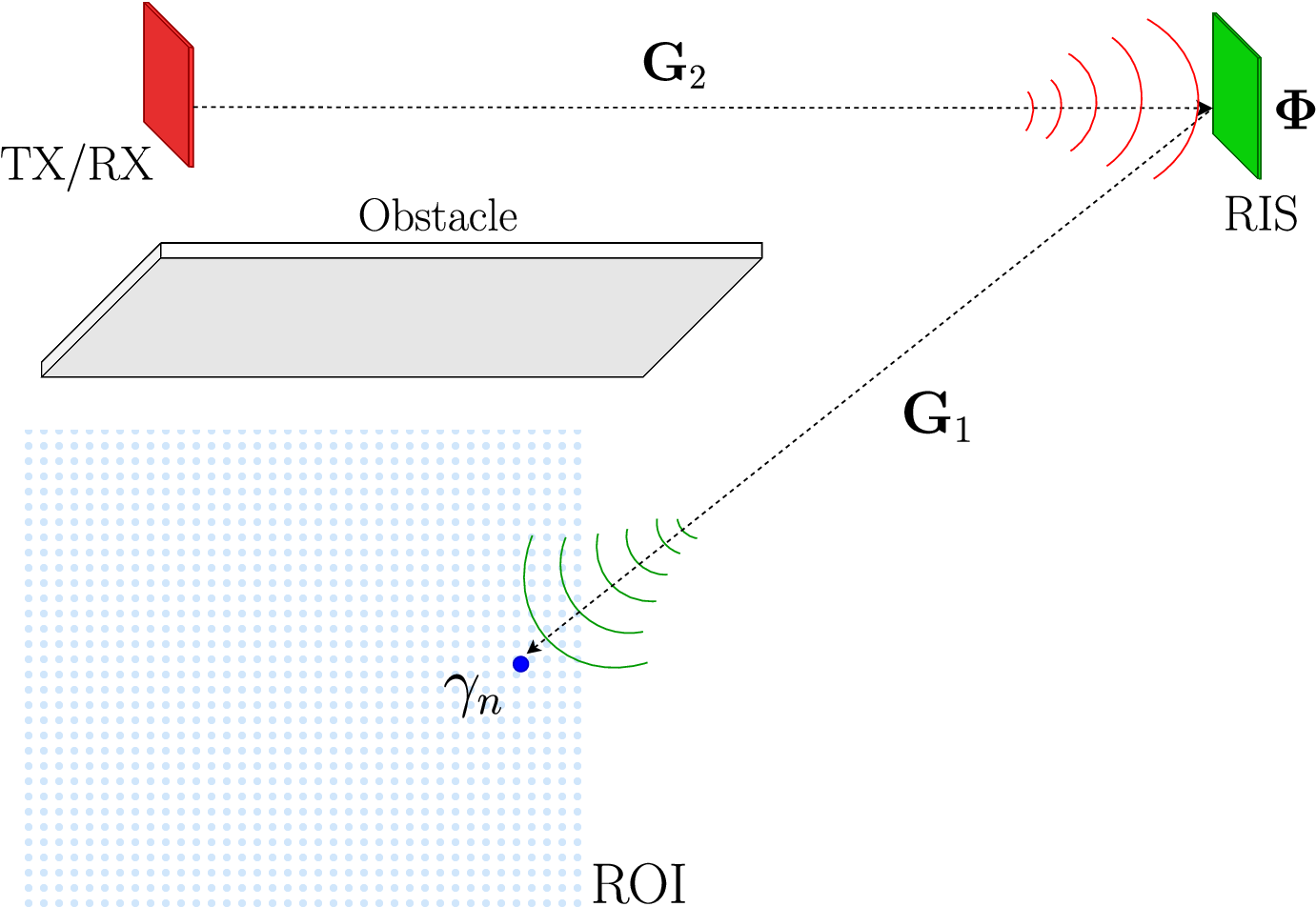}
    \caption{Monostatic \ac{RIS}-aided setup for \ac{NLOS} imaging of a \ac{ROI}, where
    the XL-MIMO transceiver (TX/RX) illuminates the \ac{RIS}, the \ac{RIS} reflects the impinging \ac{EM} signal toward the  \ac{ROI} which reflects it back towards the \ac{RIS} and the XL-MIMO transceiver. The TX/RX-RIS and RIS-ROI channels are indicated with $\mathbf{G}_1$ and $\mathbf{G}_2$, and the RIS reflection matrix with $\mathbf{\Phi}$, respectively.  }
    \label{fig:SystemScenariowithRIS}
\end{figure}

Consider a holographic imaging system as illustrated in Fig.~\ref{fig:SystemScenario}. We discuss three distinct configurations: (\textit{i}) a \textit{monostatic setting} (depicted in Fig.~\ref{fig:monostatic_NORIS}), wherein the transmitting and receiving discrete antenna arrays are colocated and are in a free-space \ac{LOS} condition in relation to the \ac{ROI}\footnote{Throughout the remainder of the paper, \textit{\ac{LOS} imaging} will henceforth denote the imaging procedure conducted under free-space \ac{LOS} conditions.}; (\textit{ii}) a \textit{bistatic setting} (depicted in Fig.~\ref{fig:bistatic_LOS}), where the two \ac{XL-MIMO} antennas are spatially separated while still maintaining \ac{LOS} with the \ac{ROI} under investigation; (\textit{iii}) a \textit{\ac{RIS}-aided monostatic} scenario in \ac{NLOS} condition (shown in Fig.~\ref{fig:SystemScenariowithRIS}). Specifically, with \ac{NLOS}, we here refer to the case where the direct path between the \ac{XL-MIMO} transceiver and the \ac{ROI} to be imaged is obstructed, whereas there exists a free-space \ac{LOS} condition between the \ac{RIS} and both the \ac{ROI} and the transceiver. In this manner, we consider that the overall \ac{XL-MIMO} transceiver-\ac{ROI} link is a concatenation of two free-space \ac{LOS} channels connected through the \ac{RIS}. The latter scenario with the \ac{RIS} will be discussed in Sec.~\ref{sec: RISaidedImag} while, in the sequel of this section, we will focus on the first two configurations.

The transmitting antenna array (TX) is composed of $\Nt$ antenna elements located at $\mathbf{p}_{\text{T}, i}=\left[x_{\text{T}, i},\, y_{\text{T}, i},\, z_{\text{T}, i} \right]^T$, with $i \in \mathcal{N}_{\text{T}}=\left\{1, 2, \ldots, \Nt \right\}$,  and illuminates the \ac{ROI} with a signal having wavelength $\lambda$.
The \ac{ROI} is observed by a receiving antenna array (RX) equipped with $\Nr$ antenna elements, each of them located at $\mathbf{p}_{\text{R},r}=\left[x_{\text{R}, r},\, y_{\text{R},r},\, z_{\text{R}, r} \right]^T$, $r\in \mathcal{N}_{\text{R}}=\left\{1, 2, \ldots, \Nr\right\}$. In the \textit{monostatic case}, $\mathcal{N}_{\text{R}}=\mathcal{N}_{\text{T}}$ and $\mathbf{p}_{\text{T},i}=\mathbf{p}_{\text{R},i},\, \forall i \in \mathcal{N}_{\text{T}}, \mathcal{N}_{\text{R}}$. For both configurations, the positions of the transmitting and receiving arrays are considered as known.

The \ac{ROI} is divided into a grid of $N$ square cells of size $\Delta$, whose locations are $\mathbf{p}_{n}=\left[x_{n},\, y_{n}, \, z_n \right]^T$, with $n \in \mathcal{N}=\left\{1, 2, \ldots, N \right\}$.
The $n$th  cell is characterized by a scattering coefficient, denoted $\gamma_n$,  related to the \ac{RCS} of the scatterer included in the cell (if any), i.e., $\text{RCS}_n=|\gamma_n|^2 \lambda^2 /4 \pi$. If the cell is empty, then $\gamma_n=0$.  
The magnitude of the scattering coefficient  $\lvert \gamma_{n} \rvert$ is upper bounded by the maximum \ac{RCS} from a scatterer of area $\Delta^2$, which corresponds to the \ac{RCS} of a \ac{PEC} having the same area given by  $\text{RCS}_{\text{max}}=\frac{4\, \pi}{\lambda^2}\, \Delta^2$. Therefore, $|\gamma_n| \le \gamma_{\text{max}} = \sqrt{ \frac{4 \pi}{\lambda^2} \, \text{RCS}_{\text{max}}}$, $\forall n \in \mathcal{N}$.

The values of the scattering coefficients are unknown and should be estimated by the imaging process. Let us define $\bm{\gamma} = [ \gamma_1, \gamma_2, \ldots, \gamma_{N}]^T \in \mathbb{C}^{N\times 1}$ as the unknown vector, i.e., the state vector, associated with the selected \ac{ROI} \cite{YanZha:C20}.
In the following, we assume that each scattering cell performs a perfect isotropic reflection of the illumination signal, as assumed in most of the existing literature \cite{YanZha:C20}.\footnote{Thanks to the adoption of high frequencies, we neglect the presence of any diffusive or multiple scattering effects. Moreover, we assume no coupling occurs between the TX/RX antennas and the scatterers.}

The TX antenna emits an illumination signal $ \boldx = [ x_1, x_2, \ldots, x_i, \ldots, x_{N_T}]^T \in \mathbb{C}^{N_T \times 1}$, with $||\boldx||^2 \leq \Pt$ and $\Pt$ the available transmit power. 
At the RX antenna side, the received signal $\boldy = [ y_1, y_2, \ldots, y_r, \ldots y_{\Nr}]^T$ is expressed as
\begin{align}\label{eq:rx_sig}
 \boldy &= \Gr\,  \bm{\Gamma}\,  \Gt\,  \boldx + \boldw   \, ,
\end{align}
where $\boldw \in \mathbb{C}^{\Nr\times 1} \sim \cn (0, \sigma^2\boldI_{\Nr})$ is an \ac{AWGN} noise vector, whose elements are \ac{i.i.d.} complex Gaussian random variables and $\sigma^2$ is the noise variance. Moreover, we define
\begin{align}
\bm{\Gamma} = \diag( \bm{\gamma})=\diag(\gamma_1, \gamma_2, \ldots, \gamma_N) \in \mathbb{C}^{N \times N} 
\end{align}
as the diagonal matrix containing the \ac{ROI}'s scattering coefficients. Finally, $\Gt={\{ \gtni  \}}\in \mathbb{C}^{N \times N_T}$ and $\Gr= {\{ \grrn  \}}\in \mathbb{C}^{\Nr \times N}$ denote, respectively, the TX-\ac{ROI} and \ac{ROI}-RX channel matrices. In our study, we assume that the channel is known as typical for imaging problems. We consider the performance under Rice fading propagation conditions in Sec.~\ref{subsec:rice}. Further information for acquiring \ac{CSI} can be found \cite{zhang2023near,lu2023near,hu2022hybrid,biguesh2006training}. In the presence of \ac{RIS}, $\Gt$ and $\Gr$ denote the cascade channels for establishing a \ac{LOS} link between the TX/RX and the \ac{ROI}, as further detailed in Sec.~\ref{sec: RISaidedImag}.

In the following, we assume that the \ac{ROI} is located in the radiative near-field region of the TX and the RX, and/or the \ac{RIS}, when the distance between the \ac{ROI} and the transceiver/RIS is
\begin{align}
&  2 D\sqrt{2N} \le d \le  \frac{2\left(D\sqrt{2N}\right)^2}{\lambda} \, ,
\end{align}
where $d$ is the distance between the array/\ac{RIS} and the \ac{ROI}, $D$ is the size of the largest \ac{ROI}/array, and $N$ is the number of antenna elements composing the array or, equivalently, the number of \ac{ROI}'s pixels. This definition is the array equivalent of the Fresnel region of an antenna \cite{BjoOzlSan:C21}, which allows us to identify the spatial region where both the amplitude and phase variations of the \ac{EM} spherical wavefronts cannot be neglected when comparing local phases between antenna elements, even if the wave is locally planar at each antenna. This corresponds to a region where we are not in a strongly near-field condition, i.e., almost close to the reactive near-field behaviors, nor in an almost far-field condition, where only phase variations among antenna elements are perceivable without amplitude variations.

Hereafter, we discuss \ac{LOS} imaging in Secs.~\ref{sec: LOSImaging}-\ref{sec:OptimizProblem}, and \ac{NLOS} imaging in Sec.~\ref{sec: RISaidedImag}.

\section{\ac{LOS} Imaging: Problem Formulation}
\label{sec: LOSImaging}
The primary objective of the imaging procedure is to estimate the scattering coefficient $\bm{\gamma}$ starting from the received signal in \eqref{eq:rx_sig}. To do so, we determine the best illumination signal $\boldx$ that minimizes the estimation error with respect to $\bm{\gamma}$. In the following, we first focus on the \ac{LOS} imaging case. In Sec.~\ref{sec: RISaidedImag}, we treat the \ac{NLOS} \ac{RIS}-aided scenario.

For \ac{LOS} imaging, we can rewrite \eqref{eq:rx_sig} as
\begin{equation}\label{eq:y}
 \boldy =\Gr \, \bm{\Gamma} \, \tilde{\boldx} + \boldw = \Gr \, \tilde{\boldX}\, \bm{\gamma} + \boldw    \, ,
\end{equation}
where $\tilde{\boldX}=\diag(\tilde{\boldx}) \in \mathbb{C}^{N \times N}$, with $\tilde{\boldx} = \Gt \, \boldx \in \mathbb{C}^{N \times 1}$ being the vector describing the illumination signal as observed at the \ac{ROI} side.
In \ac{LOS} conditions, the elements of the channel matrices $\Gt$ and $\Gr$ in  \eqref{eq:rx_sig} and \eqref{eq:y} are given by
\begin{align} \label{eq:Gt}
    &\gtni = \frac{\lambda}{4\, \pi\, d_{i, n}} \, \sqrt{G_{\text{T}}\left(\boldsymbol{\Theta}_{i, n}\right) } \, e^{- j \frac{2\, \pi}{
    \lambda} \,d_{i, n}  }\\
    &\grrn = \frac{\lambda}{4\, \pi\, d_{n, r}} \,\sqrt{G_{\text{R}}\left(\boldsymbol{\Theta}_{n,r}\right)} \, e^{- j \frac{2\, \pi}{ \lambda}\,d_{n,r}  }  \, ,\label{eq:Gr}
\end{align}
where $\{n, i, r \}$ are, respectively, the $n$th cell of the \ac{ROI}, the $i$th element of the TX array, and the $r$th element of the RX array. The quantities 
\begin{align} 
\left[ d_{i,n}, \Thetain \right]&=\left[ d_{i,n}, \phi_{i,n},\, {\theta}_{i,n} \right], \\  
\left[ d_{n, r}, \Thetanr \right]&=\left[ d_{n, r}, \phi_{n, r},\, {\theta}_{n, r} \right]
\end{align}
represent the distance and angles between the $(\mathbf{p}_{\text{T}, i}, \mathbf{p}_n)$ and the $(\mathbf{p}_n, \mathbf{p}_{\text{R}, r})$ pairs of TX/RX antennas and cell location. Moreover,  $\{G_{\text{T}}\left(\boldsymbol{\Theta}_{i, n}\right),\, G_{\text{R}}\left(\boldsymbol{\Theta}_{n, r}\right) \}$ are the transmitting and receiving beam-pattern gains, evaluated in the direction of arrival (i.e., $\Thetanr$) and departure (i.e., $\Thetain$). The considered received signal model accounts for the near-field propagation regime by considering the exact distances and angles at each antenna pair.

To decouple the estimation problem with that of the illumination design, we denote by $\boldbeta=\tilde{\boldX}\bm{\gamma}\,\in \mathbb{C}^{N\times 1} $ the signal backscattered by the scatterers present within the \ac{ROI}. As a consequence, the received signal in \eqref{eq:y} can be rewritten as 
\begin{align}\label{eq:y2}
    \boldy =\Gr \, \boldbeta +\boldw   \, ,
\end{align}
which appears as a conventional linear estimation problem. 

In the following, we discuss a possible approach for estimating the parameter vector $\bm{\beta}$ and, thus, the scattering coefficients ${\bm \gamma}$. Given that the matrix $\Gr$ can be rank-deficient and create ill-posed problems, we introduce \ac{TSVD} regularization of the $\Gr$ matrix. Subsequently, we analyze its associated \ac{MSE}, which constitutes the objective function targeted for minimization in relation to the transmitted signal.

\subsection{Scattering Coefficients Estimate}
To determine an estimate of $\bm{\beta}$ from \eqref{eq:y2}, we consider a \ac{LS} estimator
\begin{align} \label{eq:inv}
    \hbbeta=\Gr^{\dagger} \, \boldy =\Gr^{\dagger} \, \Gr \, \boldbeta + \Gr^{\dagger}\, \boldw  \, ,
\end{align}
where $\hbbeta$ is the \ac{LS} estimate of $\bm{\beta}$ for $\Nr \geq N$. To provide an expression for $\Gr^{\dagger}$, we first introduce the \ac{SVD} of $\Gr$, which can be expressed as
\begin{equation}\label{eq: Gr_SVD}
\Gr = \boldU \, \bm{\Sigma} \, \boldV^H = \sum_{k=1}^{{K}} \xi_k\, \mathbf{u}_k\mathbf{v}_k^H \,\; , 
\end{equation}
where 
\begin{align}
    \mathbf{\Sigma}&=  \left[ 
\begin{array}{cc}
    \diag\left( \xi_1, \ldots, \xi_K \right) & \mathbf{0}_{K \times (N-K)}  \\
     \mathbf{0}_{(\Nr - K) \times K}   &  \mathbf{0}_{(\Nr - K) \times (N - K)}
\end{array} 
\right] \,  \in \mathbb{C}^{\Nr \times N} 
\end{align}
is a diagonal matrix with $K=\min(\Nr, N)$ diagonal elements $\xi_k=\sigma_k(\Gr) \in \mathbb{R}_0^+$, $\mathbb{R}_0^+ = \left[0, +\infty\right)$, $k\in \mathcal{K}=\left\{1,\ldots, K\right\}$, and where $\boldU \in \mathbb{C}^{\Nr \times \Nr }$ and $\boldV\in \mathbb{C}^{N \times N}$ are unitary matrices, with $\mathbf{u}_k $ and $\mathbf{v}_k$ being the columns of $\boldU$ and $\boldV$ respectively, i.e., the left-singular vectors and right-singular vectors of $\Gr$.

The \ac{SVD} of $\Gr^{\dagger}$ is given by \cite{lipschutz2001linear, golub1996matrix} 
\begin{align}\label{eq:pinvGr}
    \Gr^{\dagger}= \boldV \, \pinvsigma \, \boldU^H= \sum_{k=1}^{{K}} \xi_k^{-1}\, \mathbf{v}_k \, \mathbf{u}_k^H \, ,
\end{align}
with $ \Gr^{\dagger} \in \mathbb{C}^{N \times \Nr}$ and $\pinvsigma \in \mathbb{C}^{N \times \Nr}$. 
In our specific case, the \ac{MIMO} wireless channel, represented by the matrix $\Gr$, typically does not possess full rank. Therefore, the presence of null or very small eigenvalues $\xi_k$ within the matrix $\bsigma$ poses challenges in computing its pseudoinverse, thus leading to numerical instability \cite{BerBocDe:B21, BucFran:J89}. To address this issue, a typical approach is to introduce regularization resulting in \cite{bertero2021introduction}
\begin{align}\label{eq:sigmareg}
   \!\! \textcolor{red}{\bsigma \approx\,} \tbsigma&=  \left[ 
\begin{array}{cc}
     \diag\left( \omega_1\,\xi_1, \ldots, \omega_K\, \xi_K \right) & \mathbf{0}_{K \times (N-K)}  \\
     \mathbf{0}_{(\Nr - K) \times  K}   &  \mathbf{0}_{(\Nr - K) \times (N - K)}
\end{array} 
\right],
\end{align}
where $\left\{\omega_1, \omega_2, \ldots, \omega_K\right\}$ are weights set according to the adopted regularization technique, e.g., \ac{TSVD} \cite{MamArbMad:C19} or Tikhonov \cite{och2020high}. 

Then, by denoting the pseudoinverse of \eqref{eq:sigmareg} as $\tbsigma^{\dagger}$, we have
\begin{align}
     \tilde{\mathbf{G}}_{\text{R}}^{\dagger} &= \boldV \tilde{\bm{\Sigma}}^{\dagger} \boldU^H = \sum_{k=1}^{{K}} {\omega_k^{-1}} \xi_k^{-1}\, \mathbf{v}_k \, \mathbf{u}_k^H   \, .
\end{align}
Therefore, $\hbbeta$ can be computed as
\begin{align}\label{eq:truncatedest}
    \hbbeta&=\boldV \tbsigma^\dagger \boldU^H \boldy 
=\boldV \tbsigma^\dagger \bsigma \boldV^H\boldbeta + \boldz = \boldV \blambda \boldV^H \boldbeta + \boldz \, ,
\end{align}
where $\boldz= \tilde{\mathbf{G}}_{\text{R}}^{\dagger} \boldw =\boldV\tbsigma^\dagger \boldU^H  \boldw $, and $\blambda=\tbsigma^\dagger \bsigma \in \mathbb{C}^{N \times N} $ is a diagonal matrix with its first $K$ elements different from zero and the remaining $(N-K)$ elements are null, i.e.,  
\begin{align}\label{eq:lambdamatrix}
\blambda &= \diag\left(\omega_1^{-1},\omega_2^{-1}, \ldots, \omega_K^{-1}, 0, \ldots, 0 \right) \,.
\end{align}
Defining $\boldH=\boldV \blambda \boldV^H \in \mathbb{C}^{N \times N} $, we obtain
\begin{align}\label{eq: beta eq}
    \hbbeta&=\boldH\boldbeta+\boldz=\boldH \tilde{\boldX}\bm{\gamma}  +\boldz = \boldH\, \diag(\Gt \boldx) \, \bm{\gamma}  +\boldz \nonumber\\
&= \boldbeta +\left( \mathbf{H}- \mathbf{I} \right)\diag(\tilde{\boldx}) \, \bm{\gamma}   + \boldz\,.
\end{align}

Assume that $\tilde{\boldX}$ is a full-rank square diagonal matrix, i.e., $\tilde{\boldx}$ does not contain zero elements, which is generally reasonable. 
Then, we can easily obtain an estimate of $\bm{\gamma}$ from $\hbbeta$ by multiplying both sides of \eqref{eq: beta eq} by $\tilde{\boldX}^{-1}$, leading to  
\begin{align}
    & \hat{\bm{\gamma}}=\bm{\gamma} +\tilde{\mathbf{X}}^{-1}\left( \mathbf{H}- \mathbf{I} \right)\tilde{\mathbf{X}}\, \bm{\gamma}   + \tilde{\mathbf{X}}^{-1}\boldz \, .
\end{align}
We can identify three distinct terms contributing to the \ac{LS} estimate $\hat{\bm{\gamma}}$: (\textit{i}) the true original value of the estimated quantity (i.e., $\bm{\gamma}$); (\textit{ii}) a distortion term due to the regularization process (i.e., $\tilde{\mathbf{X}}^{-1}\left( \mathbf{H}- \mathbf{I} \right)\tilde{\mathbf{X}}\, \bm{\gamma}$); and (\textit{iii}) the noise contribution (i.e., $\tilde{\mathbf{X}}^{-1}\boldz$).

\subsection{Mean Squared Error}\label{subsec:mse}

To evaluate the accuracy in estimating $\bm{\gamma}$, we now derive a closed-form expression for the \ac{MSE}. 
Let us start by defining the covariance matrix $\boldC  \in \mathbb{C}^{N \times N}$ of the error vector $\left(  \bm{\gamma}-\hat{\bm{\gamma}}  \right)$
\begin{align}\label{eq: Cov_matrix}
   \boldC &=  \mathbb{E}\left\{ ( \bm{\gamma}-\hat{\bm{\gamma}} )( \bm{\gamma}-\hat{\bm{\gamma}} )^H\right\} \nonumber\\
&= \mathbb{E}\left\{ \left( 
 \tilde{\mathbf{X}}^{-1}\left( \mathbf{H}- \mathbf{I} \right)\tilde{\mathbf{X}}\, \bm{\gamma}   + \tilde{\mathbf{X}}^{-1}\boldz \right)\, \right. \nonumber\\
&\qquad \quad \left.\cdot \left( 
 \tilde{\mathbf{X}}^{-1}\left( \mathbf{H}- \mathbf{I} \right)\tilde{\mathbf{X}}\, \bm{\gamma}   + \tilde{\mathbf{X}}^{-1}\boldz \right)^H \right\} \, .
\end{align}

Assume that the only random vector is $\mathbf{z}$ and consider $\boldsymbol{\gamma}$ as an unknown deterministic vector.
After some computations, reported in Appendix~\ref{app: appendix_a}, the \ac{MSE} as a function of the signal $\tilde{\boldx}$ received at the \ac{ROI} side can be calculated as the trace of the covariance matrix $\boldC$, hence resulting in
\begin{align}\label{eq:MSE}
    \MSE(\bm{\gamma}; \tilde{\boldx}) &= \text{tr}(\boldC) \nonumber\\
&= \sum_{n=1}^N \left\lvert (h_{n,n}-1)\gamma_n + \tilde{x}_n^{-1} \, \sum_{\substack{i=1 \\ i \neq n}}^N  h_{n,i} \, \tilde{x}_i\, \gamma_i \right\rvert^2 \nonumber\\
    &\quad+ \sum_{n=1}^N \frac{\sigma^2}{|\tilde{x}_n|^2} \sum_{{k}=1}^K {\omega_k^{-2}}\,\xi_{k}^{-2} |v_{n,k}|^2 \, ,
\end{align}
where $h_{n,i}$ are the elements of $\mathbf{H}$ and $v_{n,k}$ are the elements of the vector $\mathbf{v}_k$.
In~\eqref{eq:MSE}, it is possible to identify two distinct contributions: the first addend is the distortion in the estimation procedure introduced by the regularization technique, whereas the second term is associated with the \ac{AWGN} noise. 

Notably, \eqref{eq:MSE} depends on $\bm{\gamma}$, which is the objective of the estimation process and depends on the physical scatterers present within the \ac{ROI}, thus being unmodifiable. At the same time, the \ac{MSE} depends on the illumination signal $\boldsymbol{x}$, and varying $\boldsymbol{x}$ might lead to different levels of estimation accuracy (i.e., of distinct values of \ac{MSE}). For this reason, in the sequel, we formulate an optimization problem aimed at minimizing the \ac{MSE} across the vector $\boldsymbol{x}$, while simultaneously constraining $\boldsymbol{\gamma}$ to its maximum value, which corresponds to the worst achievable \ac{MSE}.

\section{Optimization of the Illumination Signal for \ac{LOS} Imaging}\label{sec:OptimizProblem}

As highlighted in the previous section, a fundamental aspect of performing imaging is to identify the optimal illumination signal $\boldx^\star$ that minimizes the \ac{MSE} defined in \eqref{eq:MSE}. To facilitate the analysis, we decompose the problem into two subsequent steps. First, we estimate the illumination signal $\tilde{\boldx}^\star=\Gt\, {\boldx}^\star $ that should be received at the \ac{ROI}, i.e., after propagating through the TX-ROI channel $\Gt$. Secondly, we derive the corresponding transmit signal $\boldx^\star$ to enable a received illumination signal that closely approximates $\tilde{\boldx}^\star$.
Since the \ac{MSE} depends on the actual value of $\boldsymbol{\gamma}$, which is unknown, we formulate our problem as follows
\begin{align}\label{eq:problem0}
    \tilde{\boldx}^\star&=\arg \, \minimize{\boldx} \; \maximize{\boldsymbol{\gamma}} \;\MSE(\bm{\gamma}; \tilde{\boldx}) \\
    &\text{s.t.\,\,}  \lvert\lvert\boldx\rvert\rvert^2\leq \Pt \nonumber \\
    &\phantom{\text{s.t.\,\,} } \lvert\gamma_n\rvert\leq \gamma_{\text{max}}, \,n=1,2,\cdots,N \nonumber  \, ,
\end{align}
where we recall that $\tilde{\boldx}$ can be expressed as a function of $\mathbf{x}$ as $\tilde{\boldx}=\Gt\, \boldx$, and that the maximum magnitude of the scattering coefficient is $\gamma_{\text{max}}$ as defined in Sec. \ref{sec: SystemModel}. 

Given the complexity of globally solving the optimization problem described in \eqref{eq:problem0} due to the dependence of the \ac{MSE} on both $\bold{\gamma}$ and $\tilde{x}$, we opt to tackle a fundamentally different analytical problem. This approximates the original optimization problem, enabling us to derive a closed-form solution, whose performance will then be compared with that achieved through numerical optimization.
To this end, we first establish an upper bound of the cost function in \eqref{eq:problem0}. By applying the sub-multiplicative property, asserting that $\left\|\bf Ab\right\|_{\mathrm{F}}^2 \leq \left\|\bf A\right\|_{\mathrm{F}}^2\left\|\bf b\right\|_{\mathrm{F}}^2$ for any matrices $\bf A$ and $\bf b$, to the first term on the right hand of \eqref{eq:MSE}, we obtain the following inequality
\begin{equation}\label{eq:upper-bound}
\maximize{\boldsymbol{\gamma}}\,\, \MSE(\bm{\gamma}; \tilde{\boldx}) \leq f(\gamma_{\text{max}}; \tilde{\boldx}) + g(\tilde{\boldx}) \,,
\end{equation}
which provides the upper-bound value for the $\MSE(\bm{\gamma}; \tilde{\boldx})$ in regards to the unknown parameter $\boldsymbol{\gamma}$. 
Notably, this upper-bound encompasses two distinguishable terms, i.e., $f(\gamma_{\text{max}}; \tilde{\boldx})$ and $g(\tilde{\boldx})$, which are associated, respectively, with the  truncation error and the \ac{AWGN} noise and given by
\begin{align}\label{eq:twotermss}
&f(\gamma_{\text{max}}; \tilde{\boldx})=N\gamma_{\text{max}}^2\sum_{n=1}^N \left\lvert (h_{n,n}-1) + \tilde{x}_n^{-1} \, \sum_{\substack{i=1 \\ i \neq n}}^N  h_{n,i} \tilde{x}_i \right\rvert^2,
\\
&g(\tilde{\boldx})= \sum_{n=1}^N \frac{\sigma^2}{|\tilde{x}_n|^2} \sum_{{k}=1}^K {\omega_k^{-2}}\,\xi_{k}^{-2} |v_{n,k}|^2 \, .\label{eq:twoterms2}
\end{align}
According to \eqref{eq:upper-bound}, we can reformulate the problem \eqref{eq:problem0} as per
\begin{align}\label{eq:problem}
    \tilde{\boldx}^\star&=\arg\, \minimize{\tilde{\boldx}} \;  f(\gamma_{\text{max}}; \tilde{\boldx})+g(\tilde{\boldx})   \\
    &\text{s.t.\,\,}  \lvert\lvert\boldx\rvert\rvert^2\leq \Pt\;\;. \nonumber
\end{align}
Due to the non-linear and non-convex nature of the objective function presented in \eqref{eq:problem}, solving it directly poses a substantial challenge. To address this difficulty, we propose a two-step algorithm for its effective solution. In the first step, we temporarily ignore the term $f(\gamma_{\text{max}}; \tilde{\boldx})$ and focus  solely on $g(\tilde {\bold x})$. Given this simplification, \eqref{eq:problem} can be written as follows
\begin{align}\label{eq:problem_simplified}
    &\minimize{\tilde{\boldx}} \; g(\tilde{\boldx}) = \sum_{n=1}^N \frac{\sigma^2}{|\tilde{x}_n|^2} \sum_{{k}=1}^K {\omega_k^{-2}}\,\xi_{k}^{-2} |v_{n,k}|^2   \\
    &~~\text{s.t.\,\,}  \lvert\lvert\boldx\rvert\rvert^2\leq \Pt  \nonumber  \, .
\end{align}
While still retaining its non-convex nature, it is feasible to obtain a closed-form solution for the simplified problem in \eqref{eq:problem_simplified}, as validated by the following theorem.

\textbf{Theorem 1: }{\it The optimal solution to problem \eqref{eq:problem_simplified} is given by
\begin{equation}\label{eq:Theo}
\tildexn^\star = \sqrt{b_n} \, e^{j \phi_n}, \,n=1,2, \cdots,N,
\end{equation}
where 
\begin{equation} \label{eq:b_n}
b_n={\frac{P \sqrt{\alpha_n}}{\sum_n \sqrt{\alpha_n}}}
\end{equation} 
denotes the magnitude of the optimal $\tildexn^\star $, $\phi_n$ can be any arbitrary phase, and}
\begin{align}
&\alpha_n\triangleq \sigma^2 \, \sum_{k=1}^K\, \left( \omega_k\, \xi_k\right)^{-2} \lvert v_{n,k} \rvert^2,  \\
&P={\Pt} \cdot \sum_n \mathbf{g}_{\text{T},n} \mathbf{g}_{\text{T},n}^H  \, ,
\end{align}
{\it with $P$ denoting the upper bound of the term $\sum_n \lvert \tildexn \rvert^2$.
}

\textbf{Proof:} See Appendix \ref{app: appendix_b}. 

Theorem 1 indicates that the optimal solution to problem \eqref{eq:problem_simplified} is independent of the phase of each $\tilde{x}_n$. This independence arises because the magnitude squared operation in \eqref{eq:twoterms2} eliminates the phase information. Consequently, commencing with the solution $\tildexn^\star=\sqrt{b_n}e^{j \phi_n}$ obtained in the first step, the subsequent optimization involves adjusting the phase $\phi_n$ of each component to minimize $f(\tilde {\bold x}, \gamma_{\text{max}})$, while keeping $b_n$ fixed. Indeed, optimizing only the phase does not compromise the optimality of the solution found for $g(\tilde {\bold x})$. Thus, by substituting $\tildexn^\star=\sqrt{b_n}e^{j \phi_n}$ into \eqref{eq:twotermss}, the second step entails solving the following optimization problem
\begin{align}\label{eq:step2}
    & \minimize{\left\{\phi_n \in [0,2\pi)\right\}}  \; f(\gamma_{\text{max}}; \left\{\phi_n \right\}) \nonumber\\
     =&N\gamma_{\text{max}}^2\sum_{n=1}^N \left\lvert (h_{n,n}-1) + \frac{e^{-j \phi_n}}{{\sqrt{b_n}}} \, \sum_{\substack{i=1 \\ i \neq n}}^N  \frac{h_{n,i}} {\sqrt{b_i}}e^{-j \phi_i}\, \right\rvert^2    \, .
\end{align}

Determining directly the optimal $\phi_n$ 
for \eqref{eq:step2} is challenging due to the complexity of the absolute value operation, which introduces non-differentiable points. To overcome this, we employ an alternating optimization method, wherein we optimize each phase term $\phi_n$ while keeping all other phases fixed. After optimizing $\phi_n$, we proceed to optimize $\phi_{n+1}$, and so forth, in an alternating manner.
Specifically, for a fixed $n$, the optimization problem \eqref{eq:step2} can be formulated as
\begin{align}\label{eq:step2_solution}
     \minimize{\phi_n \in [0,2\pi)} \; \left\lvert (h_{n,n}-1) + \frac{e^{-j \phi_n}}{\sqrt{b_n}} \, \sum_{\substack{i=1 \\ i \neq n}}^N  h_{n,i} \frac{e^{-j \phi_i}}{\sqrt{b_i}} \right\rvert^2   \, .
\end{align}
By leveraging the geometric nature of the problem \eqref{eq:step2_solution}, its optimal solution can be expressed as
\begin{equation}\label{eq:optimal_phase}
\phi_n^* = \left[{\rm arg}\left(d_n\right) - {\rm arg}\left(c_n\right) -\delta_n \pi \right]~{\rm mod}~2\pi \, ,
\end{equation}
for $n=1,2,\cdots,N$, where 
\begin{align}
&c_n = (h_{n,n}-1), \\ 
&d_n = \frac{1}{\sqrt{b_n}}  \sum_{\substack{i=1 \\ i \neq n}}^N  h_{n,i} \frac{e^{-j \phi_i}}{\sqrt{b_i}},
\end{align}
and 
\begin{align}
\delta_n \triangleq 
\begin{cases} 
+1 & \text{if } {\rm arg}\left(c_n\right) \in [0, \pi) \\
-1 & \text{if } {\rm arg}\left(c_n\right) \in [\pi, 2\pi) \, .
\end{cases}
\end{align}

Notably, the combination of  \eqref{eq:Theo} and \eqref{eq:optimal_phase} provides the most suitable illumination that is required at the \ac{ROI} side to minimize the \ac{MSE} given worst-case magnitudes for $\mathbf{\gamma}$. 
\begin{algorithm}[h]
\caption{Proposed Algorithm for Solving Problem \eqref{eq:problem} }
\label{alg2}
\begin{algorithmic}[1]
\STATE Initialize the phase parameters $\left\{\phi_n\right\}_{n=1}^N$.
\STATE Calculate $\left\{b_n\right\}_{n=1}^N$ according to \eqref{eq:b_n};
\STATE \textbf{Repeat} 
\STATE \quad  Update the phases $\left\{\phi_n\right\}_{n=1}^N$ in an alternating (element-wise) manner according to \eqref{eq:optimal_phase};
\STATE \textbf{Until} the objective function value in \eqref{eq:step2} converges. 
\STATE \textbf{Output:} $\left\{\tildexn^\star = \sqrt{b_n} \, e^{j \phi_n^*}\right\}_{n=1}^N$.
\end{algorithmic}
\end{algorithm}

The proposed two-step algorithm for solving the problem in \eqref{eq:problem} is summarized as Algorithm 1, whose computational complexity is analyzed as follows. The complexity for calculating $\left\{b_n\right\}_{n=1}^N$ is $\mathcal{O}(N)$. The
complexity for updating  $N$ phases is $\mathcal{O}(TN)$, where $T$ denotes the number of iterations required for convergence.  Therefore, the overall computational complexity of Algorithm 1 is $\mathcal{O}((T+1)N)$.

As a final step, to obtain the corresponding illumination signal to be employed at the TX side, we compute
\begin{align}
    \boldx^\star = \Gt^\dagger \,  \tilde{\boldx}^\star \, . 
\end{align}
Since, in general, $\Gt$ is not a full-rank matrix, the result obtained is the minimum norm solution given the limited $N_{\text{DOF}}^{(c)}$ of the TX-ROI channel $\Gt$.

\section{RIS-aided NLOS Imaging}
\label{sec: RISaidedImag}

According to the existing literature, passive walls have mostly been employed to assist the TX in conventional \ac{NLOS} imaging problems \cite{KirEtAl:C09,CuiTri:22}. However, introducing a \ac{RIS} allows for the design of a smarter reflection, potentially enhancing imaging performance.
As illustrated in Fig.~\ref{fig:SystemScenariowithRIS}, we now extend the previous analysis to the case where the illuminating link ($\Gt$) used to illuminate the target \ac{ROI} and/or the sensing link ($\Gr$), delivering the scattered \ac{EM} field, is in \ac{NLOS} condition and is supported by a \ac{RIS} whose parameters have to be optimized. 

Denote with $\mathbf{G}$ the cascade channel of interest (therefore, it could be equally to $\mathbf{G}=\Gt$ and/or $\mathbf{G}=\Gr$).
It can be decomposed into 
\begin{equation}
    \mathbf{G}=\mathbf{G}_1\, \boldsymbol{\Phi}\, \mathbf{G}_2 \,,
\end{equation}
where $\mathbf{G}_2$ is the TX-\ac{RIS} (RX-\ac{RIS}) channel matrix, $\mathbf{G}_1$ is the \ac{RIS}-\ac{ROI} matrix, and $\boldsymbol{\Phi}$ the matrix characterizing the \ac{RIS} phase shifts configuration. We do not restrict $\boldsymbol{\Phi}$ to be diagonal to ensure general applicability.
Given the passive and lossless nature of the \ac{RIS}, it follows that $\boldsymbol{\Phi}^H \boldsymbol{\Phi}=\mathbf{I}$, indicating unitarity. The scenario in which the reflection is not attributed to the \ac{RIS} but rather arises from a natural reflector, characterized by a specular reflection (e.g., a wall), can be straightforwardly represented by defining $\boldsymbol{\Phi}=\eta \mathbf{I}$, where $0<|\eta|\le 1$ denotes the reflection coefficient ($\eta=1$ corresponds to a \ac{PEC}).

As evident from the previous analysis and the numerical results, the imaging quality depends on the rank (i.e., the \ac{DoF}) of the cascade channel matrix $\mathbf{G}$ as well as its total gain $\| \mathbf{G} \|_{\text{F}}^2$. 
In general, being 
\begin{align}
\text{rank}(\mathbf{G}_1\, \boldsymbol{\Phi} \, \mathbf{G}_2) \le \min (\text{rank}(\mathbf{G}_1) , \text{rank}(\boldsymbol{\Phi}) , \text{rank}(\mathbf{G}_2) ) \, ,
\end{align}
the rank of the cascade channel is dominated by the channel with minimum rank. Since the number of elements of the \ac{RIS} is typically larger than $\text{rank}(G_1)$ or $ \text{rank}(G_2)$ and the \ac{RIS} phase matrix is full rank, it does not concur in determining the rank of the cascade channel. More precisely, the rank of the cascade channel cannot be increased through a specific RIS design. Therefore, to improve the image quality we seek the optimal \ac{RIS}'s configuration $\boldsymbol{\Phi}$ which maximizes $\| \mathbf{G} \|_{\text{F}}^2=\sum_n \xi_n^2$,  where $\xi_n=\sigma(\mathbf{G})$ are the singular values of $\mathbf{G}$, under the constraint $\boldsymbol{\Phi}^H \boldsymbol{\Phi}=\boldI$ (passive \ac{RIS}) and for fixed $\mathbf{G}_1$ and $\mathbf{G}_2$ matrices. To this end, we recall the following Theorem \cite{Wang97}:

\textbf{Theorem 2: }{\it
Given Hermitian matrices $\mathbf{A}_1 \in \mathbb{C}^{N \times N}$, $\mathbf{A}_2 \in \mathbb{C}^{N \times N}$ and $\mathbf{A}_3  \in \mathbb{C}^{N \times N}$ with singular values, respectively,  $\sigma_n(\mathbf{A}_1)$, $\sigma_n(\mathbf{A}_2)$ and $\sigma_n(\mathbf{A}_3)$, we have
\begin{equation} \label{eq:sigma}
    \sum_n \sigma^2_n(\mathbf{A})\le  \sum_n \sigma_n^2(\mathbf{A}_1) \, \sigma_n^2(\mathbf{A}_2) \, \sigma_n^2(\mathbf{A}_3) \, ,
\end{equation}
where $\mathbf{A}=\mathbf{A}_1\, \mathbf{A}_2 \, \mathbf{A}_3$.  The equality is true when 
\begin{align}
\sigma_n(\mathbf{A})=\sigma_n(\mathbf{A}_1) \, \sigma_n(\mathbf{A}_2) \, \sigma_n(\mathbf{A}_3), \quad \forall \, n \,.
\end{align}}

Let us define 
\begin{align}
&\mathbf{A}_1=\mathbf{G}_1^H \mathbf{G}_1, \,\,\mathbf{A}_2=\boldsymbol{\Phi}^H \boldsymbol{\Phi}, \,\,\mathbf{A}_3=\mathbf{G}_2^H \mathbf{G}_2  \, .
\end{align}
We perform the \ac{SVD} decomposition $\mathbf{G}_1=\mathbf{U}_1 \boldsymbol{\Sigma_1} \mathbf{V}_1^H$ and $\mathbf{G}_2=\mathbf{U}_2 \boldsymbol{\Sigma}_2 \mathbf{V}_2^H$. Since $\mathbf{A}_2=\boldsymbol{\Phi}^H \boldsymbol{\Phi}=\mathbf{I}$, the right hand side of \eqref{eq:sigma} does not depend on the \ac{RIS}'s configuration $\boldsymbol{\Phi}$ and hence achieving the equality in \eqref{eq:sigma} ensures that $\sum_n \sigma^2_n(\mathbf{A})$ is maximized. We can thus write
\begin{equation} \label{eq:G1}
    \mathbf{G}=\mathbf{U}_1  \boldsymbol{\Sigma}_1 \mathbf{V}_1^H \boldsymbol{\Phi} \mathbf{U}_2 \boldsymbol{\Sigma}_2 \mathbf{V}_2^H \, .
\end{equation}
Let us now define $\boldsymbol{\Phi}=\mathbf{V}_1 \mathbf{Z} \mathbf{U}_2^H$, with $\mathbf{Z}$ being a generic unitary matrix. Then, we can rewrite \eqref{eq:G1} as
\begin{equation} \label{eq:G}
    \mathbf{G}=\mathbf{U}_1  \boldsymbol{\Sigma}_1 \mathbf{Z} \boldsymbol{\Sigma}_2 \mathbf{V}_2^H=\mathbf{U}_1  \boldsymbol{\Sigma}_t \mathbf{V}_2^H \, ,
\end{equation}
having defined the matrix $\boldsymbol{\Sigma}_t=\boldsymbol{\Sigma}_1 \mathbf{Z} \boldsymbol{\Sigma}_2$. Note that $\boldsymbol{\Sigma}_t$ is in general not diagonal because of $\mathbf{Z}$.

If we choose $\mathbf{Z}=\mathbf{I}$, $\boldsymbol{\Sigma}_t$ becomes diagonal and \eqref{eq:G} takes the form of the \ac{SVD} of the matrix $\mathbf{G}$. As a consequence, in \eqref{eq:sigma} it is 
 $\sigma_n^2(\mathbf{A})=\sigma_n^2(\mathbf{A}_1) \, \sigma_n^2(\mathbf{A}_3)$, $\forall n$, being $\sigma^2_n(\mathbf{A}_2)=1$. This translates in
 \begin{equation}
     \xi_n^2=\sigma^2_n(\mathbf{G})=\sigma_n^2(\mathbf{G}_1) \, \sigma_n^2(\mathbf{G}_2) \;,\;\; \forall n  \, ,
     \end{equation}
 by considering that $\sigma_n^2(\boldsymbol{\Phi})=1$ 
 being $\boldsymbol{\Phi}$ unitary. 
 According to Theorem 2, we obtain the equality in \eqref{eq:sigma}, thus maximizing the quantity $\sum_n \xi_n^2$. 
As a result, the optimal \ac{RIS} configuration is given by
\begin{equation} \label{eq:matchedRIS}
\boldsymbol{\Phi}=\mathbf{V}_1 \mathbf{U}_2^H \, .
\end{equation}
Incidentally, this is the same result achieved in \cite{BarAbrDecDarDiR:J23} through an alternative method for maximizing the mutual information between a TX and a RX communicating via a \ac{RIS}. Note that, in general, the optimal $\boldsymbol{\Phi}$ configuration is not in a diagonal form.

\section{Numerical Results}\label{sec: NumericalResults}

\begin{figure*}[th!]
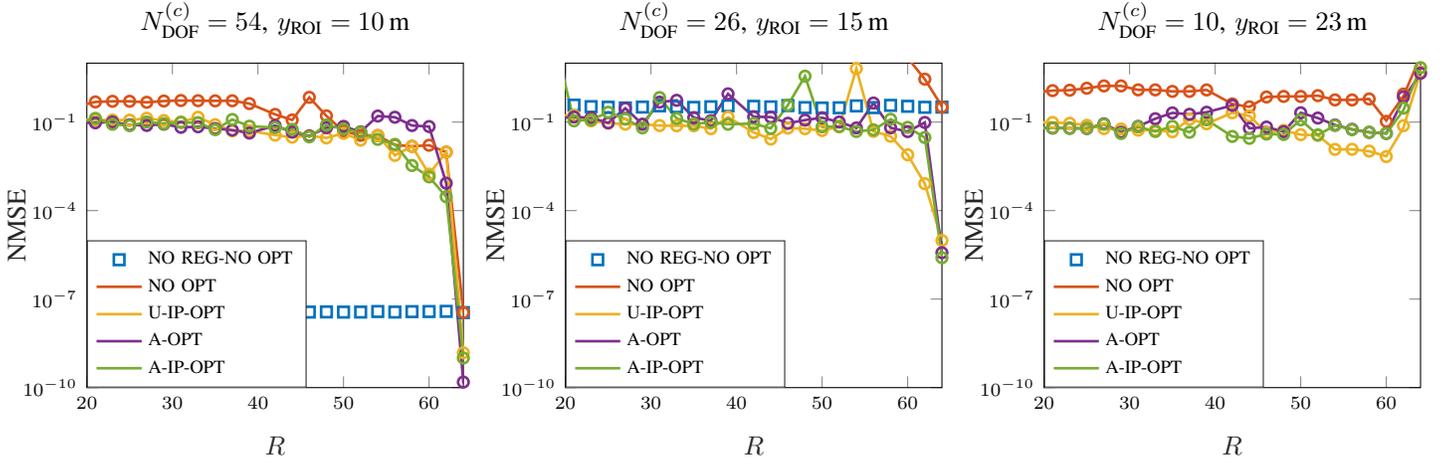

    \centering
\hspace{-10mm}
    \begin{minipage}[t]{0.3\textwidth}
        \centering
        \input{NMSE_monostatic_10m.tex}
\end{minipage}%
\hspace{8mm}
    \begin{minipage}[t]{0.3\textwidth}
        \centering
        \input{NMSE_monostatic_15m}
    \end{minipage}%
\hspace{8mm}
    \begin{minipage}[t]{0.3\textwidth}
        \centering
        \input{NMSE_monostatic_23m}
    \end{minipage}%
    \caption{NMSE as a function of the truncation index $R$ selected when applying the \ac{TSVD} to $\Gr$ and different optimization techniques for the monostatic \ac{LOS} setup. The TX/RX \ac{XL-MIMO} array is located at $(0,0, 0)\, \mathrm{m}$ and three distinct locations for the \ac{ROI} are tested, namely $y_{\text{ROI}} \in \{10, 15, 23\}\, \mathrm{m}$. Dotted ($\circ $) and continuous ($-$) lines correspond to the E-NMSE and T-NMSE, respectively, while the square markers denote the non-regularized case for benchmarking.}
    \label{fig:monostatic}
\end{figure*}

\begin{figure*}[th!]
\psfrag{Lx}[c][c][0.8]{$N_x$}
\psfrag{Ly}[c][c][0.8]{$N_y$}
\psfrag{Lz}[c][c][0.8]{$N_y$}
\psfrag{TRUE}[c][c][0.8]{Reference Image}
\psfrag{WCASE}[c][c][0.8]{NO REG-NO OPT}
\psfrag{REGNOOPT}[c][c][0.8]{NO OPT}
\psfrag{IPOPT}[c][c][0.8]{U-IP-OPT}
\psfrag{SUBOPT}[c][c][0.8]{A-OPT}
\psfrag{EIPOPT}[c][c][0.8]{A-IP-OPT}
    \centering    
    \includegraphics[width=0.3\textwidth]{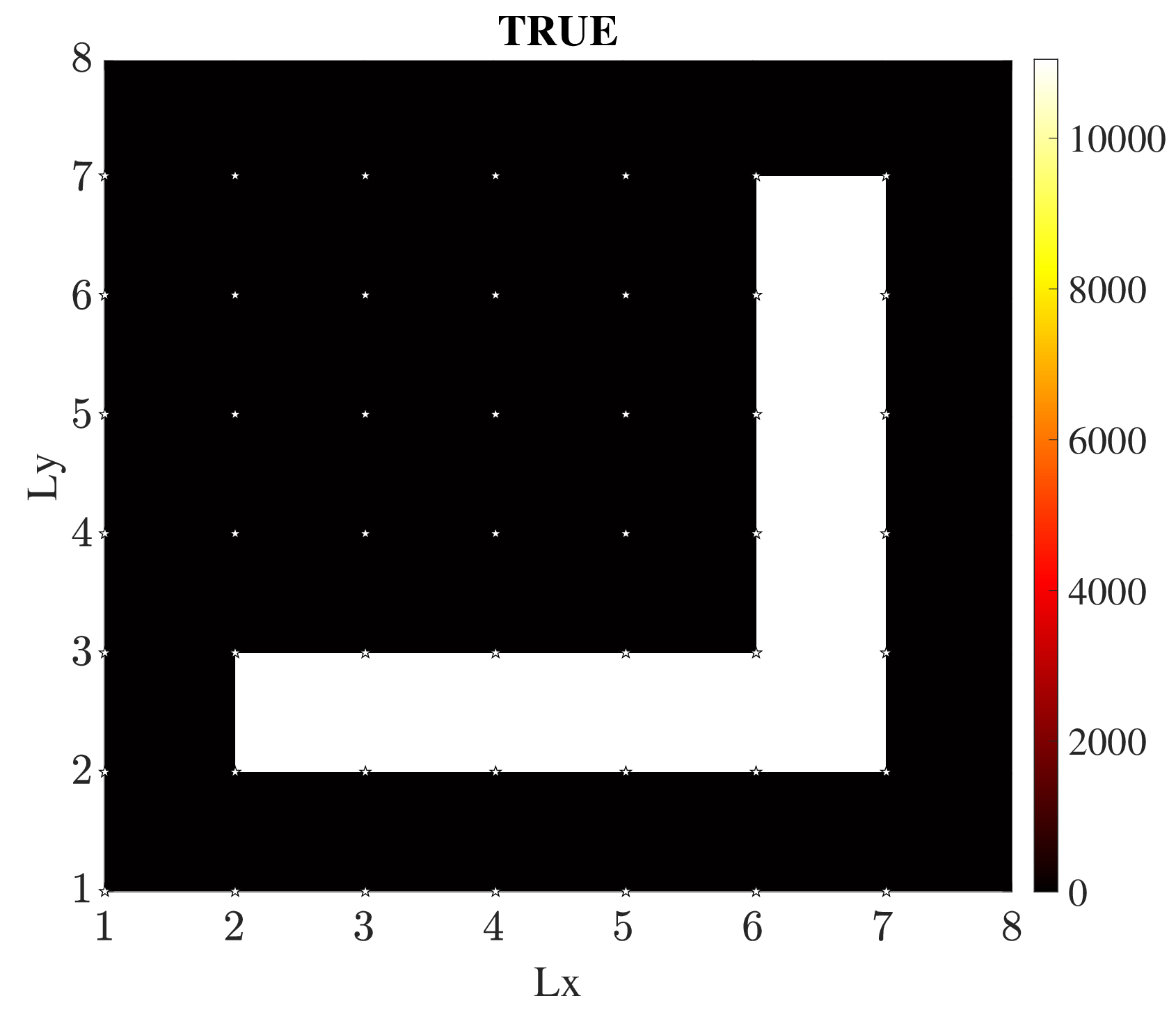}
    \includegraphics[width=0.3\textwidth]{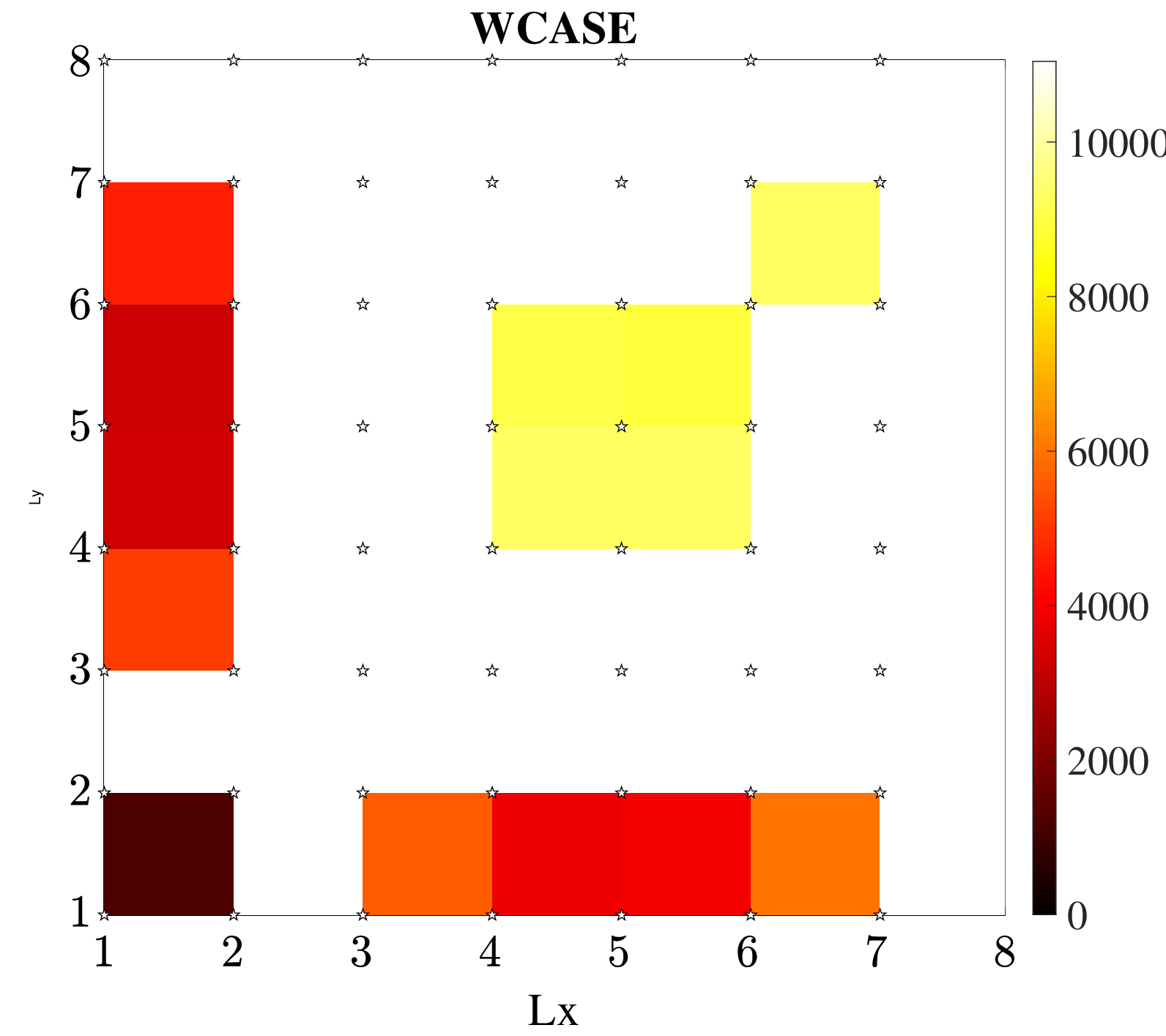}
\includegraphics[width=0.3\textwidth]{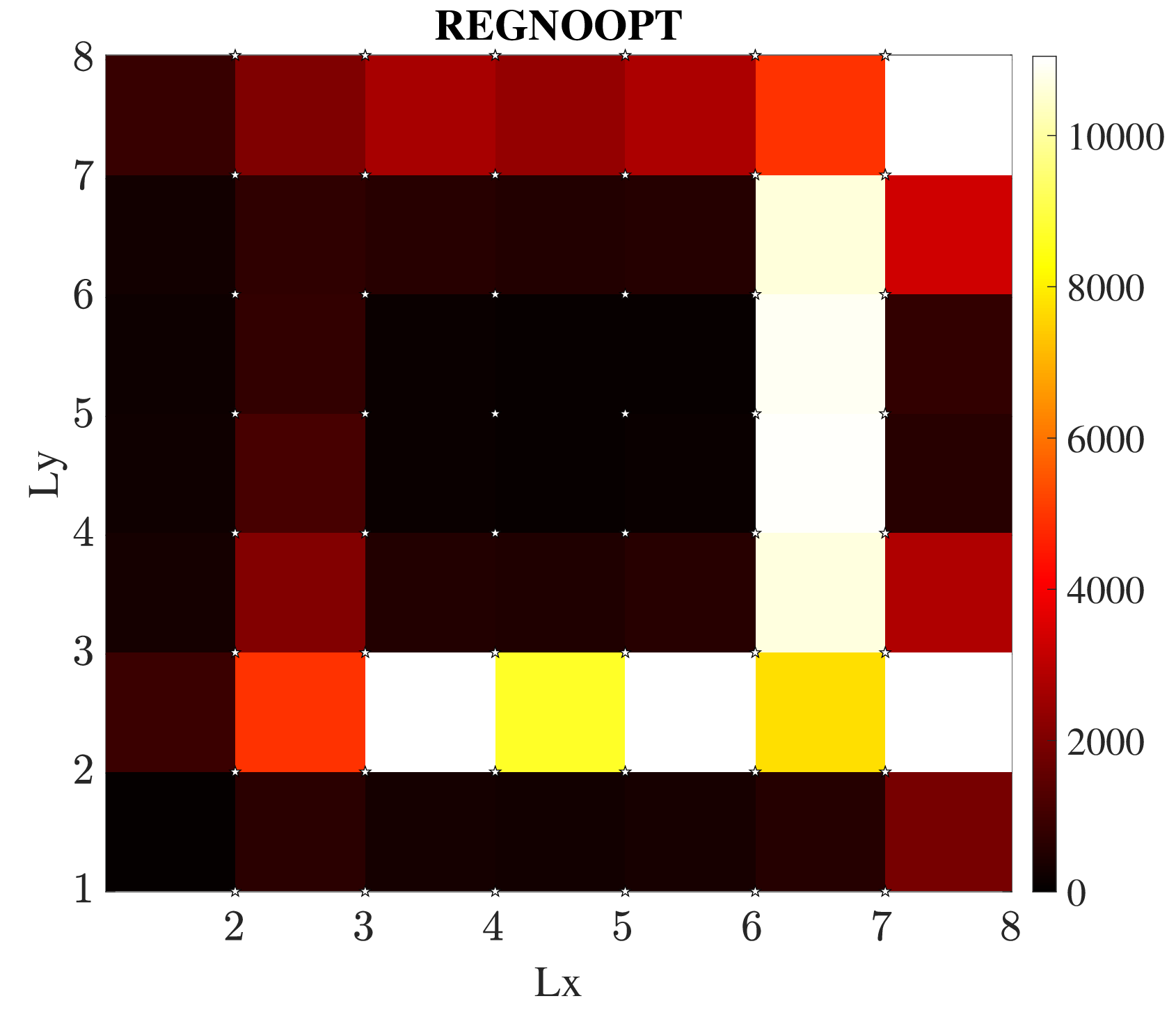}
\includegraphics[width=0.3\textwidth]{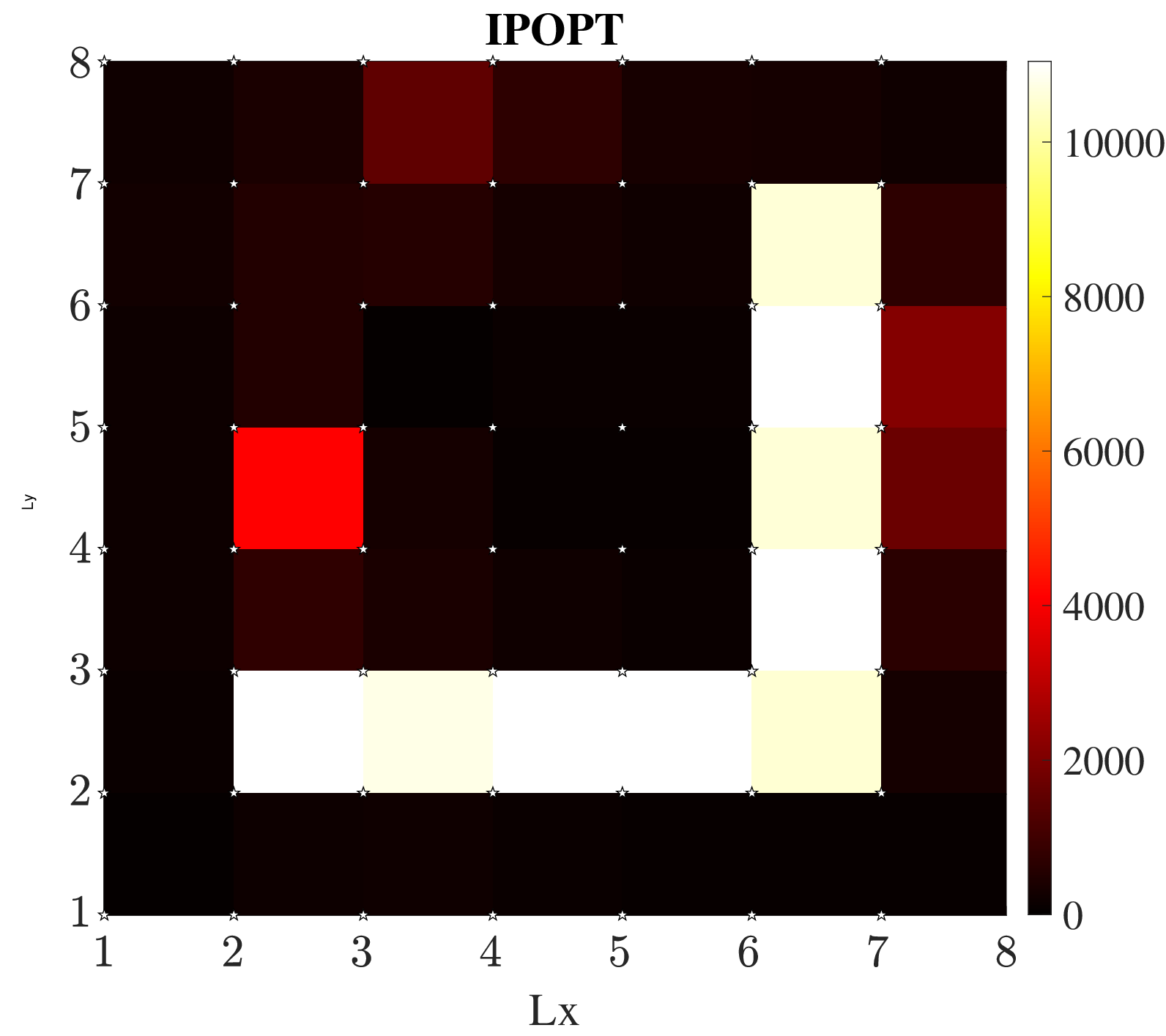}
\includegraphics[width=0.3\textwidth]{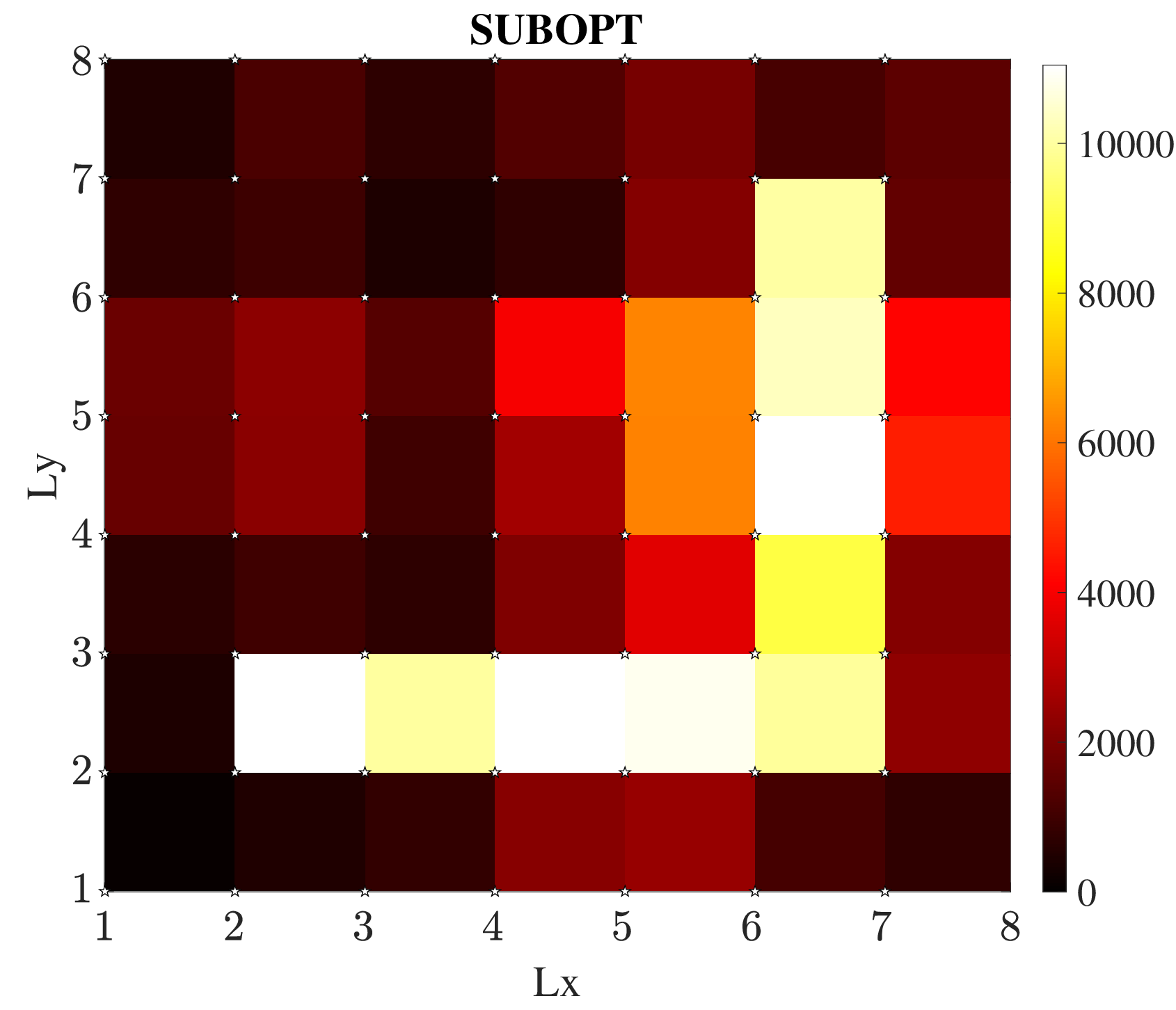}
\includegraphics[width=0.3\textwidth]{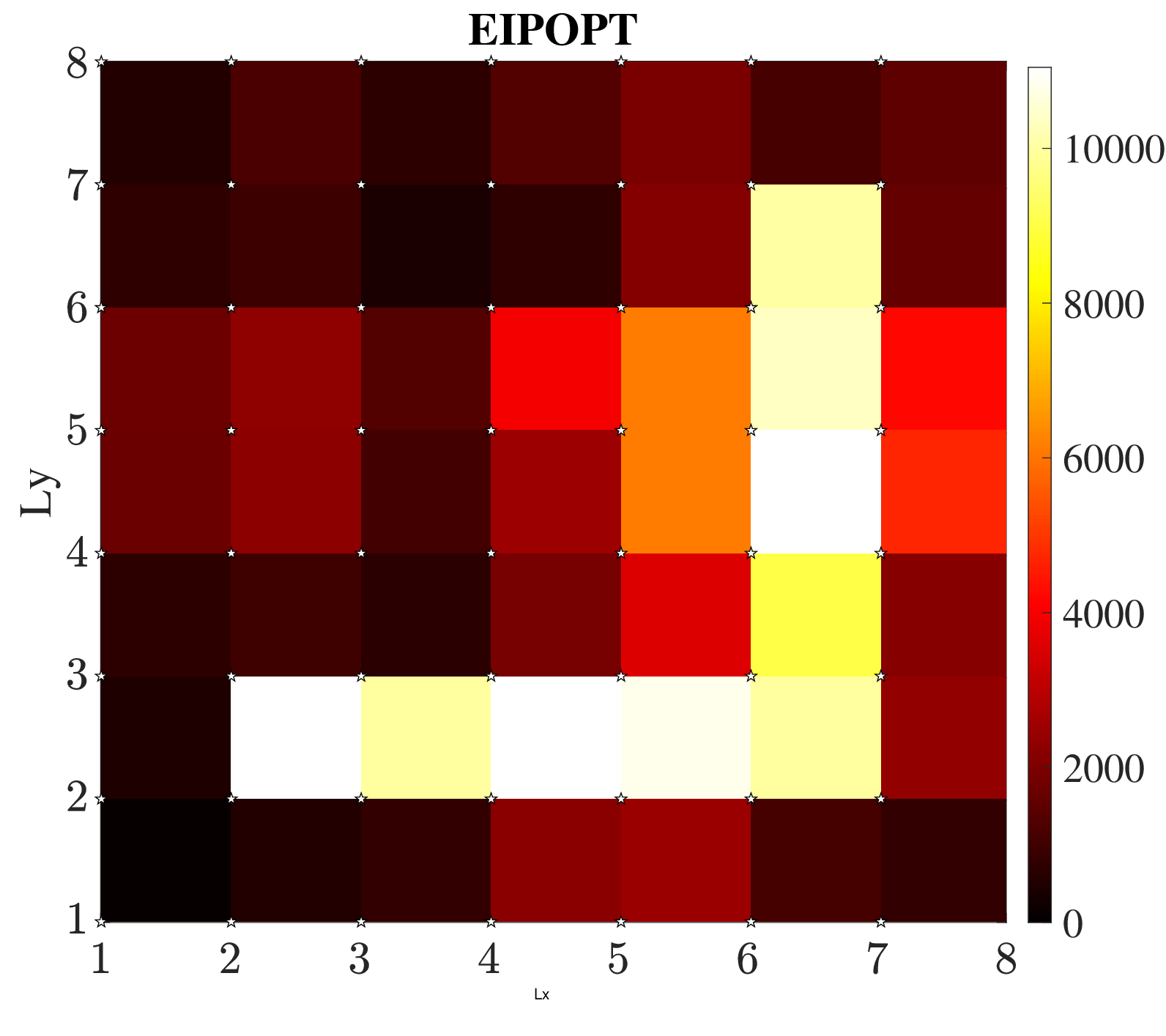}
    \caption{Estimated images ($\hat{\boldsymbol{\gamma}}$) for $R=60$, $y_{\text{ROI}}=23\, \mathrm{m}$, different optimization approaches, and reference image with $N_{\text{DOF}}^{(i)} = 2$.}
    \label{fig:monostatic_images}
\end{figure*}

\subsection{Simulation Setup}

In the proposed setup, imaging is performed over a narrow frequency band $\Delta f=120\,\mathrm{kHz}$ centered at $f_{\text{c}}=28\, \mathrm{GHz}$, with a wavelength of $\lambda\simeq 0.01\, \mathrm{m}$. This choice can correspond to adopting a sub-carrier or a resource block in a \ac{OFDM} signal used also for communication. We set the noise power spectral density to $\sigma^2=-170~\mathrm{dBm/Hz}$, and the transmitted power $\Pt$ to $30~\mathrm{dBm}$.
The \ac{XL-MIMO} TX and RX arrays are configured as uniform squared arrays with the same size of $(10\lambda\times 10\lambda)\,\text{m}^2$, with antenna elements spaced apart of $\lambda/2$ (i.e., $20\times 20$ antennas in the considered settings). The \ac{ROI} spans dimensions of $(750\lambda \times 750\lambda)~\mathrm{m}^2$, with $8 \times 8$ cells equally distributed with an inter-spacing of $\Delta = 93.75\lambda$, if not otherwise indicated.
In the simulations, in the presence of a scatterer in the $n$th cell, $\gamma_n$ has been set equal to $M = 10^{-1}\gamma_{\text{max}}$ considering that, in realistic scenarios, objects typically exhibit a \ac{RCS} smaller than those observed in the case of a \ac{PEC}. Specifically, the magnitude of the scattering coefficients $\bm{\gamma}$, i.e., the image, to be estimated is depicted in Fig.~\ref{fig:monostatic_images}-(top,left). 
When present, the \ac{RIS} has dimensions equal to $(60\lambda \times 60\lambda)\,\text{m}^2$ and is in paraxial configuration with respect to the TX/RX. The RIS's elements are separated by a distance equal to $\lambda/2$ (i.e., $120\times 120$ RIS elements in the setup in Fig.~\ref{fig:SystemScenariowithRIS}).

For each simulation, we evaluated both the empirical and theoretical \ac{MSE} together with its \ac{NMSE} version. In particular, these metrics are computed as
\begin{equation}
    \text{E-MSE}= \frac{\sum_{m=1}^{N_{\text{MC}}}\lVert \boldsymbol{\gamma} - \hat{\boldsymbol{\gamma}}_m \rVert^2}{N_{\text{MC}}}, \quad \text{E-NMSE}=\frac{\text{E-MSE}}{N M^2} \, ,
\end{equation}
where $N_{\text{MC}}$ represents the number of Monte Carlo iterations that were set to $100$. Likewise, the theoretical MSE (T-MSE) is given by \eqref{eq:MSE} and its normalized version is denoted with T-\ac{NMSE}. Notably, the relationship between the \ac{PSNR} and \ac{NMSE} is expressed as $\text{PSNR}=\text{NMSE}^{-1}$.\footnote{The \ac{PSNR}, a commonly utilized metric in image processing, denotes the ratio between the maximum attainable power of the generic pixel and the power of the estimation noise influencing the fidelity of its representation \cite{MaitEtAl:J18}.} 
In our results, the following cases were considered:

\begin{figure*}[t]
\psfrag{Nx}[c][c][0.8]{$N_x$}
\psfrag{Ny}[c][c][0.8]{$N_y$}
\psfrag{Lx}[c][c][0.8]{$N_x$}
\psfrag{Ly}[c][c][0.8]{$N_y$}
\psfrag{TRUE}[c][c][0.8]{Reference Image}
\psfrag{WCASE}[c][c][0.8]{NO REG-NO OPT}
\psfrag{REGNOOPT}[c][c][0.8]{NO OPT}
\psfrag{IPOPT}[c][c][0.8]{U-IP-OPT}
\psfrag{SUBOPT}[c][c][0.8]{A-OPT}
\psfrag{EIPOPT}[c][c][0.8]{A-IP-OPT}
    \centering    
    \includegraphics[width=0.3\textwidth]{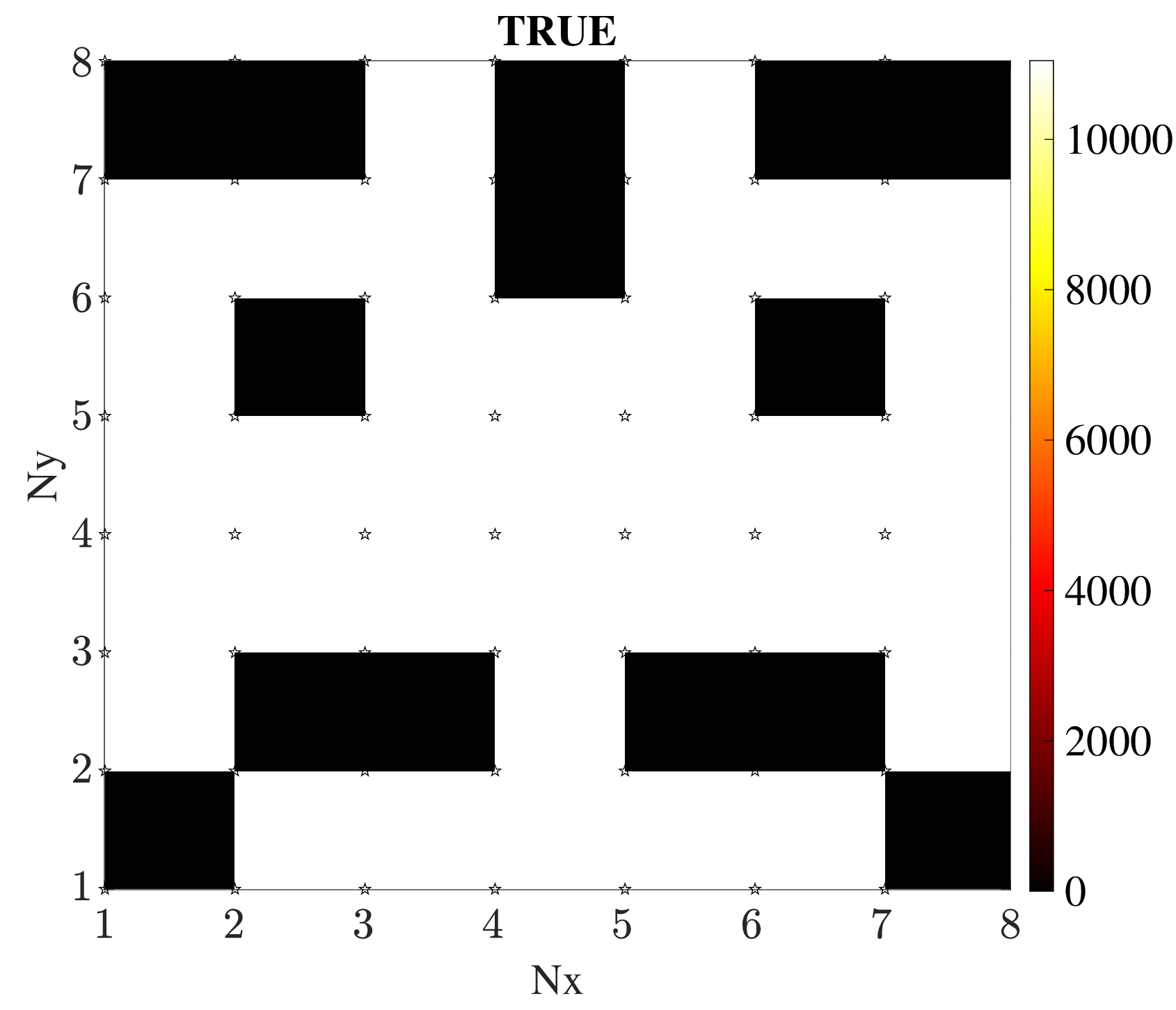}
    \includegraphics[width=0.3\textwidth]{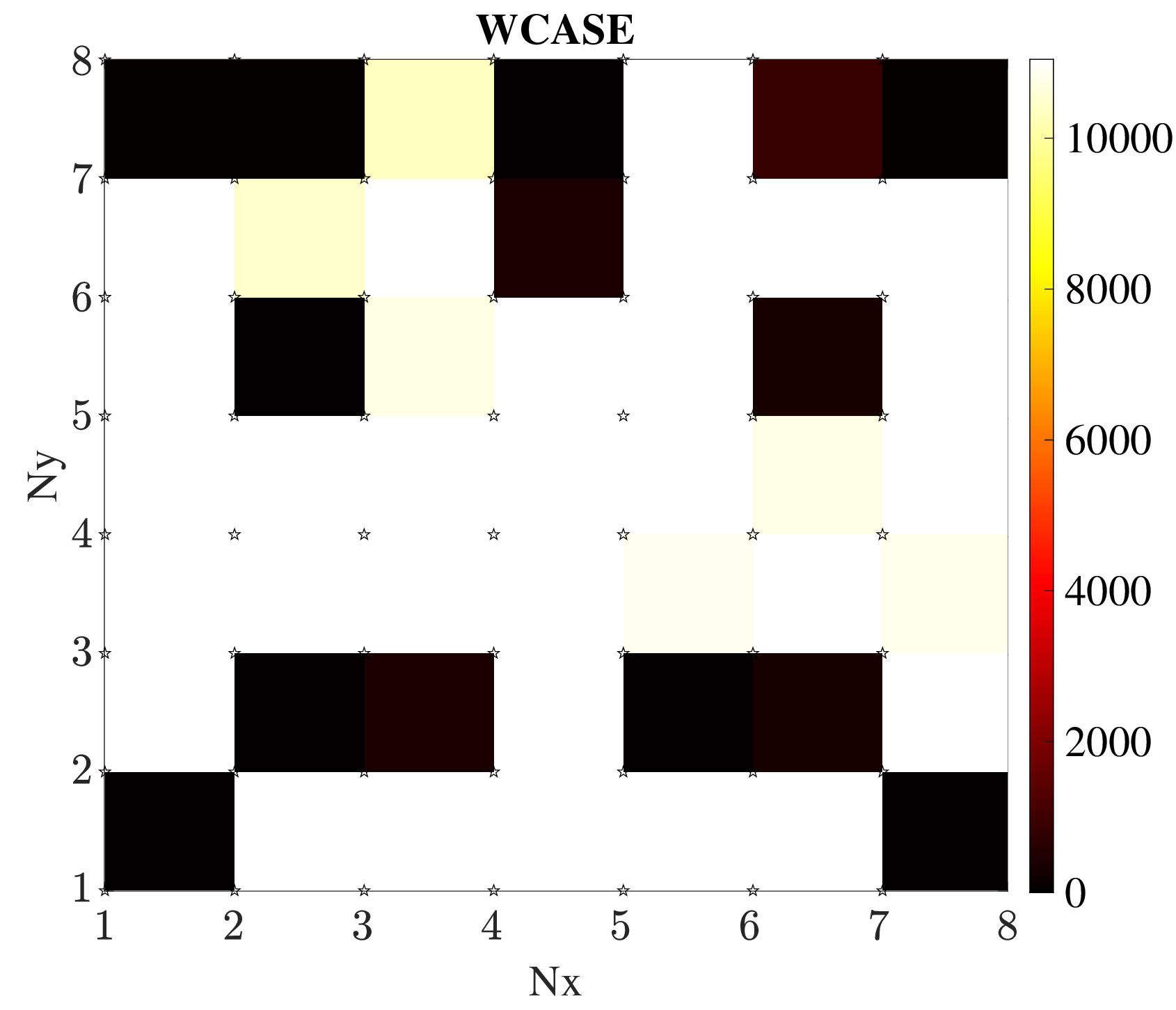}
    \includegraphics[width=0.3\textwidth]{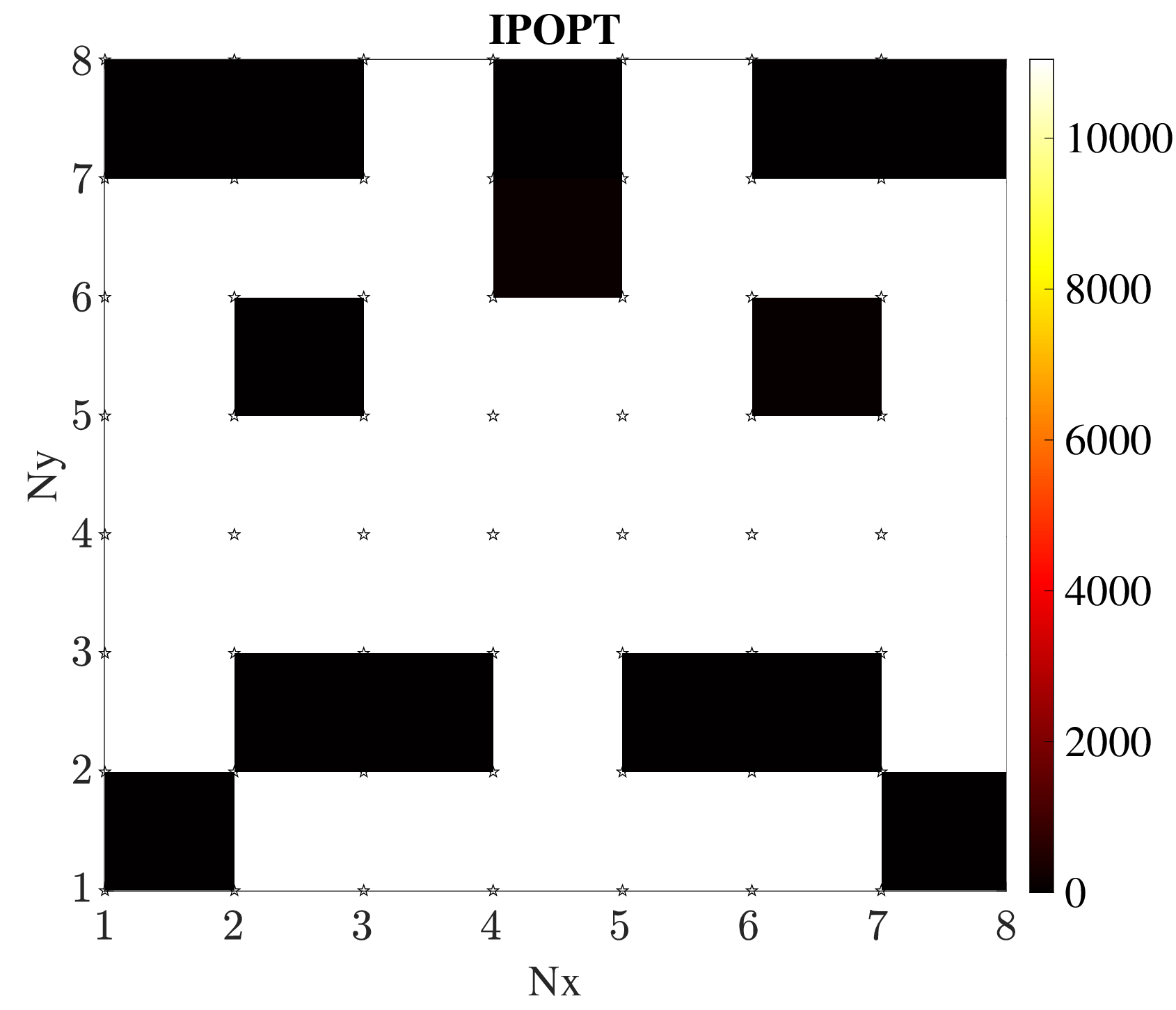}\\
    \includegraphics[width=0.3\textwidth]{TRUE_15m.eps}
    \includegraphics[width=0.3\textwidth]{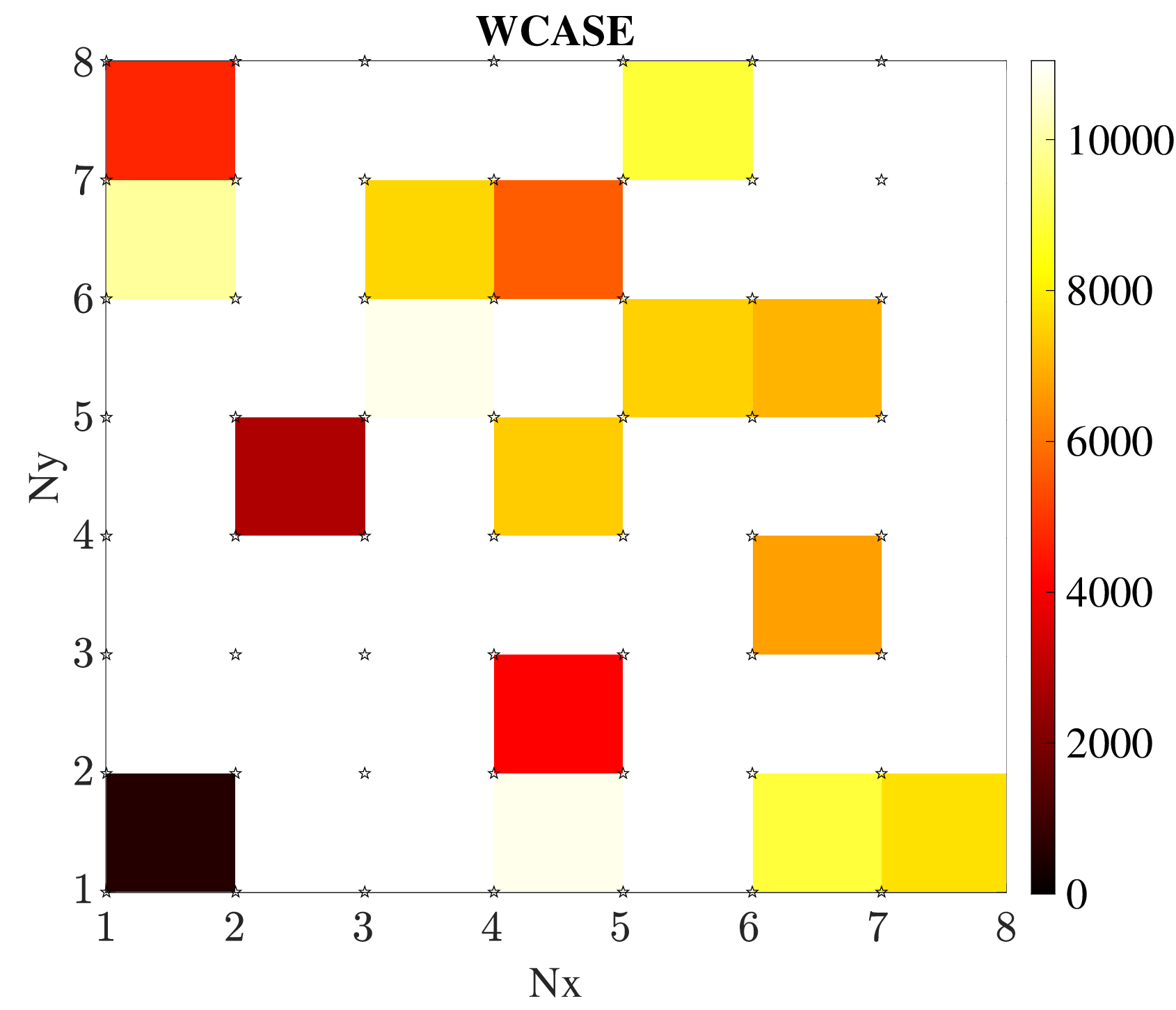}
    \includegraphics[width=0.3\textwidth]{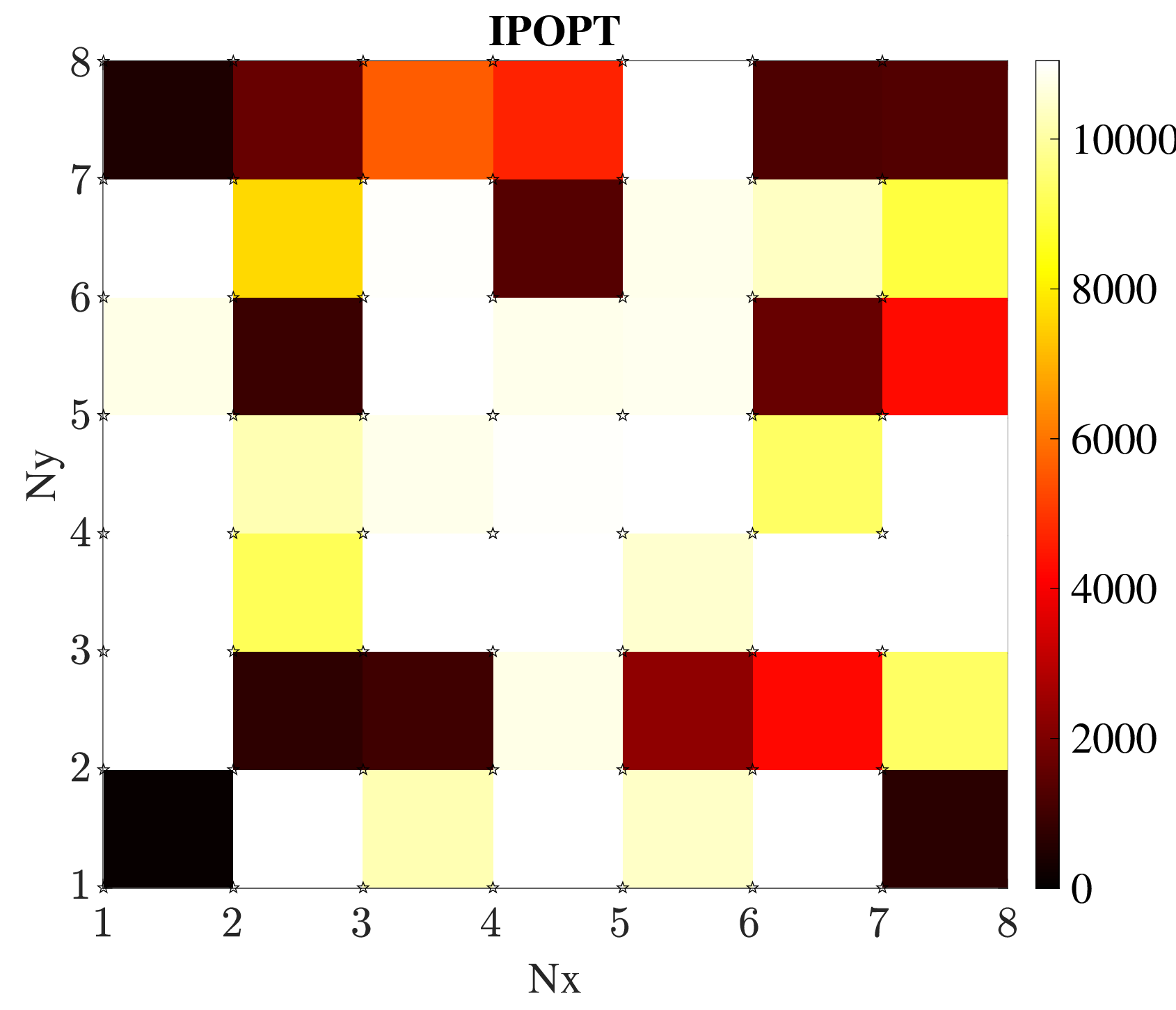}
    \caption{Estimated images ($\hat{\boldsymbol{\gamma}}$) in the NO REG-NO OPT and IP-OPT cases for the selected reference image. The first row refers to the configuration $y_\text{ROI}=15\, \mathrm{m}$ and $R=64$, and the second one to the case $y_\text{ROI}=23 \, \mathrm{m}$ and $R=58$. The reference image has $N_{\text{DOF}}^{(i)} = 4$.}
    \label{fig:ImageDOF}
\end{figure*}

\begin{figure}[h!]
    \centering
            \hspace{-15mm}
    \begin{minipage}{0.38\textwidth} 
        \centering
        \input{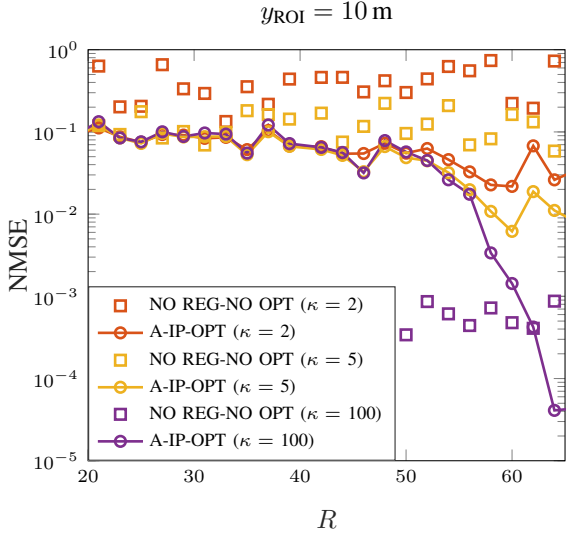}
    \end{minipage}
    \caption{NMSE as a function of the truncation index $R$ selected for \ac{TSVD} regularization for the monostatic setup and for different $\kappa$ values, i.e., $\kappa = \{2,5,100\}$. Two cases are shown, namely NO REG-NO OPT and A-IP-OPT.}
    \label{fig:fading}
\end{figure}

\begin{itemize}
    \item \textit{No regularization, No optimization (NO REG-NO OPT)}: This represents the worst-case scenario, where we considered neither the $\Gr$ matrix regularization nor the optimization of the transmit illuminating vector. This scenario is incorporated in our simulations as a benchmark. In the following cases, it is presumed that $\Gr$ regularization is always performed.
    \item \textit{Regularization, No optimization (NO OPT)}: In this case, we considered the regularization of the matrix $\Gr$ but we did not apply the optimization of the transmitting signal, setting it equal to $\mathbf{x}=\frac{\sqrt{Pt}}{2\, \Nt}\left(\mathbf{1}_{\Nt \times 1} + j \mathbf{1}_{\Nt \times 1}\right)$. This choice corresponds to a uniform \ac{ROI} illumination.
    \item \textit{Uniform Interior-Point Optimization (U-IP-OPT)}: We applied the optimization of the transmitting vector by running an interior-point method \cite{byrd1999interior} to search for $\tilde{\mathbf{x}}^\star$ starting from a guess solution equal to $\tilde{\mathbf{x}}_0=\frac{\sqrt{Pt}}{2\, \Nt} \Gt\,\left(\mathbf{1}_{\Nt \times 1} + j \mathbf{1}_{\Nt \times 1}\right)$ and $\boldsymbol{\gamma}^\star=\gamma_{\text{max}}\, \mathbf{1}_{N \times 1}$. To compute the transmitted vector, we applied ${\mathbf{x}}^\star= \tilde{\boldG}_{\mathrm{T}}^\dagger \tilde{\mathbf{x}}^\star$, with $\tilde{\boldG}_{\mathrm{T}}$ being the regularized version of $\Gt$.
    \item \textit{Analytical Solution (A-OPT)}: In this case, we applied the optimization of the transmitting vector by implementing \eqref{eq:Theo} with $\boldsymbol{\gamma}^\star=\gamma_{\text{max}}\, \mathbf{1}_{N \times 1}$. Then, as before, to derive the transmitted vector, we applied ${\mathbf{x}}^\star= \tilde{\boldG}_{\mathrm{T}}^\dagger \tilde{\mathbf{x}}^\star$.
\item \textit{Analytical Interior-Point Optimization (A-IP-OPT)}: It is the same approach as U-IP-OPT but with a different initial guess $\tilde{\mathbf{x}}_0$, which is set equal to the analytical solution in \eqref{eq:Theo}.
\end{itemize}

Notably, the number of \ac{DoF} associated with the wireless channel, particularly $\Gr$, impacts the imaging accuracy and the capability of estimating the scattering features of the \ac{ROI}. 
Specifically, recalling the results obtained in \cite[Eq. 31]{Dar:J20},  it can be shown that
\\
\begin{equation} \label{eq:N_DOF}
N_{\text{DOF}}^{(c)} \simeq \! \frac{2 L^2}{\lambda^2}\!\!\left(\frac{S \tan ^{-1}\left(\frac{S}{\sqrt{4 d^2+S^2}}\right)}{\sqrt{4 d^2+S^2}}\!+\!\frac{S \tan ^{-1}\left(\frac{S}{\sqrt{4 d^2+S^2}}\right)}{\sqrt{4 d^2+S^2}}\right) \!,
\end{equation}
where $L = L_x = L_y$ denotes the side of the device having the smallest area, e.g., the TX/RX, $S = S_x = S_y$ denotes the side of the larger entity, i.e., the \ac{ROI}, and $d$ represents the distance between their centers. 
Notably, such a number approximates the rank of the channel matrix of the considered link. Particularly, the geometric setup is favorable for holographic imaging if it is close to $K=\min(\Nr,N)$, i.e., $N_{\text{DOF}}^{(c)} \to K$. This also hints that no regularization is required to estimate $\boldsymbol{\gamma}$ if the channel matrices are full-rank. Conversely, as the number of \ac{DoF} diminishes (with the far-field condition representing the most challenging scenario, wherein $N_{\text{DOF}}^{(c)}=1$), there is a substantial deterioration in the imaging performance. In such cases, regularization is pivotal in facilitating the resolution of the ill-posed \ac{ISP}.
Nevertheless, the imaging performance is influenced not only by  $N_{\text{DOF}}^{(c)}$ but also by the complexity of the image. A rough estimate of such complexity could be given by the number of the most significant principal components, denoted as $N_{\text{DOF}}^{(i)}$, derived using the \ac{PCA} \cite{solomon2019deep}. Through simulations, we will explore the interplay between these two quantities.

To investigate the impact of regularization on imaging performance, we employ a \ac{TSVD} approach with a fixed threshold to retain only the first $R$ eigenvalues of $\Gr$. Conversely, in the case of the $\Gt$ matrix, we adopt \ac{TSVD} by retaining all eigenvalues whose cumulative sum does not surpass $99\%$ of the total transmit power $\Pt$, being the illuminating channel not subject to any noise enhancement.

\subsection{Monostatic LOS Imaging Performance}

\begin{figure*}[t]
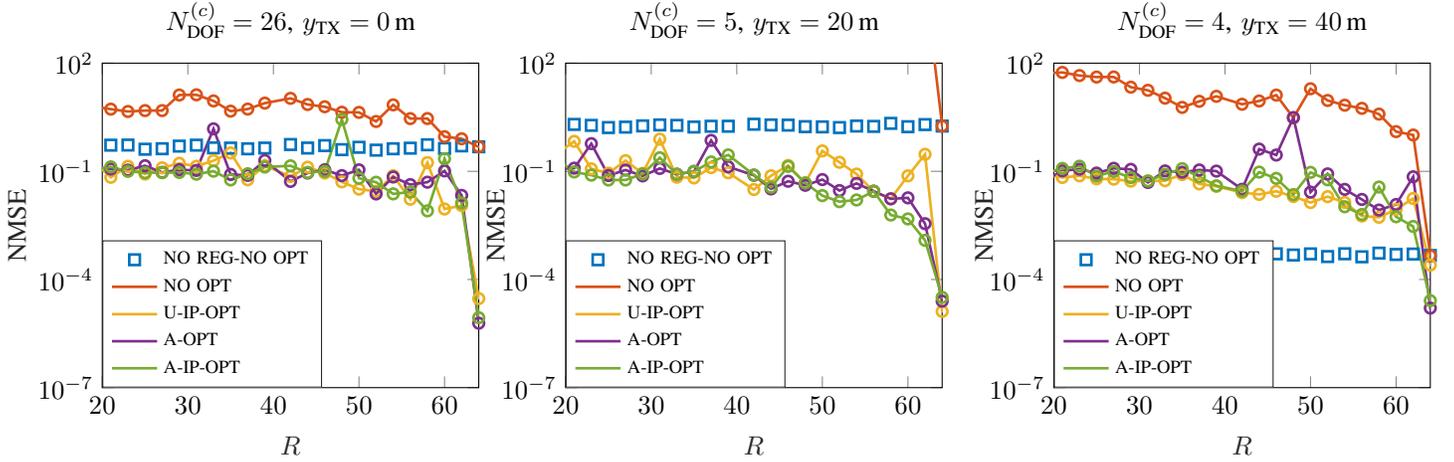

    \centering
    \hspace{-10mm}
    \begin{minipage}[t]{0.3\textwidth}
        \centering
        \input{Bistatic_0m}
    \end{minipage}%
 \hspace{8mm}
    \begin{minipage}[t]{0.3\textwidth}
        \centering
\input{Bistatic_20m}
    \end{minipage}
\hspace{8mm}
    \begin{minipage}[t]{0.3\textwidth}
        \centering
\input{Bistatic_40m}
    \end{minipage}
    \caption{\ac{NMSE} as a function of the truncation index $R$ selected when applying the \ac{TSVD} to $\Gr$ and different optimization techniques for the bistatic \ac{LOS} setup. In this case, the receiver is located at $(0,0, -2)\, \mathrm{m}$, while the transmitting device is placed at $(0, -y_{\text{TX}}, 2)\, \mathrm{m}$ and three different distances from the \ac{ROI} are tested, namely $y_{\text{TX}} \in (0, 20, 40 )\, \mathrm{m}$. Dotted ($\circ $) and continuous ($-$) lines correspond to the E-\ac{NMSE} and T-\ac{NMSE}, respectively, while the square markers denote the non-regularized case for benchmarking. $N_{\text{DOF}}^{(c)}$ refers to the illuminating channel $\Gt$.}
    \label{fig:bistatic_NF_vs_FF}
\end{figure*}

Let us consider a monostatic \ac{LOS} configuration as per Fig.~\ref{fig:monostatic_NORIS}, with the center of the TX/RX array set in $\left( 0,\, 0,\, 0 \right)\,\mathrm{m}$ and the \ac{ROI} center in $\left( 0,\, y_\text{ROI},\, 0 \right)\,\mathrm{m}$, with $y_\text{ROI} \in \left\{ 10,\,15,\,23\right\}\,\mathrm{m}$. These distances correspond to different channel \ac{DoF}, i.e., $N_{\text{DOF}}^{(c)} \in \{54,\, 26,\, 10\}$, evaluated as per \eqref{eq:N_DOF}.

In Fig.~\ref{fig:monostatic}, the curves depicting the E-\ac{NMSE} and T-\ac{NMSE} are presented for different values of the selected truncation index $R$, with $R \le K$. In this setup, it is $K=N=64$. 

Let us analyze this figure from left to right, progressing from a strong near-field to an almost far-field regime as the link distance between the TX/RX and the \ac{ROI} increases. Fig. \ref{fig:monostatic}-left corresponds to a specific scenario where regularization is unnecessary due to the full-rank nature of the channel matrix $\mathbf{G}_R$, resulting in good imaging performance ($N_{\text{DOF}}^{(c)}=54$). Similarly, as shown in Fig. \ref{fig:monostatic}-center, this geometric configuration still yields quite satisfactory imaging performance even in the absence of regularization and/or optimization. This outcome is attributed to the specific $N_{\text{DOF}}^{(i)}$ of the image evaluated through \ac{PCA} as specified above. Indeed, for the L-shaped image of Fig. \ref{fig:monostatic_images}-(top,left), $N_{\text{DOF}}^{(i)} = 2$, and hence, since $N_{\text{DOF}}^{(c)} \gg N_{\text{DOF}}^{(i)}$, regularization is unnecessary. Consequently, for truncation indices $R < N$, a decline in performance is observed for all the cases depicted through solid lines due to intrinsic information loss. In fact, when performing regularization, we are discarding $\Gr$ elements that are informative and contain valuable information for optimal \ac{ROI} image reconstruction. However, for $R = N$, some performance enhancement can still be obtained, as shown for the A-OPT, U-IP-OPT, and A-IP-OPT curves in both Fig. \ref{fig:monostatic}-left and Fig. \ref{fig:monostatic}-center. These improvements are mainly attributed to the illumination optimization, as opposed to the NO-OPT case (red curve and markers), which somehow allows us to obtain improvements in the image reconstruction procedure. Finally, Fig. \ref{fig:monostatic}-right depicts the results when the ROI is placed at a larger distance. In this case, provided that the number of channel's DOF decreases (i.e., $N_{\text{DOF}}^{(c)} \rightarrow N_{\text{DOF}}^{(i)}$), both regularization and optimization play a crucial role in improving the performance.

In fact, in this case, there is a trade-off between the joint effect of the thermal noise and image distortion leading to an optimal truncation index of $R=60$. 
Considering this last scenario and fixing $R=60$, Fig.~\ref{fig:monostatic_images} displays the estimated images, i.e., the $\hat{\boldsymbol{\gamma}}$ values, for the various optimization approaches. As anticipated, illumination design plays a pivotal role in enhancing performance for this configuration, and the proposed U-IP-OPT approach outperforms the others.

In Fig.~\ref{fig:ImageDOF}, we assess the imaging performance using a different reference image, characterized by an increased irregularity and complexity, having $N_{\text{DOF}}^{(i)} = 4$. Specifically, we examine the cases where \(y_{\text{ROI}} = 15\,\mathrm{m}\) and \(y_{\text{ROI}} = 23\,\mathrm{m}\), selecting \(R = 64\) and \(R = 58\) as the truncation indices corresponding to the minimum \ac{NMSE} values. The illustration includes the worst-case scenario (NO REG-NO OPT) and the one yielding improved reconstruction for this particular image, i.e., U-IP-OPT. Remarkably, the scenario corresponding to \(y = 23\,\mathrm{m}\), which provided satisfactory reconstructions in Fig.~\ref{fig:monostatic_images}, yields unsatisfactory results for this more complicated image having a higher $N_{\text{DOF}}^{(i)}$. This underscores the fundamental interdependence between $N_{\text{DOF}}^{(c)}$ and $N_{\text{DOF}}^{(i)}$. In particular, to reconstruct highly complex images characterized by a high $N_{\text{DOF}}^{(i)}$, a channel with a correspondingly high rank, and thus a large $N_{\text{DOF}}^{(c)}$ is essential,
e.g., $N_{\text{DOF}}^{(c)}$ = 26 as in Fig.~\ref{fig:monostatic}-center.

\subsection{Imaging Performance under Rician Fading Conditions} 
\label{subsec:rice}
Given that, in realistic radio environments, additional scattered paths coexist with the direct \ac{LOS} component, we have examined the monostatic setup under Rician fading conditions for the TX/RX-\ac{ROI} link of interest. In this specific scenario, we have incorporated the effects of the Rician fading as follows
\begin{equation}
 \boldy = \overline{\Gr} \,  \bm{\Gamma}\,  \overline{\Gt} \,  \boldx + \boldw   \, ,
\end{equation}
where
\begin{equation}
\overline{\Gt} =  \sqrt{\frac{\kappa_{\text{T}}}{\kappa_{\text{T}}+1}} \Gt + \sqrt{\frac{1}{\kappa_{\text{T}}+1}} \boldS_{\text{T}} \, ,
\end{equation}
with $\boldS_{\text{T}} =  \left\{ \stni \right\}\in \mathbb{C}^{N \times N_T}$ and $\stni \sim \cn (0, \left| \gtni\right|^2)$.
Similarly, it is
\begin{equation}
    \overline{\Gr} =  \sqrt{\frac{\kappa_{\text{R}}}{\kappa_{\text{R}}+1}} \Gr + \sqrt{\frac{1}{\kappa_{\text{R}}+1}} \boldS_{\text{R}} \, ,
\end{equation}
with $\boldS_{\text{R}} =  \left\{ \srrn \right\}\in \mathbb{C}^{\Nr \times N}$ and $\srrn \sim \cn (0, \left| \grrn\right|^2)$.
Given the monostatic setup, we set the Ricean factor $ \kappa_{\text{T}} = \kappa_{\text{R}}= \kappa \geq 0$ to be identical for the TX/RX-ROI link and the reciprocal one.
In Fig.~\ref{fig:fading}, we reported the numerical results for the monostatic setup, having the same geometrical parameters as per Sec.~VI-B and when the reference L-shaped image of Fig.~\ref{fig:monostatic_images} has to be reconstructed. The center of the TX/RX array is set in $\left( 0,\, 0,\, 0 \right)\,\mathrm{m}$ and the \ac{ROI} center is located in $\left( 0,\,  10,\, 0 \right)\,\mathrm{m}$. We assumed $\kappa \in \{2,5,100\}$ and, for each of these values, we considered the NO REG-NO OPT and A-IP-OPT cases. Notably, at lower values of $\kappa$, imaging performance is significantly influenced by strong fading. However, for $\kappa=100$, improvements are more pronounced, despite not achieving the same level of NMSE as depicted in Fig. 3-left under LOS conditions. For instance, in the case of NO REG-NO OPT for $\kappa=100$, the NMSE reaches approximately $10^{-3}$, whereas in Fig. 3-left, NMSE values around $10^{-7}$ are attained. As intuitively predictable, fading strongly impacts the image reconstruction process, given that our approach is tailored for LOS conditions.

\begin{figure*}[th!]
    \centering
\hspace{-10mm}
    \begin{minipage}[t]{0.3\textwidth}
        \centering
        \input{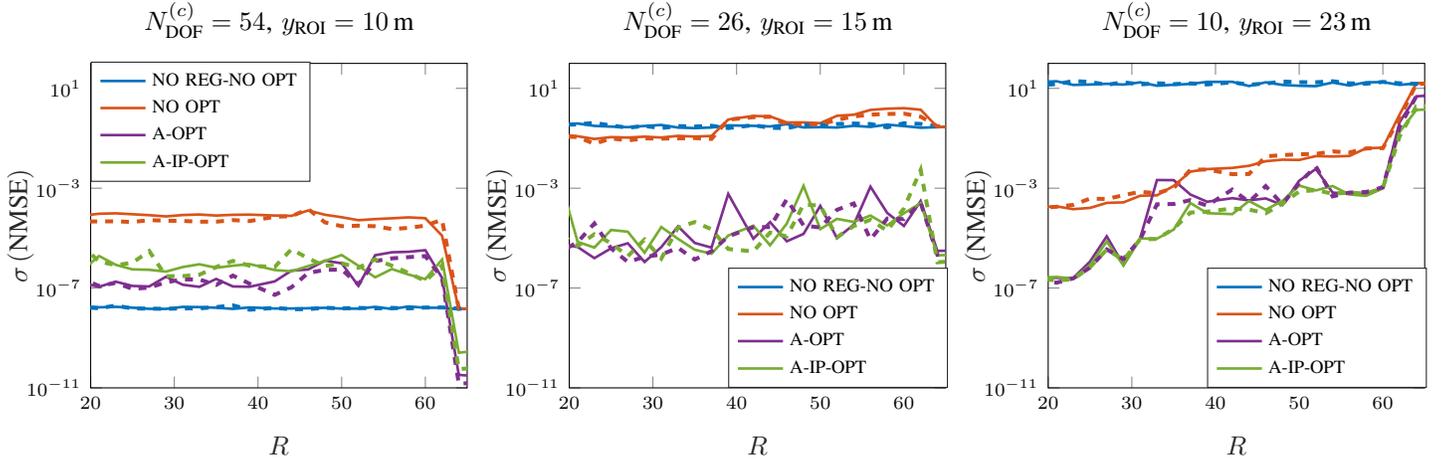}
\end{minipage}%
\hspace{8mm}
    \begin{minipage}[t]{0.3\textwidth}
        \centering
%
%
\definecolor{mycolor1}{rgb}{0.00000,0.44700,0.74100}%
\definecolor{mycolor2}{rgb}{0.85000,0.32500,0.09800}%
\definecolor{mycolor3}{rgb}{0.92900,0.69400,0.12500}%
\definecolor{mycolor4}{rgb}{0.49400,0.18400,0.55600}%
\definecolor{mycolor5}{rgb}{0.46600,0.67400,0.18800}%

\pgfplotsset{every axis/.append style={
		legend style={at={(0.99,0.99)}},legend cell align=left,font=\scriptsize,  title style={font=\normalsize}}
} %

\begin{tikzpicture}

\begin{axis}[%
ylabel style={yshift=-0.3cm},
width=0.92\linewidth,
scale only axis,
xmin=20,
xmax=65,
xlabel style={font=\color{white!15!black}},
xlabel={\shortstack{$R$\\\\(a)}},
xlabel={$R$},
ymode=log,
ymin=1e-11,
ymax=1e+2,
yminorticks=true,
ylabel style={font=\color{white!15!black}},
ylabel={$\sigma\left( \text{NMSE}\right)$},
title={$N_{\text{DOF}}^{(c)}=26$, $y_\text{ROI}=15 \,\text{m}$},
axis background/.style={fill=white},
legend style={font=\scriptsize,at={(0.422,0.0)},fill opacity=0.8, anchor=south west, legend cell align=left, align=left, draw=white!15!black}
]
\addplot [color=mycolor1, line width=1.0pt]
  table[row sep=crcr]{%
1	0.365653760451465\\
3	0.37296491347819\\
5	0.291362764669296\\
7	0.276464547808882\\
9	0.29406794825861\\
11	0.345917885109541\\
13	0.402400149888217\\
15	0.334565554742376\\
17	0.254860912114122\\
19	0.363895706673646\\
21	0.391275957778533\\
23	0.313718637973428\\
25	0.293064100352188\\
27	0.278968965769926\\
29	0.310520581380741\\
31	0.341722936219792\\
33	0.268417447126932\\
35	0.255100761699625\\
37	0.269983558039291\\
39	0.329054651254787\\
42	0.315186593632153\\
44	0.297678849732693\\
46	0.367687321102254\\
48	0.27225362076717\\
50	0.280068016669176\\
52	0.269404630360935\\
54	0.315134119484088\\
56	0.285048923642497\\
58	0.351670916473687\\
60	0.307790245056051\\
62	0.264258884471103\\
64	0.27821398996483\\
66	0.295041197457547\\
68	0.278959423476215\\
70	0.31924486665636\\
72	0.322060029303769\\
74	0.339858097069684\\
76	0.31175672944978\\
78	0.26267030814266\\
80	0.289446348637681\\
};
\addlegendentry{NO REG-NO OPT}

\addplot [color=mycolor2, line width=1.0pt]
  table[row sep=crcr]{%
1	0.00903793307858733\\
3	0.0088423865245913\\
5	0.00913808822202865\\
7	0.0615021201391508\\
9	0.0611107276707393\\
11	0.0812617091348595\\
13	0.0789366678133257\\
15	0.0906139148833047\\
17	0.198766923546104\\
19	0.137807068948021\\
21	0.120451635780151\\
23	0.0936241354792457\\
25	0.111312600574928\\
27	0.106565100077818\\
29	0.115917866974819\\
31	0.107153478403578\\
33	0.12493579091551\\
35	0.116835013468783\\
37	0.123320626566086\\
39	0.563743143855257\\
42	0.788190908587916\\
44	0.750402084849648\\
46	0.430264184584659\\
48	0.431268083982051\\
50	0.403273479594823\\
52	0.796998129507291\\
54	0.876638151188901\\
56	1.36689993955768\\
58	1.53246216582248\\
60	1.60412963059158\\
62	1.39715170683253\\
64	0.278213989964778\\
66	0.295041197457496\\
68	0.278959423476138\\
70	0.319244866656441\\
72	0.322060029303796\\
74	0.33985809706923\\
76	0.311756729450164\\
78	0.26267030814295\\
80	0.28944634863765\\
};
\addlegendentry{NO OPT}

\addplot [color=mycolor4, line width=1.0pt]
  table[row sep=crcr]{%
1	4.37482182316536e-07\\
3	4.40491428247679e-07\\
5	5.81126014227976e-07\\
7	8.4653730844265e-07\\
9	1.36096998444883e-06\\
11	9.37210575677717e-07\\
13	1.17844587338135e-05\\
15	1.14865175008703e-05\\
17	1.23837428566739e-06\\
19	5.78241013739755e-06\\
21	4.41035784635857e-06\\
23	1.57795321476707e-06\\
25	3.91039466730359e-06\\
27	6.20404701541939e-06\\
29	1.10552501298475e-06\\
31	4.37112222927612e-06\\
33	2.95842414097979e-05\\
35	6.40095669858493e-06\\
37	4.16234004605976e-06\\
39	0.000589803193887272\\
42	9.48145400988505e-06\\
44	4.38855806067089e-05\\
46	7.45714428558702e-06\\
48	1.47382054591617e-05\\
50	0.000195401905341141\\
52	9.92747058945039e-06\\
54	3.47852705837818e-05\\
56	0.00115029350358274\\
58	0.00010200161399836\\
60	3.06761569769952e-05\\
62	0.000307452088406779\\
64	3.06206058742259e-06\\
66	3.18098183817021e-06\\
68	2.99570437127241e-06\\
70	3.48099661656602e-06\\
72	3.63137051285059e-06\\
74	3.73882012182014e-06\\
76	3.36268723418542e-06\\
78	3.04861101415826e-06\\
80	3.18515272462543e-06\\
};
\addlegendentry{A-OPT}

\addplot [color=mycolor5, line width=1.0pt]
  table[row sep=crcr]{%
1	2.40343110786787e-07\\
3	1.35967594112725e-06\\
5	4.34989100944179e-06\\
7	6.24847170454167e-06\\
9	3.69624298902499e-06\\
11	1.60336358252988e-06\\
13	2.83482605438473e-05\\
15	1.72574446609149e-06\\
17	4.51879877965724e-06\\
19	0.00179170291466178\\
21	7.08733059190388e-06\\
23	4.17652297401688e-06\\
25	2.12273686066558e-05\\
27	1.80506750291817e-05\\
29	2.59838682647152e-06\\
31	2.77657692922489e-05\\
33	9.34940684962627e-06\\
35	3.42716331795781e-06\\
37	2.36596460916994e-06\\
39	1.26800588559474e-05\\
42	3.14632475268404e-05\\
44	5.1513784136192e-06\\
46	3.81983662875913e-05\\
48	0.00123486461264945\\
50	2.29578568360555e-05\\
52	5.90246912712569e-05\\
54	4.45121889703143e-05\\
56	8.01599862021618e-05\\
58	3.22682151465688e-05\\
60	9.93711843059027e-05\\
62	0.00025089139226246\\
64	2.05112525622226e-06\\
66	2.13397493952631e-06\\
68	1.9749661441762e-06\\
70	2.3296894786654e-06\\
72	2.40369809350678e-06\\
74	2.47896321183718e-06\\
76	2.24428916715886e-06\\
78	1.97918685638461e-06\\
80	2.12489638260075e-06\\
};
\addlegendentry{A-IP-OPT}

\addplot [color=mycolor1, dashed, line width=1.5pt]
  table[row sep=crcr]{%
1	0.255784961688011\\
3	0.260154396899096\\
5	0.229487960026291\\
7	0.317379641963987\\
9	0.274692081405313\\
11	0.302218512609797\\
13	0.381528125105386\\
15	0.377675881251103\\
17	0.310052210687482\\
19	0.340857014264692\\
21	0.359551821513779\\
23	0.414920600175867\\
25	0.323224768193036\\
27	0.266305206495846\\
29	0.309489488676378\\
31	0.256441764210684\\
33	0.307394827970947\\
35	0.294665513835099\\
37	0.316016865889125\\
39	0.256232657320292\\
42	0.334601144018814\\
44	0.263139624197706\\
46	0.254831964747007\\
48	0.344742268489899\\
50	0.33871886134333\\
52	0.373693136535064\\
54	0.389531630733891\\
56	0.372176585292512\\
58	0.273895962542436\\
60	0.400623870403571\\
62	0.375313820887764\\
64	0.298669229373582\\
66	0.330067268759581\\
68	0.287867018455231\\
70	0.364554858169617\\
72	0.349809104592781\\
74	0.34584101149208\\
76	0.398627723263478\\
78	0.305362115375464\\
80	0.333355111638027\\
};

\addplot [color=mycolor2, dashed, line width=1.5pt]
  table[row sep=crcr]{%
1	0.007070755328491\\
3	0.00668857954260902\\
5	0.00790646832309314\\
7	0.0671115347011215\\
9	0.0754906627980828\\
11	0.0797061421480454\\
13	0.0795277679931973\\
15	0.0753537135637354\\
17	0.201254136687099\\
19	0.117678312686093\\
21	0.112077373625139\\
23	0.0599614925506607\\
25	0.0989024121683467\\
27	0.0819907062250912\\
29	0.0990734885249215\\
31	0.0967559913526678\\
33	0.098971206627733\\
35	0.101996614004769\\
37	0.0940589322775993\\
39	0.557482141750463\\
42	0.700236072738267\\
44	0.727365378237634\\
46	0.36007893698107\\
48	0.369415532741929\\
50	0.358916567724078\\
52	0.549789006650865\\
54	0.7479651570694\\
56	0.888029037121323\\
58	0.940853812066924\\
60	0.969626597385198\\
62	0.736827756664604\\
64	0.29866922937296\\
66	0.330067268760525\\
68	0.287867018456197\\
70	0.364554858171486\\
72	0.349809104593157\\
74	0.34584101149064\\
76	0.398627723262125\\
78	0.305362115376363\\
80	0.333355111638714\\
};

\addplot [color=mycolor4, dashed, line width=1.5pt]
  table[row sep=crcr]{%
1	2.57021324418405e-07\\
3	4.21466875763131e-07\\
5	5.63506841142929e-07\\
7	6.18514885961883e-07\\
9	2.51182995408067e-06\\
11	5.63517348319255e-06\\
13	9.23558078594695e-06\\
15	6.63106689751868e-06\\
17	1.77093859560655e-06\\
19	2.87804183498758e-06\\
21	6.88053669839912e-06\\
23	3.86517481463477e-05\\
25	3.87843648633197e-06\\
27	1.11244453680812e-06\\
29	1.77486632165132e-06\\
31	2.19176768399004e-06\\
33	4.08442916603957e-06\\
35	1.38079844885289e-06\\
37	2.99260701107753e-06\\
39	1.1796862052753e-05\\
42	2.7369045760147e-05\\
44	2.78419669166473e-05\\
46	0.000361464922193519\\
48	3.21946160084305e-05\\
50	2.65217606638217e-05\\
52	0.00013595796157847\\
54	0.000243764484375854\\
56	8.88305696239771e-06\\
58	6.06727762209453e-05\\
60	5.79457566627801e-05\\
62	0.000257859049605222\\
64	2.13570981547944e-06\\
66	2.37392172033078e-06\\
68	2.02369166063483e-06\\
70	2.72036586439584e-06\\
72	2.5082805246773e-06\\
74	2.57008783944014e-06\\
76	2.86309483730151e-06\\
78	2.20284124238685e-06\\
80	2.47701546516427e-06\\
};

\addplot [color=mycolor5, dashed, line width=1.5pt]
  table[row sep=crcr]{%
1	1.4800161390476e-07\\
3	1.32843018556021e-06\\
5	1.03006522569469e-05\\
7	2.29852820965254e-05\\
9	1.83668120711042e-06\\
11	3.67860212191482e-06\\
13	5.76335710288359e-06\\
15	4.44221961653579e-06\\
17	2.26964341817058e-06\\
19	1.39292374289106e-05\\
21	4.89161523722842e-06\\
23	1.42783812259305e-05\\
25	6.25585079654346e-06\\
27	1.89339592999927e-06\\
29	8.76570583172391e-06\\
31	1.51828498115945e-06\\
33	2.64516825173303e-05\\
35	4.40898516611231e-05\\
37	2.27235935012315e-05\\
39	3.66486324836293e-06\\
42	2.93499623335543e-06\\
44	1.66453302231582e-05\\
46	1.17509185844462e-05\\
48	0.000222118617493787\\
50	0.000200897868504512\\
52	1.08990412477777e-05\\
54	3.03942398532744e-05\\
56	5.42341758901145e-05\\
58	1.96427007812815e-05\\
60	0.000155041963417577\\
62	0.00540885545376475\\
64	1.07609670470119e-06\\
66	1.20455255832781e-06\\
68	1.04603863220816e-06\\
70	1.34417520957924e-06\\
72	1.27292751061407e-06\\
74	1.2852962030165e-06\\
76	1.47025582771678e-06\\
78	1.13322624184722e-06\\
80	1.23448406410658e-06\\
};

\end{axis}

\end{tikzpicture}%
    \end{minipage}%
\hspace{8mm}
    \begin{minipage}[t]{0.3\textwidth}
        \centering
%
%
\definecolor{mycolor1}{rgb}{0.00000,0.44700,0.74100}%
\definecolor{mycolor2}{rgb}{0.85000,0.32500,0.09800}%
\definecolor{mycolor3}{rgb}{0.92900,0.69400,0.12500}%
\definecolor{mycolor4}{rgb}{0.49400,0.18400,0.55600}%
\definecolor{mycolor5}{rgb}{0.46600,0.67400,0.18800}%

\pgfplotsset{every axis/.append style={
		legend style={at={(0.99,0.99)}},legend cell align=left,font=\scriptsize,  title style={font=\normalsize}}
} %

\begin{tikzpicture}

\begin{axis}[%
ylabel style={yshift=-0.3cm},
width=0.92\linewidth,
scale only axis,
xmin=20,
xmax=65,
xlabel style={font=\color{white!15!black}},
xlabel={\shortstack{$R$\\\\(a)}},
xlabel={$R$},
ymode=log,
ymin=1e-11,
ymax=1e+2,
yminorticks=true,
ylabel style={font=\color{white!15!black}},
ylabel={$\sigma\left( \text{NMSE}\right)$},
title={$N_{\text{DOF}}^{(c)}=10$, $y_\text{ROI}=23 \,\text{m}$},
axis background/.style={fill=white},
legend style={font=\scriptsize,at={(0.422,0.0)},fill opacity=0.8, anchor=south west, legend cell align=left, align=left, draw=white!15!black}
]
\addplot [color=mycolor1, line width=1.0pt]
  table[row sep=crcr]{%
1	18.7489813976004\\
3	16.0668008464102\\
5	15.5640735907132\\
7	17.8051681895325\\
9	14.9013195166824\\
11	17.4780550017465\\
13	17.6296679845081\\
15	18.2778889014837\\
17	14.681171669742\\
19	17.9415434217727\\
21	19.2218125610127\\
23	13.7168114514549\\
25	14.2363978903338\\
27	15.2231039427396\\
29	14.1644807174723\\
31	17.391320146761\\
33	12.7463143192769\\
35	13.956267624731\\
37	15.05315304502\\
39	18.144765375854\\
42	18.2750200677658\\
44	12.3896524923339\\
46	17.4260979821667\\
48	13.7458066330515\\
50	12.7006143000024\\
52	12.2528820515926\\
54	17.143944820526\\
56	13.001645694587\\
58	17.4244941355841\\
60	17.7897141350255\\
62	13.7899147316869\\
64	15.7239565565216\\
66	16.8440481313478\\
68	12.4885287175434\\
70	15.4235128810432\\
72	17.1845182445564\\
74	15.3512047383903\\
76	14.7093455664402\\
78	13.9415425839102\\
80	13.6325534609656\\
};
\addlegendentry{NO REG-NO OPT}

\addplot [color=mycolor2, line width=1.0pt]
  table[row sep=crcr]{%
1	3.34302918318463e-05\\
3	3.56344280354417e-05\\
5	3.48872813223327e-05\\
7	0.000101482493283255\\
9	3.49075333624189e-05\\
11	3.66094604920682e-05\\
13	4.16344314635986e-05\\
15	3.91636163706095e-05\\
17	4.05694146694108e-05\\
19	0.0001717309770631\\
21	0.000187412209835584\\
23	0.000144475227161945\\
25	0.00016237575903847\\
27	0.000258055839246005\\
29	0.000270216280548627\\
31	0.000497735526626287\\
33	0.000517322738527421\\
35	0.00123106407686445\\
37	0.00462086634157396\\
39	0.00580868820384951\\
42	0.00649499013207089\\
44	0.00789687824281868\\
46	0.0120671678821191\\
48	0.0138828991023365\\
50	0.0135573141121463\\
52	0.0193076209395682\\
54	0.018617338106242\\
56	0.0195428105852069\\
58	0.0397461857666668\\
60	0.0420037724495834\\
62	0.893438359313223\\
64	15.7239565564863\\
66	16.8440481312905\\
68	12.4885287175309\\
70	15.4235128811437\\
72	17.1845182444518\\
74	15.351204738225\\
76	14.7093455663744\\
78	13.9415425840445\\
80	13.6325534609384\\
};
\addlegendentry{NO OPT}

\addplot [color=mycolor4, line width=1.0pt]
  table[row sep=crcr]{%
1	9.12301452914511e-07\\
3	3.48830952677872e-06\\
5	5.05211475862085e-06\\
7	5.28251617242149e-07\\
9	1.6676156541256e-07\\
11	2.49202322007011e-06\\
13	3.66592652286237e-07\\
15	1.14667962195418e-06\\
17	8.12721568477301e-05\\
19	2.60030770014793e-07\\
21	2.55841541768521e-07\\
23	2.5139006848827e-07\\
25	1.15000674604254e-06\\
27	1.18090149108113e-05\\
29	8.45582254453687e-07\\
31	9.9377457274258e-06\\
33	0.00212017114392667\\
35	0.00210260997471357\\
37	0.000556305691862298\\
39	0.000286121089045854\\
42	0.000397514261213606\\
44	0.00032427424494864\\
46	0.000389369370270399\\
48	0.000210977898155278\\
50	0.00198952214302251\\
52	0.00591775616204708\\
54	0.000661014587372197\\
56	0.000614890244561893\\
58	0.000572956533348234\\
60	0.00102941067322281\\
62	0.440747296051201\\
64	4.80637162132571\\
66	5.20509861691158\\
68	4.18794904794882\\
70	4.87560165577814\\
72	5.66181017889712\\
74	5.27416415317554\\
76	5.13395308028704\\
78	4.54168617219959\\
80	4.54363179521911\\
};
\addlegendentry{A-OPT}

\addplot [color=mycolor5, line width=1.0pt]
  table[row sep=crcr]{%
1	4.69190309692101e-06\\
3	1.69047607181702e-05\\
5	7.38482229180959e-06\\
7	1.92213004505805e-06\\
9	8.38959932401692e-08\\
11	2.64397923209537e-06\\
13	9.55585224925192e-08\\
15	1.93526045413868e-06\\
17	1.27795279671249e-06\\
19	2.59534636262074e-07\\
21	2.78036979717833e-07\\
23	2.40060425517582e-07\\
25	8.30364410195522e-07\\
27	7.25922064495536e-06\\
29	7.96397892316526e-07\\
31	9.64570123602759e-06\\
33	9.4472042189034e-06\\
35	2.20406420736123e-05\\
37	0.000255171960104342\\
39	9.96627470383978e-05\\
42	9.14476702402506e-05\\
44	0.000306191186697356\\
46	0.000136254480846048\\
48	0.000310918645706932\\
50	0.0012715151297362\\
52	0.000816541900459065\\
54	0.00121756998954432\\
56	0.00055953963028323\\
58	0.000494014496609333\\
60	0.000980542601713007\\
62	0.178526674711643\\
64	1.36113478770784\\
66	1.46472224010118\\
68	1.15899989405285\\
70	1.37013380450542\\
72	1.55731103045427\\
74	1.4218850271259\\
76	1.36214431325738\\
78	1.24890833741376\\
80	1.22953413170985\\
};
\addlegendentry{A-IP-OPT}

\addplot [color=mycolor1, dashed, line width=1.5pt]
  table[row sep=crcr]{%
1	15.7659584615695\\
3	14.5435988116175\\
5	13.2973424260971\\
7	17.0253457310505\\
9	14.7194089294065\\
11	15.1793280157785\\
13	14.7401159369601\\
15	18.6701757974312\\
17	14.3484102203109\\
19	13.5305105380457\\
21	14.7073035735083\\
23	19.6959681721309\\
25	18.0376639635183\\
27	13.9748201467553\\
29	17.5049480319012\\
31	14.3400910716313\\
33	16.0452238933409\\
35	14.7818383468982\\
37	15.8570803052783\\
39	14.3620203856104\\
42	17.4599776102446\\
44	13.9731325643135\\
46	15.5894108016634\\
48	18.7122313413282\\
50	17.6619519571153\\
52	17.5412430461181\\
54	19.2585668041486\\
56	15.8322301226445\\
58	15.3228656283639\\
60	18.1459044195515\\
62	15.9870645616045\\
64	14.9909972517476\\
66	15.8258480769486\\
68	14.3192093793756\\
70	17.3867611728577\\
72	18.0238789385552\\
74	17.3805069631288\\
76	18.7139450270512\\
78	17.2270867532487\\
80	16.4582344077411\\
};

\addplot [color=mycolor2, dashed, line width=1.5pt]
  table[row sep=crcr]{%
1	3.267086121032e-05\\
3	3.73838453945323e-05\\
5	4.24680770743727e-05\\
7	9.86390951192191e-05\\
9	3.52419944473205e-05\\
11	3.57669690126218e-05\\
13	3.27301978671704e-05\\
15	3.10361016561525e-05\\
17	3.27292273563208e-05\\
19	0.000162077100618623\\
21	0.000180261599366672\\
23	0.000214555797902475\\
25	0.000383023341850796\\
27	0.000367868489411055\\
29	0.000524647191388002\\
31	0.00071849252983527\\
33	0.000628344981876705\\
35	0.00100879674987222\\
37	0.00545393465105324\\
39	0.00561472724768065\\
42	0.00364503075836428\\
44	0.00369259331136338\\
46	0.0188611847087644\\
48	0.0211398345564953\\
50	0.0227709324294734\\
52	0.0236760736446118\\
54	0.0298039208053922\\
56	0.0269698719183052\\
58	0.0389884742706357\\
60	0.0401927554458709\\
62	0.633718895654434\\
64	14.99099725176\\
66	15.82584807704\\
68	14.3192093794084\\
70	17.3867611730723\\
72	18.0238789386167\\
74	17.3805069629545\\
76	18.7139450268977\\
78	17.227086753269\\
80	16.458234407763\\
};

\addplot [color=mycolor4, dashed, line width=1.5pt]
  table[row sep=crcr]{%
1	6.24543569892739e-07\\
3	3.15461041733962e-06\\
5	5.88153467133947e-06\\
7	4.08031856794976e-07\\
9	1.62745087706017e-07\\
11	1.44116339573355e-06\\
13	3.02328646660152e-07\\
15	1.15591245940809e-06\\
17	8.41079271146137e-05\\
19	2.11379105098906e-07\\
21	1.63380036314034e-07\\
23	2.44825217406983e-07\\
25	6.59308289354236e-07\\
27	3.52231299672001e-06\\
29	1.45898187595151e-06\\
31	9.13817606839073e-06\\
33	0.000226373059847391\\
35	0.000232803199087625\\
37	0.000333332941254116\\
39	0.000200984597792494\\
42	0.000878831974370478\\
44	0.000282377626631271\\
46	0.001213476702312\\
48	0.000223247847786431\\
50	0.00164581185346397\\
52	0.00748438672430058\\
54	0.000470126729417062\\
56	0.000666629460961237\\
58	0.000703287433567919\\
60	0.00109360849272808\\
62	0.289071324943332\\
64	4.80936195586592\\
66	5.27166453796656\\
68	4.85111269479381\\
70	5.4780829889545\\
72	5.92085791688968\\
74	5.78656228545393\\
76	6.23915971465155\\
78	5.75315565001325\\
80	5.43468289848101\\
};

\addplot [color=mycolor5, dashed, line width=1.5pt]
  table[row sep=crcr]{%
1	4.44486234002448e-06\\
3	1.14133759691854e-05\\
5	2.98624697020646e-06\\
7	3.23837739160866e-06\\
9	1.16902615124129e-07\\
11	5.33143054375878e-06\\
13	9.75873934289739e-08\\
15	1.01594791333685e-06\\
17	1.52360976568809e-06\\
19	2.34680225934624e-07\\
21	2.01275606298397e-07\\
23	2.25292095319048e-07\\
25	6.81284577938374e-07\\
27	3.47633204964786e-06\\
29	1.34197806027209e-06\\
31	8.98421800011467e-06\\
33	9.22955307951328e-06\\
35	2.38594485173523e-05\\
37	0.000105730662982823\\
39	0.000114792016447781\\
42	0.000153634179682241\\
44	0.0001390945115936\\
46	9.9069788614464e-05\\
48	0.000238624242880412\\
50	0.000741976233650273\\
52	0.000831317494866048\\
54	0.000600619521469203\\
56	0.000664350505692239\\
58	0.000608898841091363\\
60	0.00102739662217126\\
62	0.11781461896886\\
64	1.89883808589285\\
66	2.05520398186027\\
68	1.85483710826239\\
70	2.20498033225515\\
72	2.30676968385489\\
74	2.25318775568357\\
76	2.44157122856691\\
78	2.22563316247058\\
80	2.10179453933155\\
};

\end{axis}

\end{tikzpicture}%
    \end{minipage}%
    \caption{Standard deviation of the \ac{NMSE} as a function of the truncation index $R$ selected when applying the \ac{TSVD} to $\Gr$ and different optimization techniques for the monostatic \ac{LOS} setup. The TX/RX \ac{XL-MIMO} array is located at $(0,0, 0)\, \mathrm{m}$ and three distinct locations for the \ac{ROI} are tested, namely $y_{\text{ROI}} \in \{10, 15, 23\}\, \mathrm{m}$. The use of the random images dataset is indicated by dashed lines ($ - - $), while the use of a single reference image is denoted by solid lines ($-$).}
    \label{fig:std_dev}
\end{figure*}

\subsection{Imaging Performance for Randomly Generated Images Dataset}

To further generalize our findings, and hence mitigate potential biases stemming from the specific reference image chosen, we have conducted additional simulations mirroring the configuration depicted in Fig.~\ref{fig:monostatic}. This involved employing a monostatic \ac{LOS} setup with three distinct link distances ($y_\text{ROI} \in \left\{ 10,\,15,\,23\right\}\,\mathrm{m}$). In this additional analysis, we have generated a diverse dataset of images for reconstruction, all possessing the same \ac{DoF}, that is, $N_{\text{DOF}}^{(i)}=4$. Then, we executed $N_{\text{MC}}$ Monte Carlo experiments wherein each experiment utilized a different image from the dataset alongside a new realization of \ac{AWGN}. Consequently, at each iteration, the illumination signal optimization was repeated to accommodate the updated $\boldsymbol{\gamma}$ parameters corresponding to the current image.
For comparison, we also examined the scenario where a unique reference image is utilized, mirroring the approach adopted in Fig.~\ref{fig:monostatic}. However, in this instance, the selected reference image corresponds to the one illustrated in Fig.~\ref{fig:ImageDOF}, maintaining an equivalent number of degrees of freedom ($N_{\text{DOF}}^{(i)}=4$) as the images within the dataset.
 The findings of this study are illustrated in Fig.~\ref{fig:std_dev}, showcasing the standard deviation of the \ac{NMSE}, i.e., $\sigma\left( \text{NMSE}\right)$, as a function of the truncation index $R$ for both the image dataset case (dashed lines) and the single reference image case (continuous lines). The plot reveals a clear trend: as the distance between the TX/RX and the \ac{ROI} expands, there is a corresponding increase in the standard deviation across all depicted scenarios. Specifically, the blue markers representing the case with no regularization and no optimization (NO REG-NO OPT) exhibit substantial variability, ranging from approximately $10^{-8}$ for $y=10\,\mathrm{m}$ in Fig.~\ref{fig:std_dev}-left to around $10^1$ for $y=23\,\mathrm{m}$ in Fig.~\ref{fig:std_dev}-right. This variability underscores the significant error fluctuations observed when transitioning from a strong near-field condition to the far-field regime in the absence of regularization or illumination optimization. Conversely, the scenarios in which illumination optimization at the TX/RX is employed demonstrate minimal variations as the link distance increases. This highlights the crucial role of selecting the optimal \ac{ROI} illumination for ensuring high accuracy in the image reconstruction process and mitigating estimation errors. Remarkably, disregarding the case $y=10\,\mathrm{m}$, wherein the high rank of the channel renders both regularization and optimization futile, for the remaining two cases depicted in Fig.~\ref{fig:std_dev}-center and ~\ref{fig:std_dev}-right, it is distinctly observable how the standard deviation for cases A-OPT and A-IP-OPT is markedly lower compared to the non-optimized cases, hence highlighting the usefulness of the proposed approaches. This is particularly evident for the $y=15\,\mathrm{m}$ scenario, corresponding to a favorable propagation regime for our imaging purposes.
Apart from this, for a fixed link distance, when comparing the dataset scenario with the utilization of a single image, it becomes evident that we achieve nearly identical performance concerning $\sigma\left( \text{NMSE}\right)$, resulting in comparable —sometimes almost overlapping — accuracy levels in the estimation process. The primary disparity lies in the enhanced stability of the curves associated with the dataset employment, where the oscillations are slightly reduced compared to the single image case. However, this improvement comes at the expense of significantly heightened computational time (almost threefold) and complexity. Consequently, a trade-off emerges between performance and execution time.
Therefore, it is clear that the fundamental factor influencing the system's ability to reconstruct a given image more or less accurately is not so much associated with the number of \ac{ROI}'s images tested, but rather with the number of \ac{DoF} of the image itself that we can calculate through \ac{PCA}.

\begin{figure*}[th!]
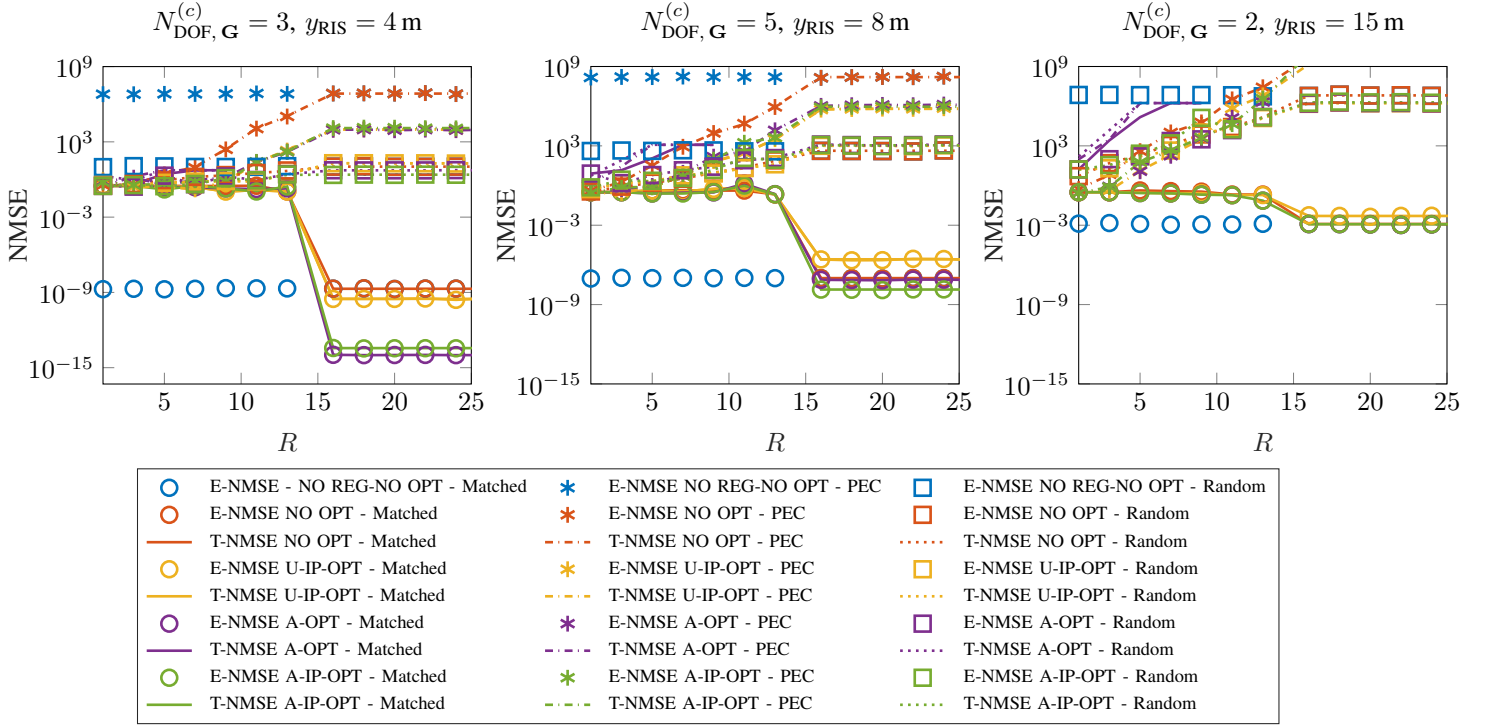

\hspace{-10mm}
\begin{minipage}{0.3\textwidth}
    \centering
\input{RIS_NMSEvsR_TX_4m}
\end{minipage}
\hspace{8mm}
\begin{minipage}{0.3\textwidth}
\centering
\input{RIS_NMSEvsR_TX_8m}
\end{minipage}
\hspace{8mm}
\begin{minipage}{0.3\textwidth}
\centering
\input{RIS_NMSEvsR_TX_15m_2}
\end{minipage}

\begin{minipage}{0.9\textwidth}
\qquad\,\,
%
%
\definecolor{mycolor1}{rgb}{0.00000,0.44700,0.74100}%
\definecolor{mycolor2}{rgb}{0.85000,0.32500,0.09800}%
\definecolor{mycolor3}{rgb}{0.92900,0.69400,0.12500}%
\definecolor{mycolor4}{rgb}{0.49400,0.18400,0.55600}%
\definecolor{mycolor5}{rgb}{0.46600,0.67400,0.18800}%
\begin{tikzpicture}
\pgfplotsset{ title style={font=\normalsize}}

\begin{axis}[%
hide axis,
    xmin=10,
    xmax=50,
    ymin=0,
    ymax=0.2,
    legend style={font=\scriptsize,draw=white!15!black,legend cell align=left, legend columns=3,legend style={draw=none,column sep=1ex},draw=white!15!black}
]

\addlegendimage{color=mycolor1, line width=1.0pt, only marks, mark size=3pt, mark=o, mark options={solid, mycolor1}}
\addlegendentry{E-NMSE - NO REG-NO OPT - Matched}

\addlegendimage{color=mycolor1, line width=1.0pt, only marks, mark size=3pt, mark=asterisk, mark options={solid, mycolor1}}
\addlegendentry{E-NMSE NO REG-NO OPT - PEC}

\addlegendimage{color=mycolor1, line width=1.0pt, only marks, mark size=3pt, mark=square, mark options={solid, mycolor1}}
\addlegendentry{E-NMSE NO REG-NO OPT - Random}


\addlegendimage{color=mycolor2, line width=1.0pt, only marks, mark size=3pt, mark=o, mark options={solid, mycolor2}}
\addlegendentry{E-NMSE NO OPT - Matched}

\addlegendimage{color=mycolor2, line width=1.0pt, only marks, mark size=3pt, mark=asterisk, mark options={solid, mycolor2}}
\addlegendentry{E-NMSE NO OPT - PEC}

\addlegendimage{color=mycolor2, line width=1.0pt, only marks, mark size=3pt, mark=square, mark options={solid, mycolor2}}
\addlegendentry{E-NMSE NO OPT - Random}

\addlegendimage{color=mycolor2, line width=1.0pt}
\addlegendentry{T-NMSE NO OPT - Matched}

\addlegendimage{color=mycolor2, dashdotted, line width=1.0pt}
\addlegendentry{T-NMSE NO OPT - PEC}

\addlegendimage{color=mycolor2, dotted, line width=1.0pt}
\addlegendentry{T-NMSE NO OPT - Random}

\addlegendimage{color=mycolor3, line width=1.0pt, only marks, mark size=3pt, mark=o, mark options={solid, mycolor3}}
\addlegendentry{E-NMSE U-IP-OPT - Matched}

\addlegendimage{color=mycolor3, dashdotted, line width=1.0pt, only marks, mark size=3pt, mark=asterisk, mark options={solid, mycolor3}}
\addlegendentry{E-NMSE U-IP-OPT - PEC}

\addlegendimage{color=mycolor3, dotted, line width=1.0pt, only marks, mark size=3pt, mark=square, mark options={solid, mycolor3}}
\addlegendentry{E-NMSE U-IP-OPT - Random}

\addlegendimage{color=mycolor3, line width=1.0pt}
\addlegendentry{T-NMSE U-IP-OPT - Matched}

\addlegendimage{color=mycolor3, dashdotted, line width=1.0pt}
\addlegendentry{T-NMSE U-IP-OPT - PEC}

\addlegendimage{color=mycolor3, dotted, line width=1.0pt}
\addlegendentry{T-NMSE U-IP-OPT - Random}

\addlegendimage{color=mycolor4, line width=1.0pt, only marks, mark size=3pt, mark=o, mark options={solid, mycolor4}}
\addlegendentry{E-NMSE A-OPT - Matched}

\addlegendimage{color=mycolor4, dashdotted, line width=1.0pt, only marks, mark size=3pt, mark=asterisk, mark options={solid, mycolor4}}
\addlegendentry{E-NMSE A-OPT - PEC}

\addlegendimage{color=mycolor4, dotted, line width=1.0pt, only marks, mark size=3pt, mark=square, mark options={solid, mycolor4}}
\addlegendentry{E-NMSE A-OPT - Random}

\addlegendimage{color=mycolor4, line width=1.0pt}
\addlegendentry{T-NMSE A-OPT - Matched}

\addlegendimage{color=mycolor4, dashdotted, line width=1.0pt}
\addlegendentry{T-NMSE A-OPT - PEC}

\addlegendimage{color=mycolor4, dotted, line width=1.0pt}
\addlegendentry{T-NMSE A-OPT - Random}

\addlegendimage{color=mycolor5, line width=1.0pt, only marks, mark size=3pt, mark=o, mark options={solid, mycolor5}}
\addlegendentry{E-NMSE A-IP-OPT - Matched}

\addlegendimage{color=mycolor5, line width=1.0pt, only marks, mark size=3pt, mark=asterisk, mark options={solid, mycolor5}}
\addlegendentry{E-NMSE A-IP-OPT - PEC}

\addlegendimage{color=mycolor5, line width=1.0pt, only marks, mark size=3pt, mark=square, mark options={solid, mycolor5}}
\addlegendentry{E-NMSE A-IP-OPT - Random}

\addlegendimage{color=mycolor5, line width=1.0pt}
\addlegendentry{T-NMSE A-IP-OPT - Matched}

\addlegendimage{color=mycolor5, dashdotted, line width=1.0pt}
\addlegendentry{T-NMSE A-IP-OPT - PEC}

\addlegendimage{color=mycolor5, dotted, line width=1.0pt}
\addlegendentry{T-NMSE A-IP-OPT - Random}
\end{axis}

\end{tikzpicture}%
\end{minipage}
\vspace{-1.5cm}
\caption{\ac{NMSE} as a function of the truncation index $R$ selected when applying the \ac{TSVD} to $\Gr$ and different optimization techniques for the \ac{NLOS} monostatic setup aided by a RIS. In this case, the TX/RX center is located at $(0,0, 0)\, \mathrm{m}$, while the \ac{RIS} is placed at $\left( 0,\, y_{\text{RIS}},\, 0 \right)\,\mathrm{m}$ with $y_{\text{RIS}} \in (4, 8, 15 )\, \mathrm{m}$. The legend delineates distinct \ac{RIS} configurations in each column, arranged sequentially from left to right as matched, \ac{PEC}, and random, respectively. 
}
    \label{fig:NLOSimaging_RIS}
\end{figure*}

\subsection{Bistatic LOS Imaging Performance}

We now assess the influence of the propagation regime on the effectiveness of the proposed optimization techniques for the illumination signal by considering a bistatic \ac{LOS} configuration. 
To this purpose, we locate the TX array center in $\left( 0,\, y_{\text{TX}},\, 2 \right)\,\mathrm{m}$, with $y_{\text{TX}} \in (0, 20, 40 )\, \mathrm{m}$ to test three distances from the \ac{ROI}, which corresponds to $N_{\text{DOF}}^{(c)} \in \{26,\, 5,\, 4\}$, with $N_{\text{DOF}}^{(c)}$ now referring to the TX-ROI link. The RX center was placed in $\left( 0,\, 0,\, -2\right)\,\mathrm{m}$, as per Fig.~\ref{fig:bistatic_LOS} and corresponding to $N_{\text{DOF}}^{(c)}=26$ for $\Gr$. In this setup, the \ac{ROI} was kept in a fixed position centered in $\left( 0,\, 15,\, 0\right)\,\mathrm{m}$. In the following, we will refer to the case where the transmitted image corresponds to the L-shaped picture as in Fig.~\ref{fig:monostatic_images}.

Fig.~\ref{fig:bistatic_NF_vs_FF} illustrates the E-NMSE and T-NMSE as a function of the truncation index $R$ and the distance between the TX and the \ac{ROI}. In these three cases, the best performance is obtained for the truncation index $R=64$. As it can be noticed, when moving from a strong near-field to almost far-field regime (from left to right), the impact of the optimization of the illuminating signal becomes decreasingly important because of the reduced $N_{\text{DOF}}^{(c)}$ of $\Gt$ available for the optimization process. This result indicates that the most critical channel for imaging purposes is $\Gr$, whose \ac{DoF} must be commensurate with the complexity of the image and, hence, it must be in the near-field. In contrast, imaging is still possible when the illuminating channel $\Gt$ experiences the far-field. However, the benefits of optimizing the illuminating signal are appreciable only when $\Gt$ is also in the near-field.

\subsection{\ac{RIS}-aided NLOS Imaging Performance} 
Finally, we analyze a monostatic \ac{NLOS} scenario aided by a \ac{RIS}. Specifically, we considered the system geometry of Fig.~\ref{fig:SystemScenariowithRIS}, where the \ac{XL-MIMO} TX/RX is placed in $\left(0,0,0\right)\, \mathrm{m}$, the \ac{RIS} is located in $\left( 0,\, y_{\text{RIS}},\, 0 \right)\,\mathrm{m}$, with $y_{\text{RIS}} \in (4, 8, 15 )\, \mathrm{m}$, while the \ac{ROI}'s center is positioned in $\left( 0,\, 0,\, -9.5\right)\,\mathrm{m}$. These distances correspond to $N_{\text{DOF},\, \boldG_2}^{(c)} \in \{25,\, 5,\, 2\}$ for the TX-\ac{RIS} link, $N_{\text{DOF},\,\boldG_1}^{(c)} \in \{3,\, 7,\, 13\}$ for the \ac{RIS}-\ac{ROI} link, hence resulting in $N_{\text{DOF},\, \boldG}^{(c)} \in \{3,\, 5,\, 2\}$ for the cascade channel, being $N_{\text{DOF},\, \boldG}^{(c)} = \min\left(N_{\text{DOF},\, \boldG_1}^{(c)}, N_{\text{DOF},\, \boldG_2}^{(c)} \right)$. In this setting, we considered a \ac{ROI} having the same size and L-shape image as before but with $N = 16$ scattering points, resulting in $\Delta = 420\lambda$.
Moreover, three distinct \ac{RIS} configurations are tested: (\textit{i}) \textit{matched}, which corresponds to the optimal configuration as per \eqref{eq:matchedRIS}, (\textit{ii}) \textit{\ac{PEC}}, i.e., the \ac{RIS} acts like a perfect reflecting mirror, (\textit{iii}) \textit{random}, according to which independent random phase shifts are imposed at each \ac{RIS} element.
Analyzing Fig.~\ref{fig:NLOSimaging_RIS}, it becomes evident that the matched case yields superior performance compared to the other two \ac{RIS} configurations, hence enabling the attainment of \ac{NMSE} values several orders of magnitude smaller.
This analysis underscores the fundamental importance of configuring the phase shift matrix $\mathbf{\Phi}$, governing the behavior of the signal reflected by the \ac{RIS}, to be perfectly adapted to wireless channels interconnecting the TX/RX to the \ac{ROI}. In particular, as depicted in Fig.~\ref{fig:NLOSimaging_RIS}-left, the A-IP-OPT case, initialized with the analytical solution found in Sec.~\ref{sec:OptimizProblem} and featuring the \ac{RIS} matched to both the incident (TX-RIS) and reflected (RIS-ROI) channels, facilitates the attainment of good imaging performance. Instead, no image reconstruction is possible when considering other \ac{RIS} configurations (i.e., \ac{PEC} or random).

Regarding the channels \ac{DoF}, $N_{\text{DOF},\, \boldG_2}^{(c)}$ substantially decreases as the \ac{RIS} approaches the far-field region, while $N_{\text{DOF},\, \boldG_1}^{(c)}$ progressively increases. This discrepancy arises because, in the former case, the TX-\ac{RIS} link approaches the far-field limit, where $N_{\text{DOF},\, \boldG_2}^{(c)} = 1$, whereas, in the latter case, the \ac{RIS} assumes a geometric configuration that enables more frontal, rather than oblique, illumination of the \ac{ROI}. Consequently, it is evident that:
(\textit{i}) in Fig.~\ref{fig:NLOSimaging_RIS}-left, despite the small $N_{\text{DOF},\, \boldG}^{(c)}$ for the cascade channel, the joint impact of the optimal illumination performed by the TX and the \ac{RIS} configuration tailored to the TX/RX-RIS and RIS-ROI channels leads to exceptionally low \ac{NMSE} values; (\textit{ii}) as the \ac{RIS} moves away and approaches the far-field region, as shown in Figs.~\ref{fig:NLOSimaging_RIS}-(center, right), optimizing the RIS illumination at the TX no longer provides substantial benefits due to the small $N_{\text{DOF},\, \boldG_2}^{(c)}$. Therefore, the RIS configuration plays a primary role in achieving low \ac{NMSE} values, which, when matched, still allows for low errors in reconstructing the \ac{ROI} scene. Given these considerations, the TX/RX-RIS channel $\boldG_2$ emerges as the most critical component within the cascade channel $\boldG$, constituting the principal bottleneck for the \ac{NLOS} imaging problem. In summary, in this \ac{NLOS} scenario, the key factor for an effective imaging reconstruction lies in optimizing the $\mathbf{\Phi}$ matrix of the \ac{RIS} to direct the reflected \ac{EM} beam accurately toward the \ac{ROI}.

\section{Conclusion}\label{sec: Conclusion}

We proposed a framework addressing the near-field imaging problem of a given \ac{ROI} in a \ac{XL-MIMO} communication scenario at millimeter-wave frequencies. 
Regularization techniques were applied to overcome ill-conditioning of the \ac{ISP}. A min-max optimization approach was introduced to find a suitable illumination waveform minimizing an upper bound of the \ac{MSE} on imaging estimation. Further, we derived the optimal \ac{RIS} configuration for handling \ac{NLOS} imaging scenarios. Numerical results demonstrated the feasibility of accurately estimating the \ac{ROI} scattering coefficients, emphasizing the crucial interplay of factors like the \ac{DoF} of the channels, system geometry (monostatic and bistatic), illumination optimization, \ac{RIS} configuration, and image complexity.

\begin{appendices}
    
\section{MSE Derivation}\label{app: appendix_a}
Assuming that the only random vector is $\mathbf{z}$ and by treating $\boldsymbol{\gamma}$ as an unknown deterministic vector, we can derive a closed-form expression of the \ac{MSE} by starting from the covariance matrix definition in~\eqref{eq: Cov_matrix}, i.e.,

 \begin{align}
 \boldC  &=\mathbb{E}\left\{   \left( 
 \tilde{\mathbf{X}}^{-1}\left( \mathbf{H}- \mathbf{I} \right)\tilde{\mathbf{X}}\, \bm{\gamma} \right) \left( 
 \tilde{\mathbf{X}}^{-1}\left( \mathbf{H}- \mathbf{I} \right)\tilde{\mathbf{X}}\, \bm{\gamma} \right)^H\right\} \nonumber\\
&+ \mathbb{E}\left\{   
 \tilde{\mathbf{X}}^{-1} \, \mathbf{z} \left( 
 \tilde{\mathbf{X}}^{-1}\left( \mathbf{H}- \mathbf{I} \right)\tilde{\mathbf{X}}\, \bm{\gamma} \right)^H \right\}  \nonumber\\
 &+\mathbb{E}\left\{ \left( 
 \tilde{\mathbf{X}}^{-1}\left( \mathbf{H}- \mathbf{I} \right)\tilde{\mathbf{X}}\, \bm{\gamma} \right)   \left( \tilde{\mathbf{X}}^{-1} \, \mathbf{z} \right)^H
  \right\} \nonumber\\
&+ \mathbb{E}\left\{\tilde{\mathbf{X}}^{-1} \, \mathbf{z}   \left( \tilde{\mathbf{X}}^{-1} \, \mathbf{z} \right)^H   \right\} \, ,
 \end{align}
where we recall that $\boldz= \tilde{\mathbf{G}}_{\text{R}}^{\dagger}\boldw =\boldV\tbsigma^\dagger \boldU^H  \boldw $.
By applying the expectation only to the random vectors and being $\mathbb{E}\left\{ \mathbf{z} \right\}=0$, $\mathbb{E}\left\{ \boldw\, \boldw^H \right\}= \sigma^2 \, \mathbf{I}$ and $\mathbf{U}\mathbf{U}^H= \mathbf{I}$, we obtain
\begin{align}\label{eq: C_computation_1}
\boldC
 &=\left( 
 \tilde{\mathbf{X}}^{-1}\left( \mathbf{H}- \mathbf{I} \right)\tilde{\mathbf{X}}\, \bm{\gamma} \right)  \left( 
 \tilde{\mathbf{X}}^{-1} \left( \mathbf{H}- \mathbf{I} \right)\tilde{\mathbf{X}}\, \bm{\gamma} \right)^H  \nonumber\\
 & \qquad +  \sigma^2\,\tilde{\mathbf{X}}^{-1} \boldV \tbsigma^\dagger\left(\tbsigma^\dagger\right)^H \boldV^H \left(\tilde{\mathbf{X}}^{-1}\right)^H  \, .
\end{align}

Let 
\begin{align}
    \mathbf{q}&= 
 \tilde{\mathbf{X}}^{-1}\left( \mathbf{H}- \mathbf{I} \right)\tilde{\mathbf{X}}\, \bm{\gamma} \\
    &=\left[ \begin{array}{c}
          (h_{11}-1)\gamma_1 + \sum_{i \neq 1} \tilde{x}_1^{-1} \, h_{1i} \tilde{x}_i\, \gamma_i\\
         (h_{22}-1)\gamma_2 + \sum_{i \neq 2} \tilde{x}_2^{-1} \, h_{2i} \tilde{x}_i\, \gamma_i \\
         \vdots \\
        (h_{NN}-1)\gamma_N + \sum_{i \neq N} \tilde{x}_N^{-1} \, h_{N,i} \tilde{x}_i\, \gamma_i
    \end{array}\right].
\end{align}
Then
\begin{align}
&\text{tr}(\mathbf{q} \mathbf{q}^H)=\sum_{n=1}^N
          \big\lvert (h_{n,n}-1)\gamma_n + \sum_{i=1 \atop i \neq n}^N \tilde{x}_n^{-1} \, h_{n,i} \tilde{x}_i\, \gamma_i \big\rvert^2 \, ,
\end{align}
where
\begin{align}
    h_{n,i}&=\left[\boldH\right]_{ni} = \left[\boldV \blambda \boldV^H \right]_{ni} = \left[\sum_{k=1}^{{K}} \omega_k^{-1} \boldv_k \boldv_k^H\right]_{ni}, 
\end{align}
with $\omega_k^{-1}$ being the elements along the diagonal of $\mathbf{\Lambda}$ as per~\eqref{eq:lambdamatrix}, i.e, its eigenvalues, and
$\mathbf{v}_k = [v_{1k}, v_{2k},\dots, v_{Nk}]^T$ being the \textit{k}-\textit{th} eigenvector of $\boldV$.

Let us now consider the second term in \eqref{eq: C_computation_1} related to the receiver noise. We can express it as
\begin{align}
    \boldN &=\sigma^2\,\tilde{\mathbf{X}}^{-1} \boldV \tbsigma^\dagger\left(\tbsigma^\dagger\right)^H \boldV^H \left(\tilde{\mathbf{X}}^{-1}\right)^H \nonumber\\
&= \sigma^2\,\tilde{\mathbf{X}}^{-1}  \left(\sum_{{k=1}}^{{K}} \frac{1}{\left({\omega_k}\,\xi_k\right)^2} \cdot \boldv_k \boldv_k^H\right)  \left(\tilde{\mathbf{X}}^{-1}\right)^H \, ,
\end{align}
and its trace is given by
\begin{align}
   \text{tr}\left(\boldN\right)&= \sum_{n=1}^N \frac{\sigma^2}{\lvert\tilde{x}_n \rvert^2} \left[\sum_{{k}=1}^{{K}} \frac{1}{{(\omega_k \xi_k)^2}} \mathbf{v}_k \mathbf{v}_k^H \right]_{n,n}\nonumber\\
 &= \sum_{n=1}^N \frac{\sigma^2}{\lvert \tilde{x}_n\rvert^2} \sum_{{k}=1}^{{K}} \frac{1}{{(\omega_k \xi_k)^2}} |v_{n,k}|^2 \, .
\end{align}
Finally, 
\begin{align}\label{eq:MSE_3}
    \MSE(\boldx) &= \text{tr}(\boldC) = \text{tr}(\boldq \boldq^H) + \text{tr}(\boldN),
\end{align}
resulting in \eqref{eq:MSE}.

\section{Optimization of the Illumination Transmit Signal}\label{app: appendix_b}

In this appendix, we solve the minimization problem in \eqref{eq:problem_simplified}.
To this end, we can write $g(\boldx)=\sum_{n=1}^N \frac{\alpha_n}{b_n}$, where $b_n=|\tilde{x}_n|^2$ and  $\alpha_n\triangleq \sigma^2 \, \sum_{k=1}^K\, \left( \omega_k\, \xi_k\right)^{-2} \lvert v_{n,k} \rvert^2$. Moreover, let us write $\tilde{x}_n$ as $\tilde{x}_n=\mathbf{g}_{\text{T},n}\mathbf{x}$, where $\mathbf{g}_{\text{T},n}$ denotes the $n$-\textit{th} row of $\mathbf{G}_{\text{T}}$.
Consequently, the sum $\sum_n\, b_n$ becomes $\sum_n\, b_n=\sum_n \lvert \tilde{x}_n \rvert^2=\sum_n  \tilde{x}_n \tilde{x}_n^* =\sum_n  (\mathbf{g}_{\text{T},n}\mathbf{x}) (\mathbf{g}_{\text{T},n}\mathbf{x})^H = \sum_n \mathbf{g}_{\text{T},n}  \mathbf{x} \mathbf{x}^H \mathbf{g}_{\text{T},n}^H  $. Hence, the optimization problem becomes 
\begin{align}
    b_n^\star&=\arg \minimize{b_n} \,g (b_n) \\
    &\text{s.t.\,\,}  \sum_n\, b_n \leq P \, ,
\end{align}
where $P$ represents the upper bound of the term $\sum_n \lvert \tilde{x}_n \rvert^2$ given by $P=\Pt \cdot \sum_n \mathbf{g}_{\text{T},n} \mathbf{g}_{\text{T},n}^H  $.

By considering the \ac{KKT} conditions, the solution is given by
\begin{equation}
b_n=\frac{P \sqrt{\alpha_n}}{\sum_n \sqrt{\alpha_n}},
\end{equation}
and it corresponds to the following MSE expression associated to the noise presence only
\begin{equation}
\text{MSE}_{\text{(noise)}}^{\text{opt}}=\frac{1}{P} \left | \sum_n \sqrt{\alpha_n} \right |^2 \,.
\end{equation}

\end{appendices}

\section*{Acknowledgments}

G. Torcolacci and D. Dardari are with the Department of Electrical, Electronic, and Information Engineering “Guglielmo Marconi” - DEI-CNIT, University of Bologna, Cesena, Italy (e-mail: \{g.torcolacci, davide.dardari\}@unibo.it). 
H. Zhang and Q. Yang are with the School of Communication and Information Engineering, Nanjing University of Posts and Telecommunications, Nanjing, China (e-mail: \{haiyang.zhang, 2020010207\}@njupt.edu.cn).\\
A. Guerra and F. Guidi are with the National Research Council of Italy, Institute of Electronics, Computer and Telecommunication Engineering, Bologna, Italy (e-mail: \{anna.guerra, francesco.guidi\}@cnr.it). \\		 
Y. C. Eldar is with the Faculty of Math and CS, Weizmann Institute of Science, Rehovot, Israel (e-mail:  yonina.eldar@weizmann.ac.il). \\
This work was partially supported by the European Union under the Italian National Recovery and Resilience Plan (NRRP) of NextGenerationEU, partnership on “Telecommunications of the Future” (PE00000001 - program “RESTART”), by the EU Horizon project TIMES (Grant no. 101096307), and by ERC-STG-2023 project CUE-GO (Grant no. 101116257). Giulia Torcolacci was funded by an NRRP Ph.D. grant.

\bibliographystyle{IEEEtran}
\bibliography{StringDefinitions,IEEEabrv, References}

\end{document}